\documentclass[12pt,preprint]{aastex}
\usepackage{lscape}

\begin{document}

\newcommand{\feh}{[Fe/H]}
\newcommand{\ebv}{$E(B-V)$}
\newcommand{\bvz}{$(B-V)_0$}
\newcommand{\dmz}{$(m-M)_0$}
\newcommand{\vint}{$\langle V\rangle_\mathrm{int}$}
\newcommand{\bint}{$\langle B\rangle_\mathrm{int}$}
\newcommand{\bvmag}{$\langle B-V\rangle_\mathrm{mag}$}
\newcommand{\bvs}{$\langle B-V\rangle_\mathrm{s}$}
\newcommand{\bvint}{$\langle B-V\rangle_\mathrm{int}$}
\newcommand{\mvrr}{$M_V(\mathrm{RR})$}
\newcommand{\meanmvrr}{$\langle M_V(\mathrm{RR})\rangle$}

\title{TOWARDS A BETTER UNDERSTANDING OF THE DISTANCE SCALE FROM
RR LYRAE VARIABLE STARS: A CASE STUDY FOR THE INNER HALO 
GLOBULAR CLUSTER NGC~6723\altaffilmark{1} }

\author{Jae-Woo Lee\altaffilmark{2,5,7}, Mercedes L\'opez-Morales\altaffilmark{3},
Kyeong-Soo Hong\altaffilmark{4,7},Young-Woon Kang\altaffilmark{2},
Brian L.\ Pohl\altaffilmark{5}, Alistair Walker\altaffilmark{6}}

\altaffiltext{1}{Partially based on observations 
made with telescopes operated by the SMARTS consortium.}
\altaffiltext{2}{Department of Astronomy and Space Science,
Sejong University, 98 Gunja-Dong, Gwangjin-Gu, Seoul, 143-747, Korea;
jaewoolee@sejong.edu}
\altaffiltext{3}{Harvard-Smithsonian Center for Astrophysics, 60 Garden Street,
Cambridge, MA, 02138, USA}
\altaffiltext{4}{Korea Astronomy and Space Science Institute, Daejon 305-348, Korea}
\altaffiltext{5}{Department of Physics and Astronomy, University of North Carolina,
Chapel Hill, NC 27599-3255, USA}
\altaffiltext{6}{Cerro Tololo Inter-American Observatory, Casilla 603, La Serena, Chile}
\altaffiltext{7}{Visiting Astronomer, Cerro Tololo Inter-American Observatory,
National Optical Astronomy Observatories, which are operated by the Association of
Universities for Research in Astronomy, Inc., under contract with the
National Science Foundation.}

\begin{abstract}
We present $BV$ photometry for 54 variables
in the metal-rich inner halo globular cluster NGC~6723.
With the discovery of new RR Lyrae variables (RRLs), we obtain
$\langle$P$_{ab}$$\rangle$ = 0.541 $\pm$ 0.066  and
$\langle$P$_{c}$$\rangle$ = 0.292 $\pm$ 0.030 day,  
n(c)/n(ab+c) = 0.167, and
$\langle V\mathrm{(RR)}\rangle_\mathrm{int}$ = 15.459 $\pm$ 0.055.
We carry out the Fourier decomposition analysis and
obtain \feh$_\mathrm{ZW}$ = $-$1.23 $\pm$ 0.11 and
\ebv\ = 0.063 $\pm$ 0.015 for NGC~6723.
By calibrating the zero-point from the recent absolute
trigonometric parallax measurements for RR Lyr,
we derive the revised \mvrr-\feh\ relation, providing
\mvrr\ = 0.52 at \feh\ = $-$1.50 
and $(m-M)_0$  = 18.54 for Large Magellanic Cloud (LMC),
in excellent agreement with others.
We obtain $(m-M)_0 = 14.65 \pm 0.05$, equivalent to the distance
from the Sun of 8.47 $\pm$ 0.17 kpc, for NGC~6723
from various distance measurement methods using RRLs.
We find that RRLs in NGC~6723 do not have
magnitude dependency on the radial distance,
due to a not severe degree of the apparent crowdedness.
Finally, we show that there exists a relation between the degree of photometric 
contamination and the apparent crowdedness of the central region 
of globular cluster systems, $|\bar{\mu} - \mu_r| \propto \log \rho_c d^2$.
The use of this relation can play a significantly role on mitigating the discrepancy
to establish a cosmic distance scale using RRLs
in resolved stellar populations in the near-field cosmology.
\end{abstract}


\section{INTRODUCTION}
Understanding the formation and evolution of our Galaxy has always been 
one of the key quests in modern astrophysics for decades \cite[e.g.][]{freeman02}. 
Since RRLs are easily identifiable and
they can provide a powerful means to probe the chemical compositions 
and dynamical properties of the old stellar populations,
RRLs in the Galactic globular cluster (GC) systems or in the field 
are of particular importance to address the question of the early history
of our Galaxy \cite[see, for example,][]{smith95}.
Also, being a primary distance indicator,
the distance to RRLs can be accurately measured and
RRLs can help to yield important insights into the structure of our Galaxy.
Recent studies by \citet{drake13} or \citet{pietrukowicz} are excellent
examples of making use of RRLs to understand the substructures
in the Galactic halo expected from the theory of the hierarchical structure formation,
and to delineate the bar structure in the central part of our Galaxy.

With the advent of large aperture telescopes, the utility of RRL will
increase further in the future in order to learn more about 
the early buildup of our Galaxy or nearby galaxies.
For example, the Large Synoptic Survey Telescope (LSST) will be capable of
detecting RRLs in the local group of galaxies out to $\approx$ 1 Mpc scale
in several years of time span \citep{lsst}, providing us wonderful opportunities
to investigate the substructures of not only the outer halo in our Galaxy
but also the nearby dwarf galaxies.
Furthermore, the future 30- to 40-m class telescopes 
will have the capability of detecting RRLs in nearby groups of galaxies
and the utility of RRLs not only as a primary distance indicator 
but also as an old stellar population tracer will become more important.
In doing so, we need to calibrate the absolute dimensions of
physical parameters of RRLs in our Galaxy using well understood
samples of RRLs and then we can apply derived transformation relations of RRLs,
such as a period-luminosity-metallicity relation,
to establish a cosmic distance scale.

In recent years, the Fourier decomposition analysis of RRLs has become very popular,
since this method can provide an efficient means
to investigate fundamental stellar parameters,
such as metallicity, absolute magnitude, intrinsic color, temperature, etc
\citep[see for example,][]{simon81, kovacs96, jurcsik98, kk98, kovacs99,
kovacs01, kk07}.
The practical application of Fourier parameters to characterize
the RRLs in our Galaxy or in the Large Magellanic Cloud
appears to be successful, but some worrisome signs
also started to emerge lately.
In their study of RRLs in the globular cluster M3,
\citet{cacciari05} noticed that some fundamental physical parameters,
such as the intrinsic color or the absolute visual magnitude, from Fourier or
pulsational parameters do not agree with the observed ones.

Naturally, both the intrinsic color and the absolute visual magnitude
of RRLs can affect the derived distance scale of RRLs.
The uncertainty in the intrinsic color of RRLs can affect
the foreground interstellar reddening value estimates and,
subsequently, the distance scale of RRLs under consideration will be affected too,
although defining a proper color of RRLs, especially that of RRab type variables
with a large pulsational amplitude,
is not a trivial task \citep[e.g.][]{carney92}.
Achieving an accuracy of a 0.1 mag level for the absolute magnitude of RRLs
is non-trivial and has been a difficult task for decades \citep{smith95}.
Putting aside the discontinuity in the metallicity-luminosity relation
of RRLs, $M_V$ = $a$ + $b$[Fe/H], between the Oosterhoff I and II GCs 
seen in our Galaxy \citep{lc99b},
the zero-point $a$ in the above relation is poorly determined
and is a major source of uncertainties in the distance measurement based on RRLs.
Note that an uncertainty of 0.1 mag in the absolute magnitude results
in a $\approx$ 5\% error in the distance measurement.
The recent results from the absolute trigonometric parallaxes of the five
nearby RRLs using Fine Guidance Sensor (FGS) 
on Hubble Space Telescope (HST) prefer a brighter absolute 
visual magnitude level of RRL than those previously known, 
with $M_V$ = 0.45 $\pm$ 0.05 at [Fe/H] = $-$1.50 dex\footnote{
It should be noted that the field RRLs also appear to show
the Oosterhoff dichotomy, which is thought to be originated from
the hierarchical formation of our Galaxy \citep{lc99b}.
If we consider metallicity as a first approximation for the Oosterhoff classes
of the field RRLs at the boundary of [Fe/H] $\approx$ $-$1.6 dex, 
XZ Cyg, UV Oct and RR Lyr correspond to the Oosterhoff group I  
while RZ Cep and SU Dra correspond to the Oosterhoff group II.}
\citep{benedict11}, which is 0.07 mag brighter than those from
others, i.e.\ $M_V$ = 0.52 mag at [Fe/H] = $-$1.5 dex
\citep[e.g. see][and references therein]{walker12,cacciari13}.
Hence, if the recent measurement by \cite{benedict11} are true,
the previous calibrations from the Fourier parameters or  
the theoretical horizontal-branch (HB) models
based on the fainter absolute visual magnitude of RRL are in error.
It would be very desirable to re-calibrate the zero-point in old transformation
relations and to set up new distance scale of RRLs.
It is expected that the zero-point of the absolute visual magnitude and
the accuracy of the measurement will be dramatically improved with Gaia
during the next decade \citep[e.g.][]{cacciari09}.

This is the first of a series of papers addressing the distance scale from RRLs.
In this paper, we present a high precision $BV$ CCD photometry of RRLs in NGC~6723.
Our main goal is to study the RRL population in NGC~6723
and scrutinize the distance scale from RRLs of the cluster
by employing various methods.
In section 2, we address the interesting aspects of the metal-rich
inner halo GC NGC~6723.
In section 3, data acquisition and data reductions are discussed,
including our photometric calibration and the color-magnitude diagram (CMD).
Section 4 describes the methods of period searching and the methods of
searching for new variables.
In section 5, we discuss the fundamental physical properties of RRLs
from Fourier and pulsational parameters.
In particular, we discuss the distance scale from RRLs in detail
and we derive the new zero-point correction terms for the absolute visual
magnitude of RRLs.
In section 6, we show a relation between the degree of photometric
contamination of RRLs
and the apparent crowdedness of the GC systems.
In the appendix, we discuss the evolutionary effect and the photometric
contamination, such as blending, on the three RRLs in NGC~6723, V14, V15 and V16.
We also show that two $\delta$ Sct variable stars
and five W UMa type eclipsing binaries are off-cluster field stars
based on the distance from the center of NGC~6723 and the true distance
from the Sun.

\section{THE METAL-RICH INNER HALO GC NGC~6723}
NGC~6723 is an old metal-rich GC
located $\approx$ 2.6 kpc from the Galactic center \citep{harris96}.
It has long been thought that NGC~6723 is a genuine Galactic bulge
GC owing to its rather high metallicity and 
its proximity to the Galactic center.
For example, \citet{vdB93} claimed that NGC~6723 is 
on a circular orbit about the Galactic center and 
it is a true resident of the central region of the Galaxy. 
However, by employing the absolute proper motion study of the cluster,
\citet{dinescue03} showed that NGC~6723 has a highly inclined polar orbit.
They suggested that the kinematics of NGC~6723 is more likely that of a halo object 
and is not likely a member of the rotationally supported system,
although the bulge/disk/halo decomposition is not a trivial task
in the central region of our Galaxy.

Nevertheless, several aspects still make NGC~6723 a very interesting
metal-rich inner halo GC in investigating the formation of the Galactic inner halo:
(i) It is metal-rich (\feh\ $\approx -$1.2) and its metallicity is
similar to (or slightly more metal-poor than) the mean metallicity of RRLs
in the Baade's Window \citep{walker91,kunder08,pietrukowicz}.
(ii) It is very old \citep{fulltonphd,ACS}.
(iii) It is rich in the RRL population \citep{carney92}.
(iv) It suffers only small amount of foreground interstellar reddening,
\ebv\ = 0.05 mag \citep{harris96} and, therefore, is very suitable
to investigate photometric parameters of RRLs.
(v) Since NGC~6723 has been confined in the Galactic bulge (or the inner halo)
region and never left the bulge (or the inner halo) 
due to lower total kinetic energy \citep{dinescue03},
NGC~6723 can play a role as the GC counterpart
for the numerous field inner halo RRLs in Baade's Window.
Therefore, one may infer the formation epoch of 
the inner halo RRLs from that of NGC~6723, whose RRLs share common physical properties 
as the inner halo RRLs.

In spite of its importance, only a few
studies have been done for the cluster.
\citet{menzies74} studied the variable star population in NGC~6723 and he
presented $BV$ light curves for 29 RRLs.
He suggested that NGC~6723 has a heavily populated blue HB.
\citet{ldz94} also found that NGC~6723 has a rather blue HB morphology, 
$(B-R)/(B+V+R)$\footnotemark[1]\footnotetext[1]{$B$, $V$, and $R$ 
represent the numbers of blue HB stars, RRLs, 
and red HB stars, respectively.} 
= $-$0.08 $\pm$ 0.08, leading them to conclude that NGC~6723 
is a very old GC. 
\citet{rosenberg99} carried out a homogenous $VI$ photometry study
of 34 GCs. They measured the magnitude difference
between the turn-off (TO) and HB, $\Delta V_{TO}^{HB}$, and
reached the same conclusion that NGC~6723 is one the oldest GCs
in our Galaxy.
Various studies by others \citep{fulltonphd,fullton96,alcaino99,ACS}
using the model isochrone fitting method, for example,
also showed that NGC~6723 has a very old age.

The previous metallicity measurements of NGC~6723 suggest that
the metallicity of the cluster ranges from \feh\ $\approx$ $-$0.7 
to $\approx$ $-$1.4.
Smith (1981 and references therein) measured the metallicity of 
NGC~6723 using the $\Delta S$ method of five RRLs 
and he obtained \feh\ $\approx$ $-$0.7 on Butler's abundance scale.
\citet{zw84} derived \feh\ = $-$1.09 from 
the CMD morphology and the Q39 integrated light index.
Later, \citet{smith86} obtained \feh\ = $-$1.14
from the DDO photometry system and 
\citet{geisler86} obtained \feh\ = $-$1.35 from
the Washington photometry system.
Later, \citet{fullton96} performed a high resolution
spectroscopic study of three red-giant branch (RGB) stars
and they obtained \feh\ = $-$1.26 $\pm$ 0.09 for NGC~6723.
They also found that NGC~6723 appears to have 
enhanced $\alpha$-elemental abundances, consistent with its old age.
\citet{rutledge97} measured the metallicity of the cluster using the Ca~II triplet lines 
of the RGB stars in near infrared passband and they obtained
\feh\ = $-$1.12 $\pm$ 0.07 on the Zinn \& West's abundance scale
and $-$0.96 $\pm$ 0.04 on the Carretta \& Gratton's abundance scale 
\citep{zw84,cg97}.

\section{OBSERVATIONS AND DATA REDUCTION}
Our observations of the cluster were made with seven runs (29 nights in total)
spanning nearly ten years, from August 2002 to July 2012,
using the CTIO 0.9m and the CTIO 1.0m telescopes.
The journal of observations is presented in Table~\ref{tab:obs}.
Also shown in the table are the number of observations per filter and per night.
The CTIO 0.9m telescope was equipped with the Tektronix 2048 No. 3 CCD,
providing a plate scale of 0.40 arcsec pixel$^{-1}$
and a field of view of 13.5 $\times$ 13.5 arcmin.
We used the CTIO 0.9m telescope for four runs in 2002, 2004 and 2012.
The CTIO 1.0m telescope was equipped with an STA 4k$\times$4k CCD camera,
providing a plate scale of 0.289 arcsec pixel$^{-1}$ and a field of view of
20 $\times$ 20 arcmin.
We used the CTIO 1.0m telescope for three runs in 2008, 2009 and 2010.
The typical exposure times for the cluster were 300 s for Johnson $B$-band and
200 s for $V$-band filters for most runs.
In total, we collected 383 frames in $B$ and 401 frames in $V$.

The raw data were processed using the standard
IRAF\footnotemark[2] \footnotetext[2]{IRAF (Image Reduction
and Analysis Facility) is distributed by the National Optical
Astronomy Observatories, which are operated by the Association of
Universities for Research in Astronomy, Inc., under contract with the
National Science Foundation.} packages.
The raw image frames were trimmed and bias-corrected.
For 2002 and 2004 seasons, dome and sky flat images were applied to remove
pixel to pixel variations and illumination corrections using 
sky flats were applied to eliminate the large scale structure.
For other seasons, only sky flat images were applied.

Since both telescopes are equipped with an iris type shutter,
the illumination across the CCD chips is not uniform and 
the shutter shading correction is non-trivial especially for short exposures.
Therefore, we applied the shutter shading correction for science frames with
less than 10 s exposure time.
During the 2002 season, we obtained the series of dome flats with the
exposure times of 1 s and 60 s using the CTIO 0.9m telescope.
We normalized these two sets of dome flats and then
we calculated the residual in the 60 s dome flats in the time unit
and calculated shutter delay time across the CCD chip.
The exposure time with the CTIO 0.9m telescope is $\approx$ 70 ms longer
in the center than the upper-left or lower-right corners of the CCD chip
and the shutter delay time is not uniform across the chip.
For the CTIO 1.0m telescope, we also obtained shutter delay time correction
images in each of three individual seasons.
The shutter delay time for the CTIO 1.0m telescope is slightly shorter
than that of the CTIO 0.9m telescope.
The exposure time with the CTIO 1.0m telescope is $\approx$ 50 ms longer
at the center of the CCD chip.
We also found that the shutter delay time for the CTIO 1.0m telescope
does not appear to have any temporal variations.

\begin{figure}
\epsscale{1}
\figurenum{1}
\plotone{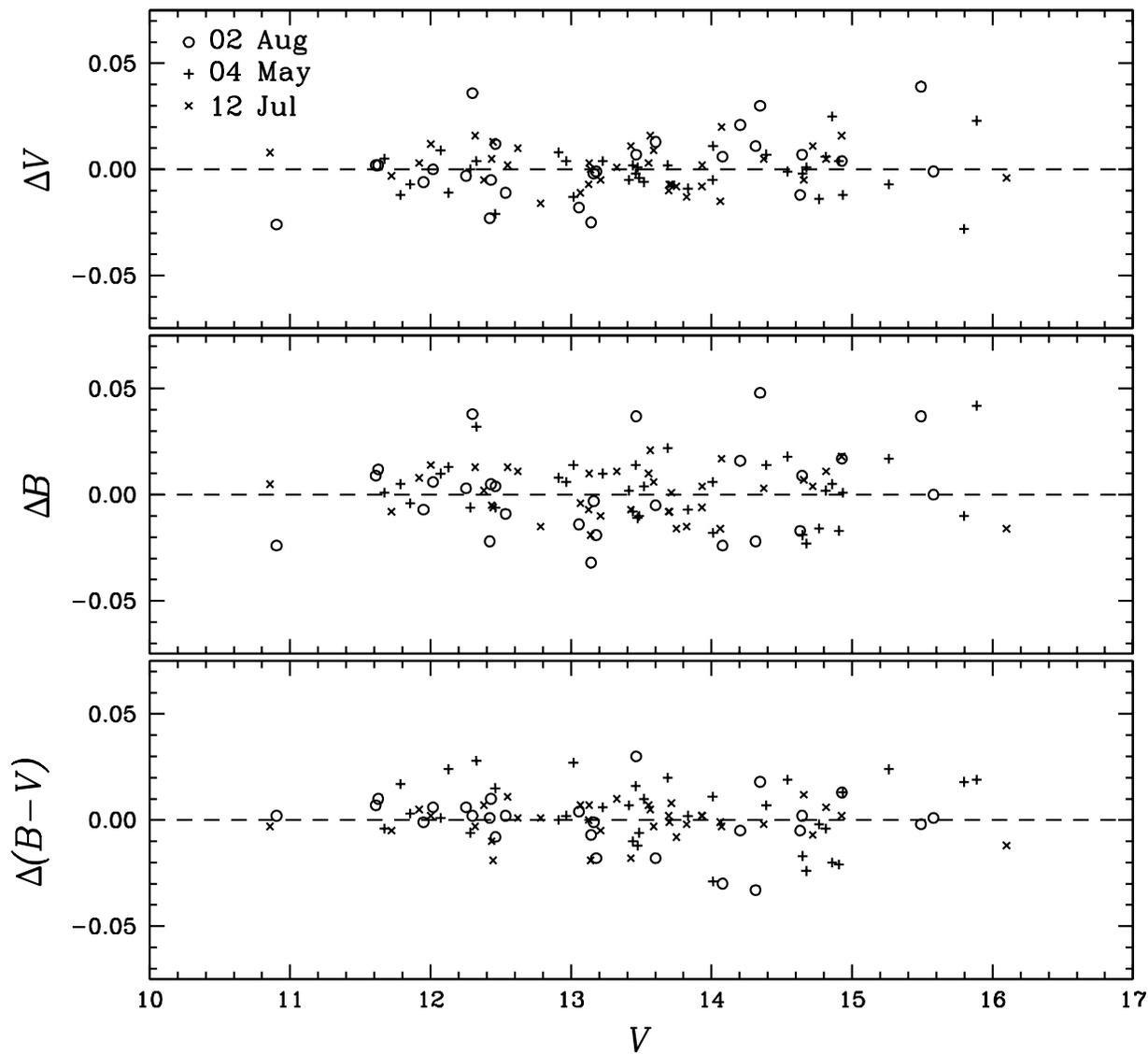}
\caption{Transformation residuals of Landolt standard stars as a
function of the magnitude. The differences are in the sense
Landolt minus our work.}
\label{fig:landolt}
\end{figure}

During August 2002, May 2004 and July 2012 seasons, we observed 26, 35 and 
36 photometric standard stars from \citet{landolt92,landolt07,landolt09}.
We selected stars with a wide range of colors and magnitudes.
All standard frames were analyzed using the PHOTOMETRY task in DAOPHOTII, 
DAOGROW and CCDSTD \citep{pbs90,pbs93,pbs95}.
In order to derive the transformation relations
from the instrumental system to the standard system, we adopt the
following equations:
\begin{eqnarray}
v &=& V + \alpha_v (B-V) + \beta_v X + \gamma_v, \\
b &=& B + \alpha_b (B-V) + \beta_b X + \gamma_b, \nonumber
\end{eqnarray}
where $X$ is the airmass and $V$, $B$ and ($B-V$) are 
the magnitudes and color on the standard system.
In Figure~\ref{fig:landolt}, we show the residuals of $V$ and ($B-V$)
in the sense \citet{landolt92,landolt07,landolt09} minus our measurements and
we show our transformation coefficients and residuals in $V$ magnitude
and $(B-V)$ color in Table~\ref{tab:coeff}.

\begin{figure}
\epsscale{1}
\figurenum{2}
\plotone{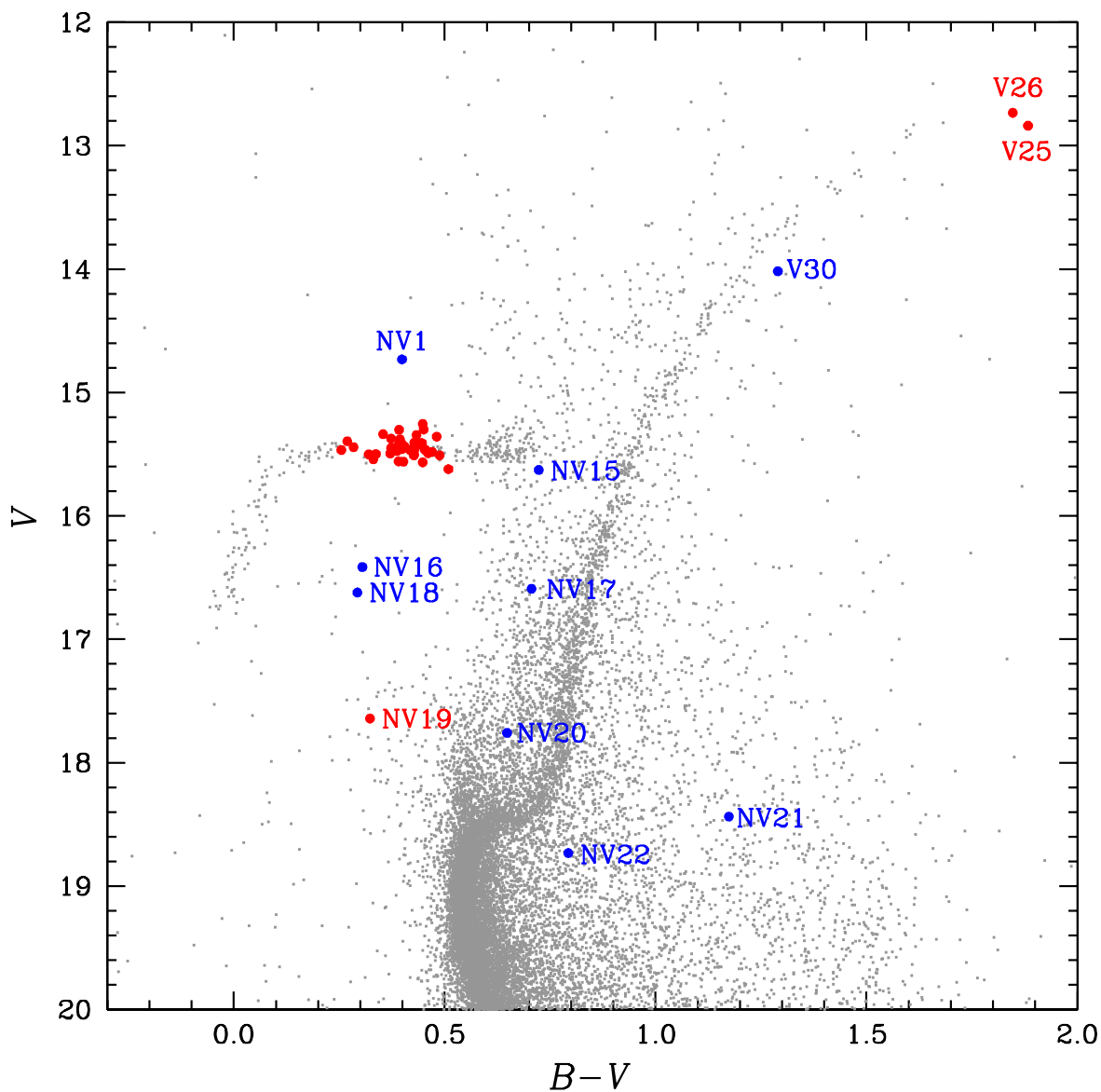}
\caption{Color-magnitude diagram of NGC~6723.
The variables are plotted with filled circles.
V25, V26 and V30 are variables listed in Menzies (1973)
and those with ``NV'' are new variables discovered in this work.
The cluster member variables are denoted with red and
the non-member variables with blue.
The SX Phe type variable NV19 is most likely a cluster member while
the RRc type variable NV1, the two $\delta$ Sct type variables NV16 and NV18,
and the five W UMa type eclipsing binaries NV15, NV17, NV20, NV21 and NV22 are
most likely non-member variables.
Note that V30 is not a cluster member (see Appendix).}
\label{fig:cmd}
\end{figure}

Point-spread function (PSF) photometry for all NGC~6723 science frames
was performed using DAOPHOTII-ALLSTAR, ALLFRAME, COLLECT-CCDAVE-NEWTRIAL
packages \citep{pbs87,pbs93,pbs94,pbs95,turner95}.
To achieve good PSFs on individual science frames,
we used about 100 to 250 isolated bright stars, depending on the seeing conditions
and exposure times, and at least three iterations were required to remove
neighboring stars from our PSF stars.
For the PSF calculation, we adopted a quadratically variable PSF for most cases.
The total number of stars measured from our ALLFRAME run
was more than 31,000 and our final CMD is shown in Figure~\ref{fig:cmd}.

Finally, astrometric solutions for individual stars in our field
have been derived using the positional data extracted from
the Naval Observatory Merged Astrometric Dataset \citep[NOMAD,][]{nomad}. 
We achieved rms scatter in the residuals of 0.05 arcsec using the IRAF IMCOORS package.
Then the astrometric fit was applied to calculate the equatorial coordinates
for all stars measured in our science frames.

\begin{figure}
\epsscale{1}
\figurenum{3}
\plotone{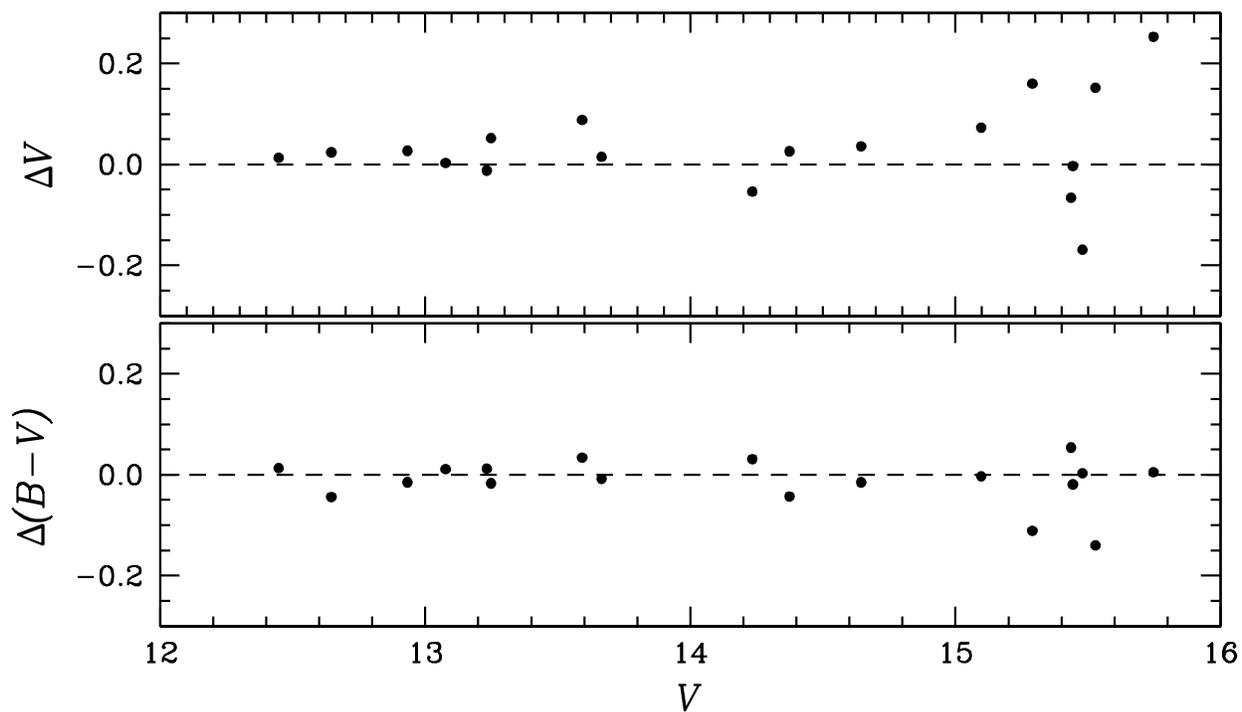}
\caption{Residuals of photoelectric photometry of Menzies (1974) as a function
of the magnitude. The differences are in the sense Menzies minus our work.
On average, our $V$ magnitude is 0.034 $\pm$ 0.093 mag (18 stars)
brighter and our ($B-V$) color is 0.014 $\pm$ 0.048 mag
bluer than those of Menzies.}
\label{fig:menzies}
\end{figure}

\begin{figure}
\epsscale{1}
\figurenum{4}
\plotone{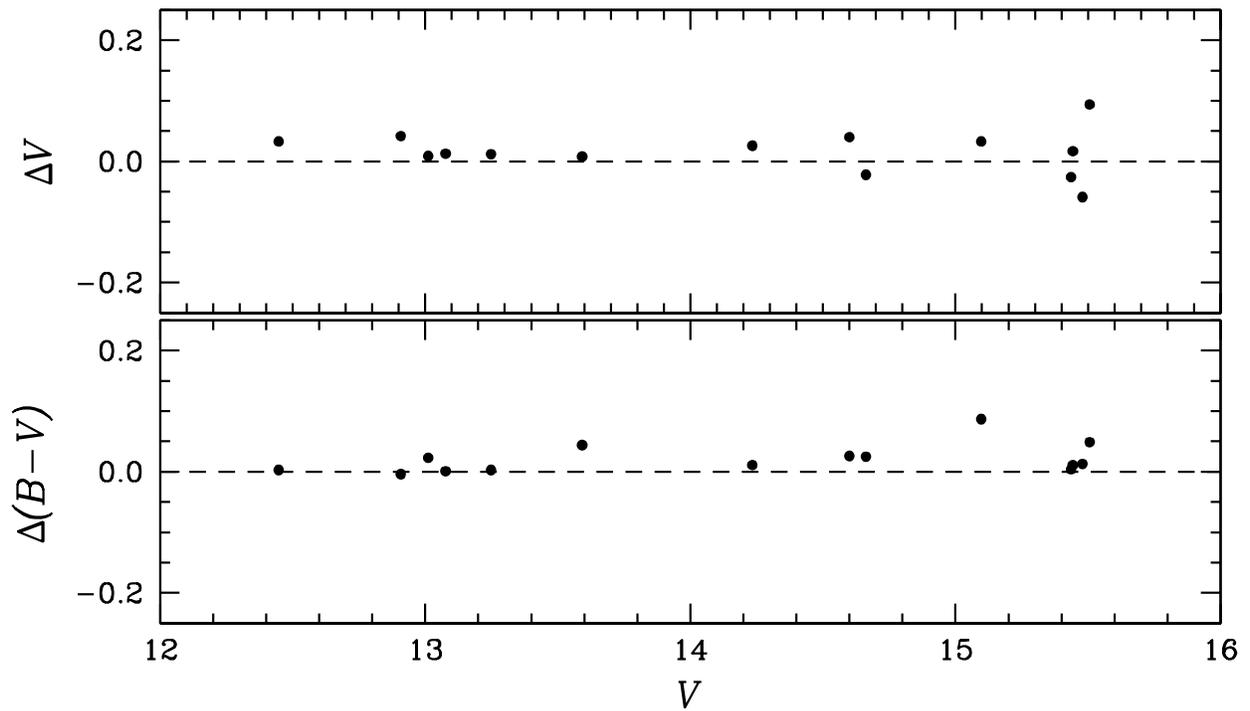}
\caption{Residuals of the photoelectric photometry of Alvarado et al. (1994)
as a function of the magnitude. The differences are in the sense Alvarado et al.
minus our work.
On average, our $V$ magnitude is 0.015 $\pm$ 0.035 mag (15 stars)
brighter and our ($B-V$) color is 0.020 $\pm$ 0.024 mag
redder than those of Alvarado et al.}
\label{fig:alvarado}
\end{figure}

\begin{figure}
\epsscale{1}
\figurenum{5}
\plotone{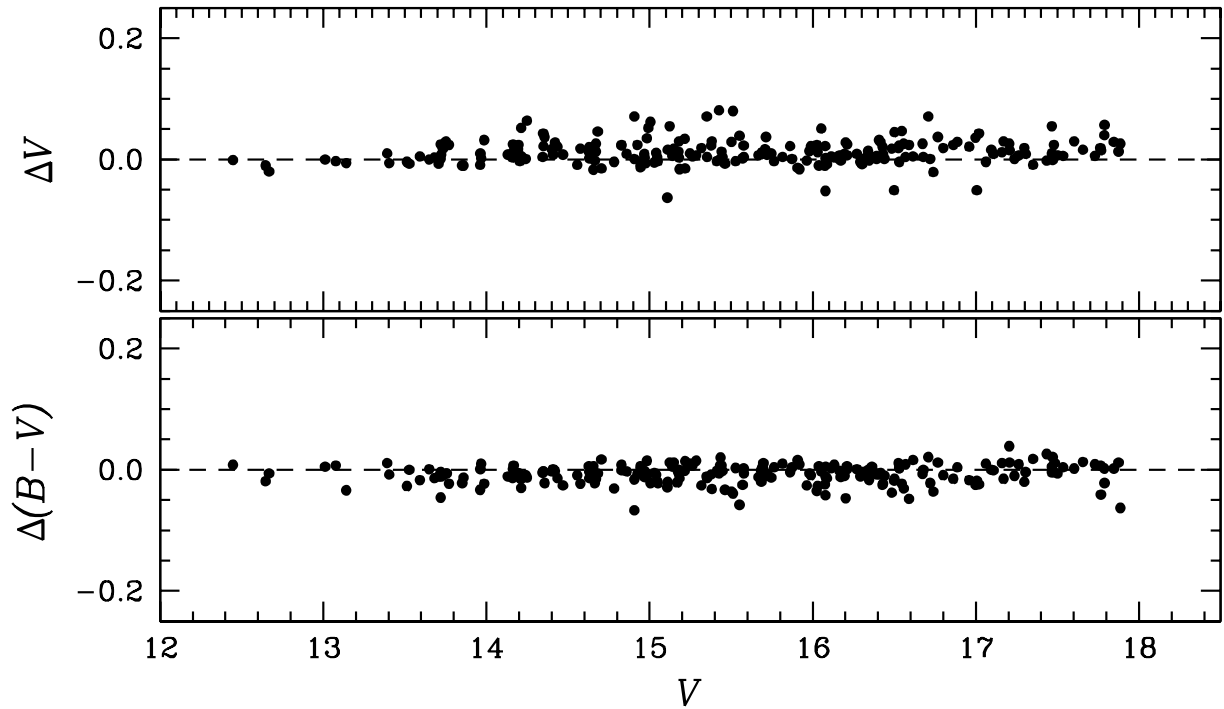}
\caption{Residuals of the photoelectric photometry of Stetson (2000)
as a function of the magnitude. The differences are in the sense Stetson
minus our work.
On average, our $V$ magnitude is 0.012 $\pm$ 0.021 mag (217 stars)
brighter and our ($B-V$) color is 0.008 $\pm$ 0.016 mag
bluer than those of Stetson.}
\label{fig:pbs}
\end{figure}

In Figures \ref{fig:menzies}, \ref{fig:alvarado} and \ref{fig:pbs},
we show comparisons of our results with the photoelectric photometry data of 
\citet{menzies74} and \citet{alvarado94} and the CCD photometry data 
of \citet{pbs_stds}.
As can be seen in the figures, our photometry is in excellent agreement with
previous measurements by others.

\section{RR LYRAE VARIABLES}
\subsection{Periods and light curves}
\citet{menzies74} and \citet{clement01} list 32 variables in NGC~6723: 
29 RRLs, two red variables and one $H_\alpha$ variable.
We used the finding charts provided by \citet{menzies74} to identify
those variables. Then we used his periods 
to estimate the photometric phases for our list of new standard
$V$ and $B$ magnitudes and Heliocentric Julian Day (HJD),
and drew light curves for each variable star.
For all variables, the light curves showed phase shifts 
in their maxima or minima indicating that period adjustments were needed.
Therefore, we determined new ephemerides and periods 
for the 29 previously known RRLs.
To determine the periods, we used three methods:
the string-length method \citep{lafler65},
the phase dispersion minimization method \citep{pdm},
and the information entropy minimization method \citep{cincotta95}.
In all cases, the periods were determined from both $B$ and $V$ data. 
In Figure~\ref{fig:rrlc}, we show new light curves for previously known 
RRLs \citep{menzies74,clement01}.
In Table~\ref{tab:rrlyrae}, we show positions, new periods,
pulsational amplitudes in $B$ and $V$ bandpasses and mean magnitudes
for individual variable stars.
In the table, $\langle V \rangle_\mathrm{mag}$ and
$\langle B \rangle_\mathrm{mag}$ denote magnitude-weighted integrated magnitudes
and $\langle V \rangle_\mathrm{int}$ and $\langle B \rangle_\mathrm{int}$
denote intensity-weighted integrated magnitudes of variables.
For the RRLs, we also provide static mean magnitudes and
colors as defined by \citet{marconi03};
\begin{eqnarray}
\langle V \rangle_\mathrm{s} &=& -0.345 \langle V \rangle_\mathrm{mag} + 1.345 \langle V \rangle_\mathrm{int}, \\
\langle B-V \rangle_\mathrm{s} &=& 0.488 \langle B-V \rangle_\mathrm{mag} + 0.479 \langle B-V \rangle_\mathrm{int}.
\end{eqnarray}

In Figure~\ref{fig:cmdHB}, we show the CMD of the HB region of NGC~6723.
For RRLs, we adopted magnitude-weighted integrated colors,
\bvmag, and the intensity-weighted integrated magnitudes, \vint.
It should be noted that the results derived from Fourier parameters presented in
Section \ref{s:fc} are for \bvmag\ and \vint.
Since the energy generated by nuclear reaction inside
the variable star is unaffected by the pulsation of the variable star,
the \vint\ should be the same as would be observed
were the star not pulsating \citep{smith95}.

\begin{figure}
\epsscale{1}
\figurenum{8}
\plotone{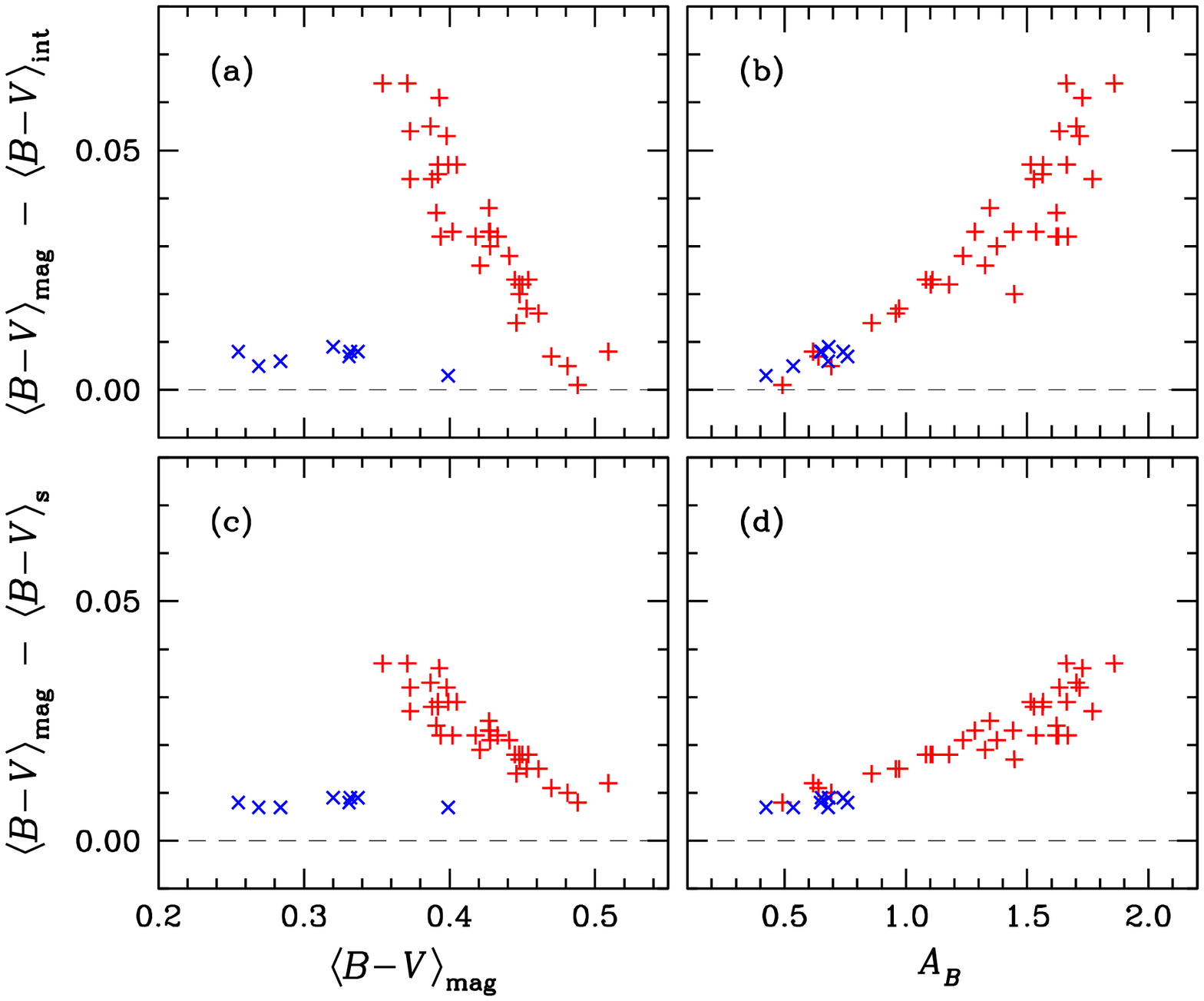}
\caption{Comparisons of integrated colors for RRLs.
The RRab type variable stars are represented by red plus signs
and the RRc type variable stars by blue crosses.
(a) and (c) show the difference in color of RRLs as a function
of the magnitude-weighted integrated color, \bvmag\
and (b) and (d) show the difference in color as a function
of the blue amplitude, $A_B$.
Panels (b) and (d) show that the difference in the color of RRLs
is closely correlated with the blue amplitude, in the sense that
as the blue amplitude increases the discrepancy between 
different color systems increases.}
\label{fig:compBV}
\end{figure}

It is known that the \bvmag\ gives the closest approximation to the
color-temperature relation of the ``equivalent static star'' \citep{sandage90}.
However, a caution should be advised on the average color of RRLs, 
in particular that of RRab type variable stars with large pulsational amplitude.
As noted by \citet{carney92}, either \bvmag\ or  \bvint\ 
(and \bvs) do not represent the correct temperature
measure for the RRLs due to excess emission 
in the $B$ passband during the rise from minimum to maximum radius.
In Figure~\ref{fig:compBV}, we show comparisons of integrated colors 
of RRLs in NGC~6723. In the Figure, we show differences in colors, 
\bvmag\ $-$ \bvint\ and \bvmag\ $-$ \bvs,
as functions of \bvmag\ and the blue amplitude, $A_B$.
As can be seen in the Figure, the differences in color between different color
systems are very small, $\leq$ 0.01 mag, for RRc type variable stars with
symmetric sinusoidal light curves.
However, the difference in color can be as large as 0.07 mag
for RRab type variable stars with asymmetric light curves with large $A_B$ values.
Figure~\ref{fig:compBV} (b) and (d) suggest that the difference in the color of RRLs
is closely correlated with the blue amplitude $A_B$, in the sense that
as the blue amplitude increases the discrepancy between 
different color systems increases.

\subsection{New variables}\label{s:nv}
Recent CCD photometry surveys of GCs reveal many new variables, 
especially RRLs with low amplitudes or in the crowded central part 
of GCs and we attempt to search for new variables 
in NGC~6723 following the method similar to \citet{lc99a}.

We collected PSF magnitudes for about 460 variable candidates
returned from NEWTRIAL based on the variability index
and calculated magnitude variations as a function of the HJD.
We then calculated light curves of each variable candidate
with the best estimated period.
We employed the information entropy minimization method
with an 8$\times$8 grid size and the string-length method for this purpose.

\begin{figure}
\epsscale{1}
\figurenum{10}
\plotone{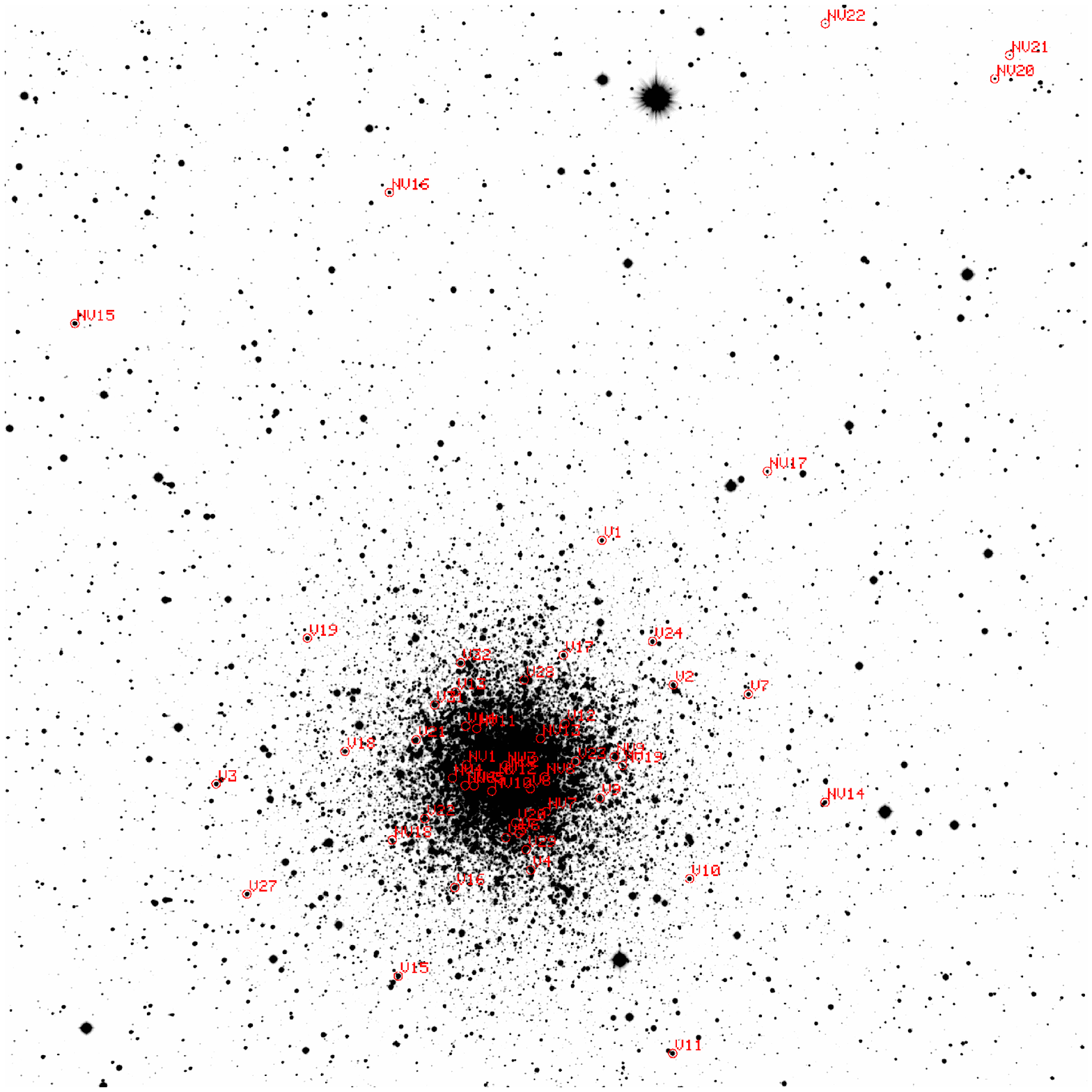}
\caption{Finding chart for new variables in NGC~6723.
The field of view is 15 $\times$ 15 arcmin.
East is to the left and north is to the bottom.
The variables by Menzies (1974) are marked with ``V'' and
the new variables with ``NV''.}
\label{fig:chart1}
\end{figure}

\begin{figure}
\epsscale{1}
\figurenum{11}
\plotone{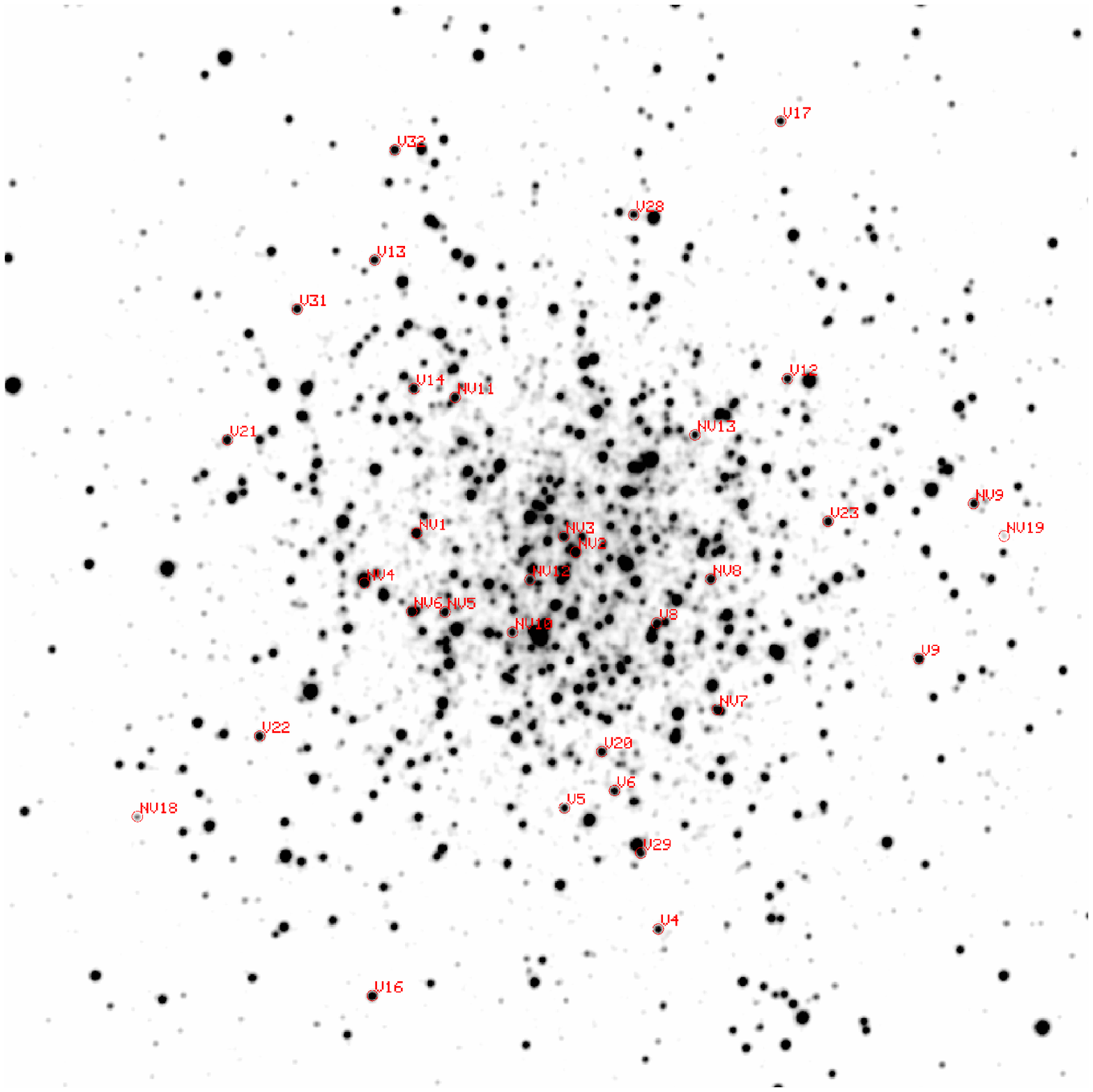}
\caption{Same as Figure~\ref{fig:chart1}, but the field of view is 
4 $\times$ 4 arcmin.}
\label{fig:chart2}
\end{figure}

Visual examinations of plots of magnitude variations against the HJD and
the phase-folded light curves for all variable candidate stars
revealed 22 new variables;
14 RRLs, one SX Phe type  and two $\delta$ Scuti type variable stars and
five W UMa type eclipsing binaries.
We also ``rediscovered" all 29 previously known RRLs.
Positions and photometric parameters for new variables
are presented in Table~\ref{tab:rrlyrae} and light curves
for the new variables are shown in Figure~\ref{fig:newrrlc}.
As expected, most of the newly found variables lie near 
the central part of the cluster.
Identification charts for the new variables are presented 
in Figures~\ref{fig:chart1} and \ref{fig:chart2} for the outer
(with a field of view of 15 $\times$ 15 arcmin) and inner part
(4 $\times$ 4 arcmin) of the cluster, respectively.
We also marked the previously known RRLs in the figures.
Note that all new RRLs lie within the tidal radius of NGC~6723.

\begin{figure}
\epsscale{1}
\figurenum{7}
\plotone{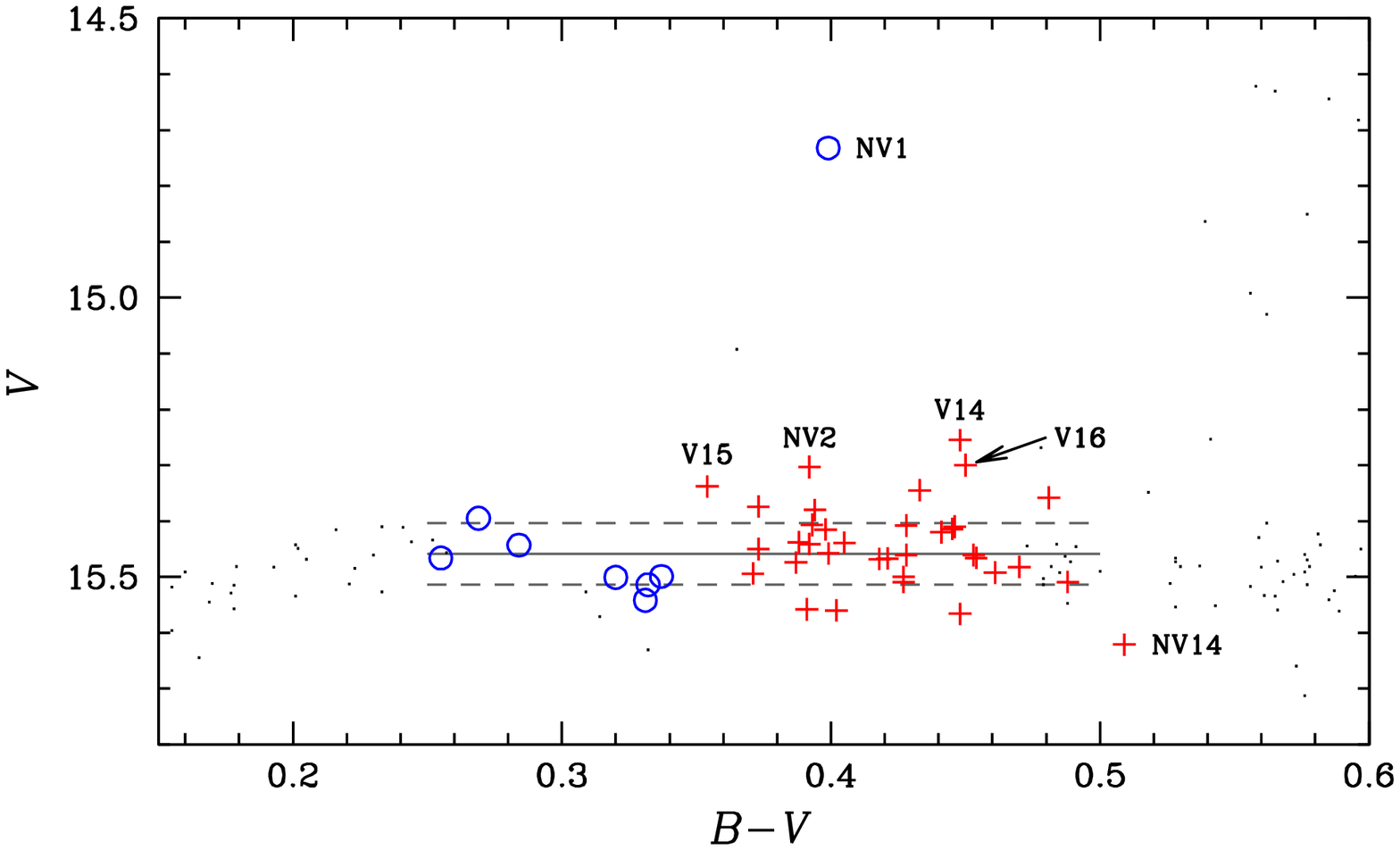}
\caption{Color-magnitude diagram of the HB region.
The RRab variables are represented by red plus signs and the RRc by blue open circles.
The  magnitude-weighted integrated colors and 
the intensity-weighted integrated magnitudes, \bvmag\ and \vint,
are used for the RRLs in the figure.
The grey dashed horizontal line indicates the average intensity-weighted
integrated magnitude of the cluster and the grey dashed lines show 
the $\pm \sigma$ level, 
$\langle V\mathrm{(RR)}\rangle_\mathrm{int}$ = 15.459 $\pm$ 0.055 mag.}
\label{fig:cmdHB}
\end{figure}

In an attempt to confirm the membership of the new RRLs,
we examined the CMD near the HB region (see Figure~\ref{fig:cmdHB})
and the period-amplitude diagram (see Figure~\ref{fig:PvsA_B}),
since there is no previous proper motion study or
radial velocity measurements are available for the new variables.

As shown in Figure~\ref{fig:cmdHB}, the RRc type variable NV1
appears to be too bright to be a regular RRc type variable star of NGC~6723.
The average intensity-weighted integrated magnitude \vint\
of the remaining seven RRc type variables including NV7 and NV11
is 15.480 $\pm$ 0.019 mag and NV1 is 0.748 mag brighter than the mean magnitude.
However, it should be noted that
NV1 is located within the core radius of the cluster\footnote{The radial distance
of NV1 is 36 arcsec from the center of the cluster and the core radius
of the cluster is 56.4 arcsec \citep{harris96}.}
and its photometric metallicity, [Fe/H] = $-$1.28 dex, from the Fourier analysis
presented in section~\ref{s:feh} is consistent with the mean metallicity of the cluster,
[Fe/H] $\approx$ $-$1.2 dex.\footnote{As discussed above, 
the Galactic bulge RRLs are slightly more metal-rich than NGC~6723 is.
Therefore, metallicity alone may not be a proper indicator to distinguish 
the cluster member star.}

It is suspected that NV1 may be a member of a binary system.
We attempted to find out the secondary periodicity for NV1 using pre-whitened
data but we failed to get any reliable secondary period of the variable.
Since it lies about 36 arcsec from the center of the cluster,
it may suffer from blending effect with nearby bright stars.
We visually examined the point-spread function of NV1 in individual
science frames and found no sign of blending.
We also found no evidence of blending for NV1 from the images taken with
the HST Advanced Camera for Surveys (ACS) by \citet{sarajedini07} which were retrieved 
via the Mikulski Archive for Space Telescopes (MAST)\footnote{STScI is operated
by the Association of Universities for Research in Astronomy, Inc., 
under NASA contract NAS5--26555. Support for MAST for non-HST data is provided
by the NASA Office of Space Science via grant NNX09AF08G and by other grants 
and contracts.},
leading us to believe that NV1 is most likely a non-member of the cluster
and is most likely a foreground RRc type variable.
However, follow-up monitoring of NV1 in the future would be desirable to reveal
the potential binary nature of NV1.

The RRab type variable NV14 is slightly fainter
and redder than the rest of RRab type variables of the cluster.
However, the radial distance of NV14 from the center of the cluster
is 264 arcsec and NV14 lies well within the tidal radius of the cluster,
$\approx$ 630 arcsec.
NV14 is in good agreement with other RRab type variable stars in NGC~6723
on the period-amplitude diagram in Figure~\ref{fig:PvsA_B},
and on the Wesenheit function-period diagram in Figure~\ref{fig:W}.
Also other physical parameters of NV14, such as metallicity,
derived from the Fourier analysis do not disagree with those from
other variable stars.
This leads us to believe that NV14 is a cluster member with
an excess differential foreground reddening effect by $\gtrsim$ 0.03 mag.

In appendix, we present the membership status of other types of new variable
stars in NGC~6723. The SX Phe type variable star NV19 is most likely a cluster
member star while the two $\delta$ Sct type variable stars, NV16 and NV18, and
the five W UMa type eclipsing binaries NV15, NV17, NV20, NV21 and NV22
do not appear to be cluster members, based on the radial distance from
the center of the cluster and the true distance modulus.

With the discovery of new variables, the new mean periods of RRab and RRc variables are
$\langle$P$_{ab}$$\rangle$ = 0.541 $\pm$ 0.066 day (35 stars) and
$\langle$P$_{c}$$\rangle$ = 0.292 $\pm$ 0.030 day (7 stars without NV1), respectively.
The errors are those of the mean.
The new RRL distribution is n(c)/n(ab+c) = 0.167.
Our mean periods are in good agreement with those of \citet{menzies74},
who obtained 0.540 day and 0.291 day for $\langle$P$_{ab}$$\rangle$
and $\langle$P$_{c}$$\rangle$, respectively.
However, our new n(c)/n(ab+c) value is slightly smaller than
that of Menzies, n(c)/n(ab+c) = 0.172.

\section{THE FOURIER ANALYSIS}\label{s:fc}
During the last two decades, Fourier decomposition methods have been
successfully applied to characterize the RRLs and other types of variables
\citep[see for example,][]{simon81, kovacs96, jurcsik98, kovacs99,
kovacs01,cacciari05}.
We performed the Fourier analysis and derived Fourier coefficients and
relevant physical parameters of the variables from the new photometry of the cluster.
For our analysis, we used the FORTRAN programs kindly provided by Dr.\ Kov\'acs.

\subsection{Fourier Coefficients}
Fourier coefficients of RRLs in NGC~6723 were derived 
for the $V$ band photometry by assuming that the variations 
in the  observed $V$ magnitude can be reproduced by a Fourier sine series;
\begin{equation}
V = A_0 + \sum^{N}_{i=1}A_i\sin(i\omega t + \phi_i),
\end{equation}
where $\omega$ = 2$\pi/P$ is the angular frequency, $t$ is the epoch
of observations, and the $A_i$ and $\phi_i$ are the Fourier coefficients.
We reconstructed artificial light curves using the derived
Fourier coefficients for each variable.
Each artificial light curve was visually inspected by comparing it to the observed data.
We show the Fourier coefficients for RRLs
in Table~\ref{tab:rrabfc}, where
$R_{i1} = A_i/A_1$, and $\phi_{i1} = \phi_i - i\phi_1$
($0 \leq \phi_{i1} < 2\pi$).

\subsection{The Metallicity}\label{s:feh}
\citet{jk96} derived an empirical relation to estimate metallicity of
RRab type variables from the Fourier parameters, 
\begin{equation}
\mathrm{[Fe/H]_\mathrm{JK96}} = -5.038 - 5.394P + 1.345\phi_{31}.
\end{equation}
To apply this relation, \citet{jk96} and \citet{kk98}
introduced the deviation parameter $D_F$
$ = |F_{obs} - F_{calc}|/\sigma_F$ to measure relative accuracy of
the prediction, where $F_{obs}$ is the observed parameter, $F_{calc}$
is the calculated value from the other observed parameters, $\sigma_F$
is the respective standard deviation in their Table 6 \citep[see also][]{kk98}.
The compatibility condition parameter $D_m$, which is defined to be the maximum
of the deviation parameters $D_F$, represents a compatibility test
on the regularity of the shape of the light curves.
\citet{jk96} suggested that physical parameters, such as metallicity,
of RRab type from the Fourier decomposition can be securely applicable
if $|D_m|$ $<$ 3.
In the last column of Table \ref{tab:rrabfc}, we show deviation parameters $D_m$
for all RRab variables in NGC~6723 based on \citet{jk96}.
It should be noted that V13, V18, NV2, NV3, NV5 and NV13 
appear to be Blazhko variables in their light curves
as shown in Figures~\ref{fig:rrlc} and \ref{fig:newrrlc},
but their $|D_m|$ values are less than 3,
suggesting that the compatibility condition may not be reliable.
\citet{cacciari05} also found that the compatibility condition by
\citet{jk96}, {\em i.e.} $|D_m|$ $<$ 3, may not be suitable to differentiate
Blazhko variables. They noted that they detected as much as $\approx$ 40\% of
Blazhko variables with $|D_m|$ $<$ 3 and $\approx$ 60\% with $|D_m|$ $<$ 5.

Using the original relation by \citet{jk96}, we obtained a mean metallicity of 
\feh$_\mathrm{JK96}$ = $-$0.93 $\pm$ 0.22 dex from all 18 RRab type variable stars
and $-$0.94 $\pm$ 0.12 dex from the 10 regular RRab variable stars\footnote{
We do not use Blazhko variables V13, V18, NV2, NV3, NV5, NV13 and NV14 in our calculations.
In addition to these, we do not include V14, which appear to suffer from
blending effect with nearby faint companions. See Appendix~\ref{ap:s:notes}.} for NGC~6723.
To calculate more widely used metallicity scales for GC systems by \citet{zw84},
we used the transformation relation given by \citet{jurcsik95},
\begin{equation}
\mathrm{[Fe/H]_\mathrm{JK96}} = (1.431 \pm 0.006)\mathrm{[Fe/H]}_\mathrm{ZW} + 
(0.880 \pm 0.010), \label{eq:zw}
\end{equation}
and we obtained \feh$_\mathrm{ZW}$ = $-$1.27 $\pm$ 0.09 dex from 10 RRab 
type variable stars in NGC~6723.
We show the metallicity of each variable in column (2) of Table~\ref{tab:rrabpam}.

We also used an empirical relation given by \citet{morgan07}
to obtain Zinn \& West metallicity scale for RRc type variables,
\begin{eqnarray}
\mathrm{[Fe/H]_\mathrm{ZW}} &=& 52.466P^2 - 30.075P + 0.131\phi^{(c)2}_{31} + \nonumber\\
 & & 0.982\phi^{(c)}_{31} - 4.198\phi^{(c)}_{31}P + 2.424,
\end{eqnarray}
where $\phi^{(c)}_{31}$ is the phase for cosine series.
This cosine phase was transformed
from our sine series phase using the following relation,
$\phi^{(c)}_{31} = mod(\phi_{31} + \pi,  2\pi)$, where
$mod$ is the modulus operator, and therefore,
$0 \leq \phi^{(c)}_{31} < 2\pi$.
We obtained \feh$_\mathrm{ZW}$ = $-$1.18 $\pm$ 0.15 dex for all 7 RRc variable
stars and $-$1.18 $\pm$ 0.20 dex for 4 regular RRc variable, 
consistent with that from RRab type variables of the cluster.
We show our results for RRc type variable stars in Table \ref{tab:rrcpam}.

The unweighted mean of the photometric metallicity of NGC~6723
in the Zinn \& West metallicity scale becomes
$\langle$[Fe/H]$_\mathrm{ZW}\rangle$ = $-$1.23 $\pm$ 0.11 dex.
We conclude that our metallicity from the Fourier analysis of RRLs
of NGC~6723 is in excellent agreement with 
that of \citet{fullton96}, who obtained \feh\ = $-$1.26 $\pm$ 0.09 
using high resolution spectroscopic abundance study
of the three red-giant branch stars in the cluster.

\subsection{Intrinsic Color $(B-V)_0$ of RRab type Variable Stars
and Interstellar Reddening $E(B-V)$}
The interstellar reddening values of GCs are of great interest
since, for example, they affect the derived photometric temperature
or metallicity from the observed color, the true distance modulus, etc.
The previous estimates of interstellar reddening values of NGC~6723
range from $\approx$ 0.0 \citep{menzies74} to 0.11 \citep{alcaino99}.
We explore the interstellar reddening by comparing the observed colors and
intrinsic colors of RRLs in NGC~6723.

We calculated the intrinsic color of RRab type variables using the 
empirical relations given by \citet{jurcsik98},
\begin{equation}
(B-V)_0 = 0.308 + 0.163P - 0.187A_1, \label{eq:bv0:1}
\end{equation}
and by \citet{kovacs01},
\begin{equation}
(B-V)_0 = 0.460 + 0.189\log P - 0.313A_1 + 0.293A_3. \label{eq:bv0:2}
\end{equation}
Note that the above relations are for \bvmag\ and these relations are
based on the same program.
Not surprisingly, the intrinsic colors \bvz\ of individual variable stars
from both relations [Equations (\ref{eq:bv0:1}) and (\ref{eq:bv0:2})]
are in excellent agreement to within 0.002 $\pm$ 0.004 and
we show average $\langle (B-V)_0\rangle$ of
RRab type variable stars in column (3) of Table~\ref{tab:rrabpam}.
In column (4) of the table, we also show our interstellar reddening \ebv\ estimates
for each variable by comparing the intrinsic color with the observed 
magnitude-weighted integrated color of RRab type variables.
For the mean interstellar reddening value of NGC~6723,
we obtained \ebv\ = 0.087 $\pm$ 0.014 mag from all 18 RRab type variable stars
and 0.086 $\pm$ 0.007 mag from 10 regular RRab type variable stars .

We also make use of an empirical period-amplitude-color-metallicity
relation given by \citet{piersimoni},
\begin{equation}
(B-V)_0 = 0.507 - 0.052A_B + 0.223\log P + 0.036\mathrm{[Fe/H]},\label{eq:bv0:3}
\end{equation}
assuming \feh\ = $-$1.23 dex for NGC~6723 and
we obtained \ebv\ = 0.089 $\pm$ 0.013 for 35 RRab stars
and \ebv\ = 0.086 $\pm$ 0.010 for 23 regular RRab type variable stars
in NGC~6723.

The interstellar reddening values from Equations (\ref{eq:bv0:1}), 
(\ref{eq:bv0:2}) and (\ref{eq:bv0:3}) are in good agreement with
the mean value of \ebv\ = 0.086 $\pm$ 0.017.
Note that our mean interstellar reddening estimate
appears to be slightly larger than the previous
interstellar reddening value\footnote{\citet{schlegel98} have provided
tools to estimate total reddening along essentially any line of sight,
which should reflect the total reddening to the cluster.
Their result is \ebv\ = 0.16 mag for the cluster and it is about three
times larger than that of \citet{harris96}.
\citet{dutra00} noted that interstellar reddening estimates
by \citet{schlegel98} for clusters near the Galactic plane
appear to be larger than those based on stellar contents,
possibly due to background dust.
\citet{arce98} have cautioned their readers that \citet{schlegel98}
may over-estimate reddening by a factor of 1.3 to 1.5
in regions with \ebv\ $> 0.2$ or so. Our results appear to
confirm these cautionary remarks about the use of the \citet{schlegel98}
reddening maps in regions of high reddening
\citep[see also][]{lcs01,lc06}.} by \citet{harris96}, \ebv\ = 0.05 but
smaller than that of \citet{alcaino99}, who obtained \ebv\  = 0.11 $\pm$ 0.01
based on $UBVRI$ photometry of the cluster.

\begin{figure}
\epsscale{1}
\figurenum{14}
\plotone{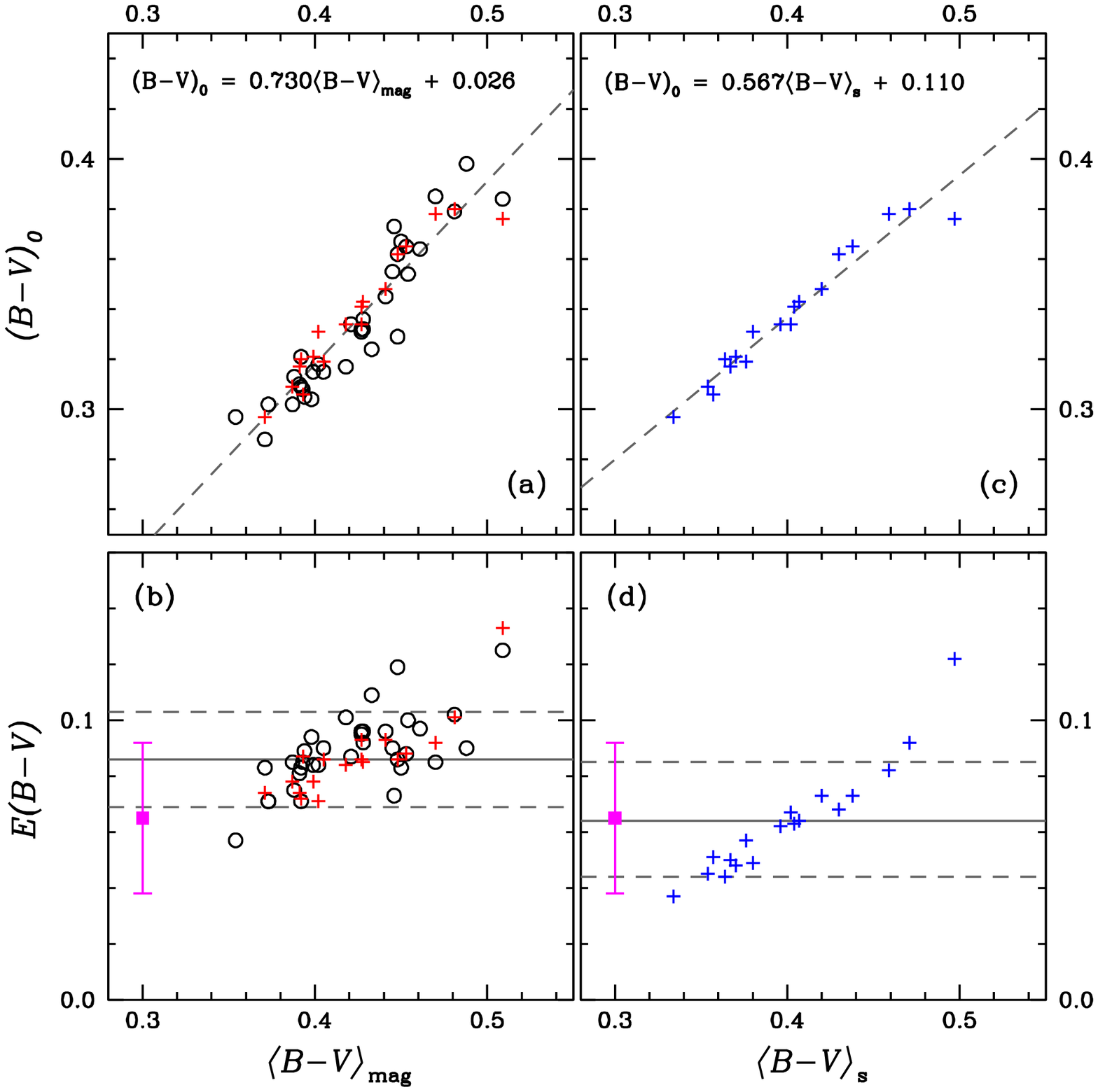}
\caption{(a) The calculated \bvz\ as a function of the observed \bvmag.
The red plus signs are for \bvz\ values from transformation relations
by Jurcsik (1998) and Kov\'acs \& Walker (2001)
and the open circles are for those by Piersimoni et al.\ (2002).
The grey dashed line is the linear fit to the data.
The slope of the relation between \bvz\ and \bvmag\
significantly deviates from the unity more than 7 $\sigma$ level.
(b) \ebv\ [= $\langle B-V\rangle_\mathrm{mag} - (B-V)_0$]
as a function of \bvmag.
The grey solid line is for the average \bvz\ and the grey dashed lines
show the $\pm \sigma$ levels.
The filled magenta box represents \ebv\ from Dotter et al. (2011).
(c) Same as (a) but for \bvs\ by Marconi et al. (2003).
The slope of the relation between \bvz\ and \bvs\
significantly deviates from the unity more than 13 $\sigma$ level.
(d) Same as (b) but for \bvs.}
\label{fig:ebv}
\end{figure}

It should be noted that the calculated intrinsic colors of RRab type
variable stars from Equations (\ref{eq:bv0:1}) -- (\ref{eq:bv0:3})
have a significant gradient against the observed colors of RRab variables 
and weak gradients against periods and blue amplitudes $A_B$, suggesting that 
Equations (\ref{eq:bv0:1}) -- (\ref{eq:bv0:3}) are slightly wrong.
In Figure~\ref{fig:ebv} (a), we show the calculated intrinsic
color of individual RRab type variable stars using 
Equations (\ref{eq:bv0:1}) -- (\ref{eq:bv0:3}) as a function of
the observed \bvmag\ and we obtained the following relation,
\begin{equation}
(B-V)_0 = (0.730 \pm 0.036) \langle B-V\rangle_\mathrm{mag} + (0.026 \pm 0.015).\label{eq:bv0_bvmag}
\end{equation}
The slope of this relation is not unity and, as a consequence, 
\ebv\ [ = $\langle B-V\rangle_\mathrm{mag} - (B-V)_0$]
has a substantial gradient against the observed color
as shown in Figure~\ref{fig:ebv} (b).
Our result strongly suggests that the intrinsic color estimation for
RRab type stars from the Fourier analysis or the pulsational properties
fails to work. We propose that the previous empirical relations
to derive the intrinsic color of RRab type variables 
using their pulsational parameters, such as Equations
(\ref{eq:bv0:1}) -- (\ref{eq:bv0:3}), should be re-examined in the future.

In their study of RRLs in M3,
\citet{cacciari05} also found that there existed 
significant discrepancies between the intrinsic colors derived from
Fourier parameters and the observed ones.
They found that the distribution of the intrinsic colors \bvz\ derived from
Fourier parameters using the transformation relation by \citet{kovacs99}
is somewhat compressed and slightly blue-shifted with
respect to the observe \bvmag\ distribution.
The narrowness of the range of their calculated intrinsic color
can be naturally understood if our derived relation between 
the calculated \bvz\ and the observed \bvmag\ [Equation (\ref{eq:bv0_bvmag})]
is still valid for M3 RRLs.

As stated above, Equations (\ref{eq:bv0:1}) and (\ref{eq:bv0:2})
are for \bvmag\ but we also investigate the interstellar reddening value
from \bvs, i.e.\ \ebv\ = \bvs\ $-$ \bvz.
We obtained the mean interstellar reddening value of \ebv\ = 0.064 $\pm$ 0.021 
from 18  RRab type variables and 
\ebv\ = 0.061 $\pm$ 0.014
from 10 regular RRab type variables.
We show our results in the column (5) of Table~\ref{tab:rrabpam}.
Note that this value is in good agreement with that of \cite{harris96}.
However, as can be seen in the relation between the observed \bvmag\ and the calculated \bvz\
in Equation (\ref{eq:bv0_bvmag}), the slope of the relation between
the observed \bvs\ and the calculated \bvz\ deviates significantly from unity 
[see Figure~\ref{fig:ebv} (c) and (d)].
We obtained the following relation;
\begin{equation}
(B-V)_0 = (0.567 \pm 0.033) \langle B-V\rangle_\mathrm{s} + (0.110 \pm 0.013).\label{eq:bv0_bvs}
\end{equation} 

Recently \citet{ACS} presented interstellar reddening values of about 60 GCs
in the HST ACS photometric system, $E(6-8) [= E(F606W-F814W)]$,
using isochrone fitting and they presented $E(6-8)$ values for individual
GCs in their Table 2. 
We compared $E(6-8)$ by \citet{ACS} to \ebv\ by \citet{harris96}
for 61 GCs and we obtained the following transformation relation,
\begin{equation}
E(B-V) = (0.959 \pm 0.027) E(6 - 8) - (0.005 \pm 0.006).
\end{equation}
For NGC~6723, \citet{ACS} obtained $E(6-8)$ = 0.073 and their value
is equivalent to \ebv\ = 0.065 $\pm$ 0.027.
This result is in excellent agreement with our interstellar reddening estimate from
the static color of RRab variables of the cluster, \ebv\ = 0.061 $\pm$ 0.014,
and is in agreement with that from \bvmag, \ebv\ = 0.087 $\pm$ 0.014,
within measurement errors.

For the foreground interstellar reddening value of
NGC~6723, we adopt the unweighted mean of those 
from the static color \bvs\ of RRLs and from the isochrone fitting by \cite{ACS},
\ebv\ = 0.063 $\pm$ 0.015.

\subsection{The visual magnitude of RR Lyrae variables}
The average intensity-weighted integrated magnitude of the 7 RRc and
30 RRab type variables (excluding V14, V15, V16, NV1, NV2 
and NV14)\footnote{In Appendix~\ref{ap:s:notes}, we discuss that 
V16 is a more evolved RRab type variable while V15 suffers from
blending with nearby faint objects.
V14 appears to suffer both effects. }  in NGC~6723 is
$\langle V\mathrm{(RR)}\rangle_\mathrm{int}$ = 15.459 $\pm$ 0.055 mag.
Our new $\langle V\mathrm{(RR)}\rangle_\mathrm{int}$ is in good agreement with
$\langle V\mathrm{(HB)}\rangle$ magnitudes
by \citet{menzies74}, \citet{fulltonphd}  and \citet{rosenberg99},
$\langle V\mathrm{(HB)}\rangle$  = 15.48, 15.47 $\pm$ 0.02,
and 15.45 $\pm$ 0.05, respectively.
It should be noted that our $\langle V\mathrm{(RR)}\rangle_\mathrm{int}$ magnitude
is not for the zero-age HB (ZAHB) but includes RRLs evolved away from
the ZAHB as shown in Figure~\ref{fig:cmdHB}.
Therefore, the maximum \vint\ of regular RRLs of NGC~6723,
which is \vint\  = 15.514 mag for V7, should be close to the ZAHB.

\subsection{The absolute visual magnitude of RR Lyrae variables}
\subsubsection{Old calibrations based on the fainter \mvrr}
\citet{jurcsik98} provided an empirical relation to calculate
the absolute visual magnitude of RRab type variables employing
Fourier parameters,
\begin{equation}
M_V\mathrm{(RRab)} = 1.221 - 1.396P - 0.477A_1 + 0.103\phi_{31}.\label{eq:mv:1}
\end{equation}
Using this relation, we obtained an average absolute visual magnitude 
$\langle M_V\mathrm{(RRab)}\rangle$ = 0.854 $\pm$ 0.029 mag
from 18 RRab type variables and 
$\langle M_V\mathrm{(RRab)}\rangle$ = 0.852 $\pm$ 0.020 mag
from 10 regular RRab type variables in NGC~6723.
We show the absolute visual magnitude for each star 
in column (6) of Table~\ref{tab:rrabpam}.

To estimate the absolute visual magnitude of RRc type variables,
we used the relation given by \citet{kovacs98},
\begin{equation}
M_V\mathrm{(RRc)} = 1.261 - 0.961P - 0.044\phi_{21} - 4.447A_4. \label{eq:mv:2}
\end{equation}
We obtained 
$\langle M_V\mathrm{(RRc)}\rangle$ = 0.809 $\pm$ 0.053 mag
for the seven RRc type variables and 
$\langle M_V\mathrm{(RRc)}\rangle$ = 0.820 $\pm$ 0.021 mag
for the four regular RRc type variables in NGC~6723.
The absolute magnitudes of each variable star are listed in Table~\ref{tab:rrcpam}.
The average absolute visual magnitude from the four regular RRc type variables is 
0.032 mag brighter than that from 10 regular RRab type variables in NGC~6723
but both absolute magnitude scales are in good agreement within the errors.
The combined absolute visual magnitude of NGC~6723 
becomes \meanmvrr\
= 0.843 $\pm$ 0.025 mag from 10 regular RRab and four regular RRc variable stars.

As a consistency check, we compare our \meanmvrr\ for NGC~6723
derived from Fourier parameters with the recent estimates of \mvrr.
As nicely summarized by \cite{cacciari13}, recent theoretical or empirical
\mvrr\ versus \feh\ relations prefer \mvrr\ = 0.52 mag at \feh\ = $-$1.50 dex.
As a consequence, \mvrr\ at \feh\ = $-$1.23 dex (i.e.\ the metallicity of NGC~6723)
should be 0.578 mag or 0.588 mag depending
on the slope of the \mvrr\ versus \feh\ relation, 0.214 or 0.250, respectively.
Our \meanmvrr\ for NGC~6723 from Equation (\ref{eq:mv:1}) or (\ref{eq:mv:2})
is about 0.26 mag fainter than the current standard value.

\citet{cacciari05} also found that the derived absolute visual magnitude of
RRLs from Fourier parameters is somewhat faint and
they used the empirical relation by \citet{kovacs98}
[Equation (\ref{eq:mv:2})] to derive
RRc type variable stars in M3, with a brighter zero point
by 0.2 mag [i.e.\ from 1.261 to 1.061 in their Equation (16)].

These results suggest that the \mvrr\ transformation relations 
using Fourier parameters based on the \mvrr\ from 
Baade-Wesselink method by \citet{jurcsik98} 
and \citet{kovacs98} may not be reliable and a revision
of the transformation relations should be performed in the future.

\subsubsection{Is RR Lyrae an evolved star?}
Recently, \citet{benedict11} measured absolute trigonometric
parallaxes and proper motions of five nearby RRLs using the HST FGS.
They presented  the absolute visual magnitude of RRLs as a function of metallicity,
\begin{equation}\label{eq:hstmv}
M_V\mathrm{(RR)} = (0.214 \pm 0.047)(\mathrm{[Fe/H]} + 1.5) + (0.45 \pm 0.05).
\end{equation}
The metallicity-luminosity relation by \citet{benedict11} gives
$M_V\mathrm{(RR)}$ = 0.45 mag at \feh\ = $-$1.50 dex, which is about 0.07 mag
brighter than those from others 
\citep[see for examples][]{federici12,cacciari13,kollmeier13}.
Although \citet{benedict11} claimed that their relation gives the absolute
visual distance modulus of LMC of 
$(m-M)_\mathrm{0}$ = 18.55 $\pm$ 0.05 mag,
the zero-point of \citet{benedict11}, 0.45 $\pm$ 0.05 mag, 
and the dereddened mean visual magnitude of RRLs in LMC
by \cite{clementini03}, 19.064 $\pm$ 0.064 mag at \feh\ = $-$1.50 dex, 
can be translated into $(m-M)_\mathrm{0}$ = 18.61 $\pm$ 0.08 mag for LMC, 
significantly larger than those from other methods \citep{walker12,cacciari13}.

As kindly noted by the referee, the evolutionary status of RRLs studied by 
\cite{benedict11} does not appear to be well established.
\cite{fernley98} suspected that SU Dra is likely an evolved RRL.
More recently \cite{catelan08} claimed that RR Lyr is indeed an evolved RRL.
They performed a careful study of the star using Str\"omgren photometry and
they claimed that RR Lyr is overluminous by 0.06 mag for its metallicity.
In their Equation (4a), \cite{catelan08} provided the following 
metallicity-luminosity relation;
\begin{eqnarray}\label{eq:catelan08}
M_V\mathrm{(RR)}_\mathrm{ZW} &=& (0.23 \pm 0.04)\mathrm{[Fe/H]} + 
(0.984 \pm 0.127). \nonumber
\end{eqnarray}
Using this relation, one can find \mvrr\ = 0.64 mag at \feh\ = $-$1.50 dex, 
which is 0.12 mag fainter than the current standard value, 0.52 mag.
It should be noted that the result by \cite{catelan08} is based
on the theoretical HB models by \cite{catelan04}, whose RRL magnitude levels
are 0.08 mag fainter than those from other works 
\citep[for example, see Table 1 of][]{cacciari13}.
At \feh\ = $-$1.50 dex in the Zinn-West metallicity scale,
the theoretical models by \cite{catelan04} predicts \mvrr\ = 0.60 mag
and their models are 0.08 mag fainter than the current standard value, 
\mvrr\ = 0.52 mag.

It is suspected that if one can make the zero-point of 
the theoretical models of \cite{catelan04} 0.08 mag brighter,
then RR Lyr could have been thought as a normal RRL near the ZAHB.

Among 5 RRLs studied by \cite{benedict11}, RR Lyr is of particular importance
at least for two reasons.
First, RR Lyr is the nearest RRL in their sample.
Their absolute trigonometric parallax of RR Lyr using the HST is 
$\pi_\mathrm{abs}$ = 3.77 $\pm$ 0.13 mas, 
equivalent to $M_V$ = 0.54 $\pm$ 0.07 mag.
Second, the duration of their HST observations of RR Lyr is over 13 years
and they claimed that their parallax measurement for RR Lyr 
has been significantly improved compared to their previous study of the star 
\citep{benedict02}, $\pi_\mathrm{abs}$ = 3.82 $\pm$ 0.2 mas 
(equivalent to $M_V$ = 0.61 $\pm$ 0.10 mag).
If we take their new absolute magnitude ($M_V$ = 0.54 $\pm$ 0.07 mag),
the metallicity of RR Lyr  (\feh\ = 1.41 $\pm$ 0.13 dex)
and the slope in the \mvrr-\feh\ relation
by \cite{clementini03} or \cite{benedict11} (0.214 $\pm$ 0.047),
the absoulte visual magnitude of RRLs becomes
\mvrr\ = 0.52 $\pm$ 0.13 mag\footnote{The error in 
\mvrr\ is dominated  by the uncertainty in the metallicity measurement 
of RR Lyr, $\pm$0.13 dex.} at \feh\ = $-$1.50 dex.
This zero-point is in excellent agreement with the current standard value
adopted by others.
Therefore, Equation (\ref{eq:hstmv}) can be re-written as
\begin{equation}\label{eq:rev_hstmv}
M_V\mathrm{(RR)} = (0.214 \pm 0.047)(\mathrm{[Fe/H]} + 1.5) + (0.52 \pm 0.13).
\end{equation}
Our revised metallicity-luminosity relation for RRLs gives 
$(m-M)_0$ = 18.54 $\pm$ 0.13 mag for LMC, 
if we use $\langle V_0(\mathrm{RRL})\rangle$ = 19.064 $\pm$ 0.064
at \feh\ = $-$1.50 for RRLs in LMC by \cite{clementini03}.

\subsubsection{\mvrr\ from Fourier parameters}
Using the same method by \citet{kinman02},
we re-derive the $M_V\mathrm{(RRab)}$ of NGC~6723.
\citet{kinman02} used RR Lyrae and
the empirical transformation relation by \citet{kovacs01},
\begin{equation}
M_V\mathrm{(RR)} = -1.876\log P - 1.158 A_1 + 0.821 A_3 + K,\label{eq:mv:3}
\end{equation}
where $K$ is a zero-point constant which should be determined.
Using $P$ = 0.566837 day, $A_1$ = 0.31539 mag, $A3$ = 0.09768 mag
and $M_V(\mathrm{RR~Lyr})$ = 0.61 $\pm$ 0.10 of \cite{benedict02},
\cite{kinman02} obtained $K$ = 0.43.

As discussed above, the absolute magnitude of RR Lyr from \cite{benedict02}
is 0.07 mag fainter than that of \cite{benedict11}.
Using the same $P$, $A_1$, $A_3$ values, 
but $M_V(\mathrm{RR~Lyr})$ = 0.54 $\pm$ 0.07 
of \cite{benedict11} for RR Lyr, we obtained $K$ = 0.36.

\begin{figure}
\epsscale{1}
\figurenum{15}
\plotone{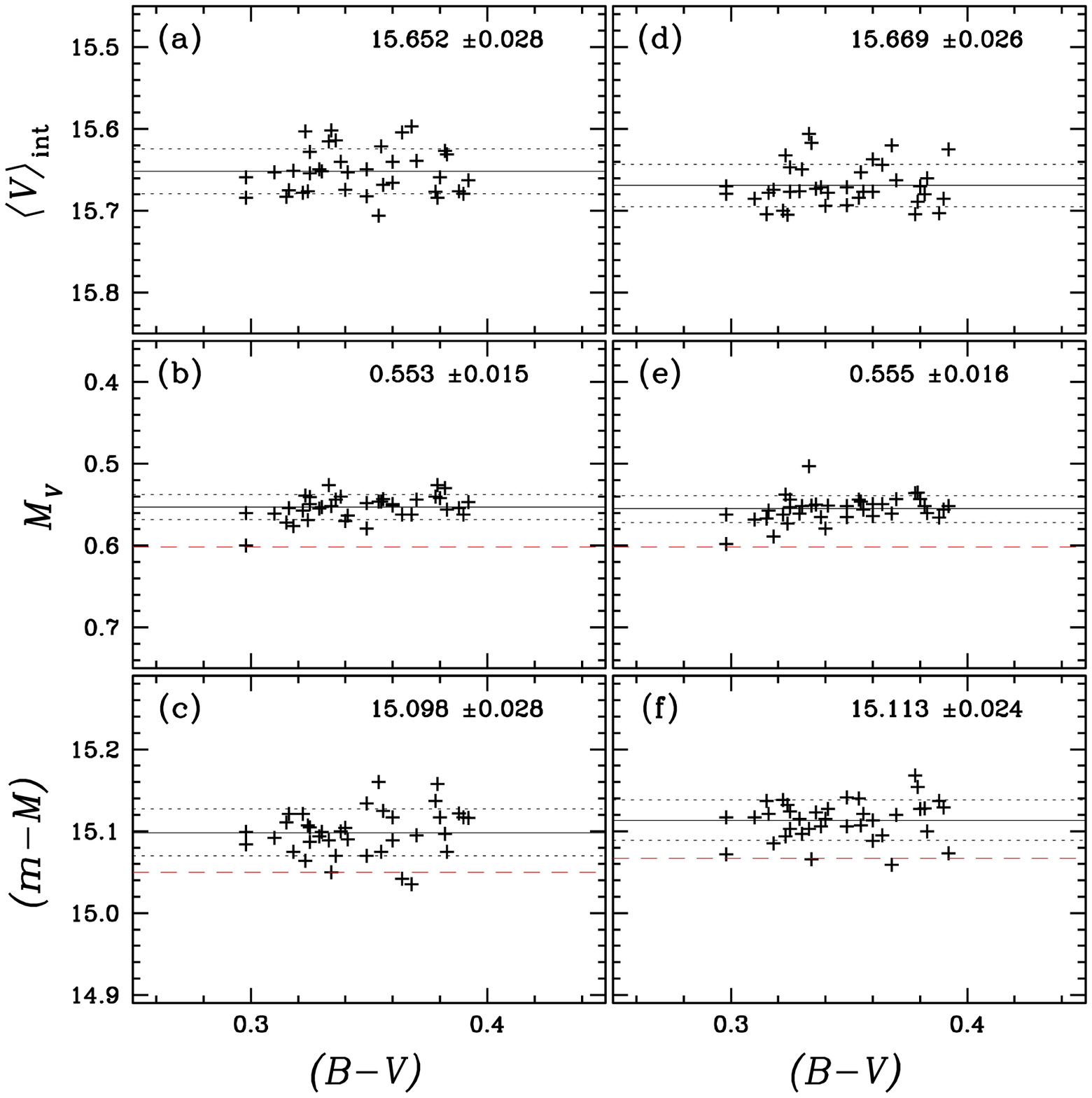}
\caption{A consistency check of Equation (\ref{eq:mv:3}) with $K$ = 0.36 mag using RRLs
in the globular cluster M3. (a) \& (d) Observed \vint\ of RRab type variables in M3
by Cacciari et al. (2005) and Jurcsik et al. (2012), respectively. 
The grey solid lines are for the mean values and the grey dashed lines
show the $\pm \sigma$ levels.
(b) \& (e) Calculated \mvrr\ using Equation (\ref{eq:mv:3}) with our zero-point,
$K$ = 0.36.
(c) \& (f) Apparent visual distance modulus for M3.
The red dashed lines indicate \mvrr\ and ($m-M$) for M3 RRab type variables
using the metallicity-luminosity zero-point by Catellan \& Cort\'es (2008).}
\label{fig:M3RRab}
\end{figure}

As a consistency check, we show \mvrr\ of the globular cluster M3 using 
Equation (\ref{eq:mv:3}) with our revised $K$ value and the apparent
visual distance modulus in Figure~\ref{fig:M3RRab}.
In panels (a) and (d), we show observed \vint\ of RRab type variables in M3
by \cite{cacciari05} and \cite{jurcsik12}, respectively.
As can be seen in the figure, both results are in good agreement.
Panels (b) and (e) show the calculated \mvrr\ using Equation (\ref{eq:mv:3})
with our revised zero-point, $K$ = 0.36. 
Note that \cite{cacciari05} obtained $M_V$ = 0.57 $\pm$ 0.02 mag for 
M3 RRab stars and 0.57 $\pm$ 0.04 mag for M3 RRc stars in their analyses,
and their results are in good agreement with our results,
0.553 $\pm$ 0.015 mag and 0.555 $\pm$ 0.016 mag from
RRL data by \cite{cacciari05} and \cite{jurcsik12}, respectively.
In panels (c) and (f), we show apparent visual distance modulus for M3
by comparing the observed \vint\ and the calculated $M_V$ of RRab stars.
We obtain apparent visual distance modulus values of 15.098 $\pm$ 0.028 mag
and 15.113 $\pm$ 0.024 mag from \cite{cacciari05} and \cite{jurcsik12}, respectively.
Our apparent visual distance modulus values for M3 are in excellent
agreement with that of \cite{federici12}, who presented $(m-M)$ = 15.11
for the cluster, indicating that our revised $K$ value works correctly.
Also shown in the figure with red dashed lines are \mvrr\ and $(m-M)$ for 
M3 RRab type variables using the metallicity-luminosity zero-point 
by \cite{catelan08}.
If we use the zero-point by \cite{catelan08}, \mvrr\ is fainter and
$(m-M)$ is smaller by about 0.05 mag than our results for M3.

\subsubsection{\mvrr\ for NGC~6723}
We show $M_V$ of individual RRab variable stars in NGC~6723
using Equation (\ref{eq:mv:3}) with $K$ = 0.36
in column (10) of Table~\ref{tab:rrabpam}.
The mean absolute visual magnitude in this case is 
$\langle M_V\mathrm{(RRab)}\rangle$ = 0.579 $\pm$ 0.041 mag
from all 18 RRab type variables and 
$\langle M_V\mathrm{(RRab)}\rangle$ = 0.581 $\pm$ 0.038 mag
from 10 regular RRab type variables in NGC~6723.
This value is in excellent agreement with our that expected from
our revised metallicity-luminosity relation, 
$\langle M_V\mathrm{(RRab)}\rangle$ = 0.578 $\pm$ 0.133 mag.
The difference in the absolute magnitude between that from the original
relation by \citet{jurcsik98} and that from the revised relation of \citet{kovacs01}
with our new zero-point constant becomes 0.270 $\pm$ 0.020 mag.
Note that if we use the zero-point constant by \citet{kinman02}, $K$ = 0.43,
then $\langle M_V\mathrm{(RRab)}\rangle$ = 0.651 $\pm$ 0.038 mag
for 10 regular RRab type variables in NGC~6723.

Our results also support the idea that the zero-point constant
in Equation (\ref{eq:mv:2}) should be decreased 
by about 0.2 mag as suggested by \citet{cacciari05}. 
If we use the reduced zero-point constant by 0.2 mag 
for RRc type variable stars in NGC~6723,
$\langle M_V\mathrm{(RRc)}\rangle$ = 0.620 $\pm$ 0.021 mag for 
four regular RRc variable stars is in marginal agreement with
that from RRab type stars within the measurement errors.

\subsection{The distance modulus}\label{s:dm}
\subsubsection{Fourier parameters}
We derive the distance modulus for NGC~6723 by comparing the observed \vint\ 
with the absolute visual magnitude of RRab variables $M_V\mathrm{(RRab)}$.
Using Equation (\ref{eq:mv:1})
and our interstellar reddening value, \ebv\ = 0.063 $\pm$ 0.015 mag,
we obtained $\langle (m-M)_0\rangle$ = 14.405 $\pm$ 0.083 mag from 18 RRab stars
and $\langle (m-M)_0\rangle$ = 14.418 $\pm$ 0.032 mag from 10 regular RRab stars
in NGC~6723.
We show our true distance modulus $(m-M)_0$
for individual variables in columns (7) of Table~\ref{tab:rrabpam}.
As discussed previously, Equation (\ref{eq:mv:1}) is based
on the fainter RRab absolute magnitude scale, hence
our $\langle (m-M)_0\rangle$ value is under-estimated.
Our derived $\langle (m-M)_0\rangle$ value should be corrected 
by 0.270 $\pm$ 0.020 mag.
Therefore, if we apply this correction term,
our true distance modulus of NGC~6723 using 10 regular RRab stars
becomes $\langle (m-M)_0\rangle$ = 14.681 $\pm$ 0.038 mag,
corresponding to a distance from the Sun of 8.63 $\pm$ 0.15 kpc.

We also make use of Equation (\ref{eq:mv:3}) with $K$ = 0.36.
We obtained $\langle (m-M)_0\rangle$ = 14.476 $\pm$ 0.074 mag 
from 18 RRab stars and 14.684 $\pm$ 0.036 mag from 10 regular RRab stars
using \ebv\ = 0.063 $\pm$ 0.015 mag and
the distance of the cluster from the Sun is 8.65 $\pm$ 0.14 kpc.
We show our results in column (11) of Table~\ref{tab:rrabpam}.

Our true distance modulus values from RRab type variable stars in NGC~6723
using the the relation by 
\cite{jurcsik98} with our correction term and
by \cite{kovacs01} with our revised zero-point
are in excellent agreement
with that of \citet{harris96}, $(m-M)_0$ = 14.69 mag.

We calculated the apparent and the true distance moduli
from RRc variable stars and
we obtained the average apparent distance modulus
$\langle (m-M)\rangle$ = 14.671 $\pm$ 0.096 mag
from all seven RRc variable stars and
14.643 $\pm$ 0.076 mag from four regular RRc variable stars.
Using \ebv\ = 0.063 mag, we obtained 
$\langle (m-M)_0\rangle$ = 14.476 $\pm$ 0.096 mag
from all seven RRc variable stars and
14.448 $\pm$ 0.076 mag from four regular RRc variable stars,
which are equivalent to distances from the Sun of 7.86 $\pm$ 0.35 kpc
and 7.76 $\pm$ 0.27 kpc, respectively.
If we apply the zero-point correction term for the absolute magnitude
of RRc type variables from Fourier parameters suggested by \citet{cacciari05},
0.20 $\pm$ 0.02 mag, the true distance modulus from four RRc variable stars
becomes \dmz\ = 14.648 $\pm$ 0.079 mag, corresponding to 8.50 $\pm$ 0.31 kpc,
for four regular RRc variable stars.
If we apply the same zero-point correction term as for RRab type variable stars
in Equation (\ref{eq:mv:3}), 0.270 $\pm$ 0.020,
the true distance modulus for NGC~6723 from 4 regular RRc variable stars becomes
$\langle (m-M)_0\rangle$ = 14.718 $\pm$ 0.079 mag, 
corresponding to the distance from the Sun of 8.78 $\pm$ 0.32 kpc.
Applying the correction term of 0.270 mag for RRc type variable stars
appears to make the distance modulus over-estimated
and the correction term by \cite{cacciari05}, 0.20 mag, appears to
be more adequate for RRc type variables.

\subsubsection{The Wesenheit function from Fourier parameters: $W(BV)_0$} 
\citet{kovacs98} introduced an alternative empirical method to estimate
true distance modulus of RRab type variables using the Wesenheit function;
\begin{equation}
(m-M)_0 = W(BV) - W_0(BV),
\end{equation}
where $W(BV)$ is the apparent Wesenheit function in $BV$ photometry defined to be
\begin{equation}
W(BV)  = V - R_V(B-V),
\end{equation}
and we adopt $R_V$ = 3.10.
$W_0(BV)$ is the true Wesenheit function and
\citet{kovacs98} derived the following transformation relation,
\begin{equation}
W_0(BV) = 0.676 - 1.943 P + 0.315 A_1 + 0.068 \phi_{41}. \label{eq:W0}
\end{equation}
To apply this transformation relation, we used \vint\ and \bvmag. 
We show the calculated $W_0(BV)$ for individual RRab variable stars
in column (8) and the true distance modulus based on this method, \dmz\,
in column (9) of Table~\ref{tab:rrabpam}.
We obtained $\langle (m-M)_0\rangle$ = 14.302 $\pm$ 0.037 mag,
which is equivalent to the distance from the Sun of 7.25 $\pm$ 0.12 kpc.
Our result is significantly smaller than that of \citet{harris96}.
However, the original calibration of Equation (\ref{eq:W0})
by \citet{kovacs98} was also based on the fainter magnitude of RRab variable stars
and the zero-point correction by 0.270 $\pm$ 0.020 mag is required
as we discussed above.
If we apply the zero-point correction term in Equation (\ref{eq:W0}),
the true distance modulus of NGC~6723
using the Wesenheit function by \citet{kovacs98} becomes
$\langle (m-M)_0\rangle$ = 14.572 $\pm$ 0.042 mag,
resulted in the distance from the Sun of 8.21 $\pm$ 0.16 kpc.

As can be seen in Table~\ref{tab:dm}, the distance modulus of NGC~6723 from the
Wesenheit function by \cite{kovacs98} is significantly smaller than those
from other methods. The improvement of Equation~(\ref{eq:W0}) in the future
would be very desirable.

\subsubsection{The Wesenheit function from theoretical models: 
$W(BV)_{th}$ }\label{ss:W_th}
\begin{figure}
\epsscale{1}
\figurenum{13}
\plotone{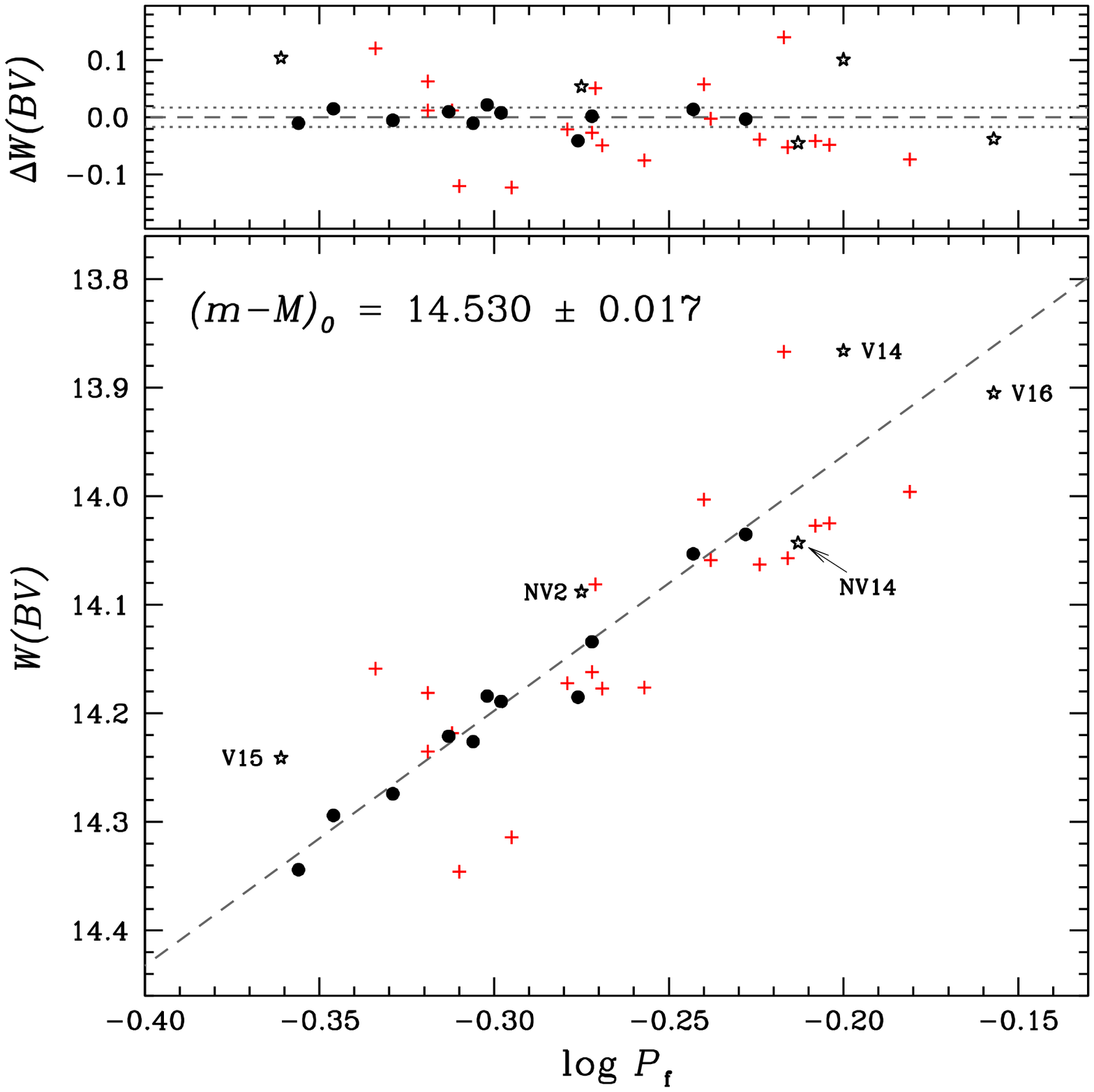}
\caption{A plot of the observed Wesenheit function $W(BV)$ as a function
of $\log P_f$ for RRab variable stars in NGC~6723.
The filled circles are for regular RRab variable stars and
the plus signs are for Blazhko RRab variable stars noted in Table~3.
The regular variables with large deviations from the theoretical Wesenheit
function, V14 and V15, the one suspected to suffer from large differential
reddening effect, NV14, and two more evolved variables, V16 and NV2,
are marked with star signs.
The grey dashed line is the theoretical Wesenheit function
for [Fe/H] = $-$1.23 dex and HB type of $-$0.08
from Cassisi et al.\ (2004) with the true distance modulus
of \dmz\ = 14.530 $\pm$ 0.017 mag,
resulted in the distance from the Sun of 8.05 $\pm$ 0.06 kpc.
If the zero-point correction is applied,
the true distance modulus is
\dmz\ = 14.610 $\pm$ 0.131 mag and
the distance from the Sun becomes 8.36 $\pm$ 0.50 kpc. 
Also note that the location of NV14
strongly suggests that NV14 is a cluster member.
The upper panel shows the residuals to the fit.
The grey horizontal dashed line is for the mean value and 
grey dotted lines show the $\pm \sigma$ level.}
\label{fig:W}
\end{figure}

\citet{cassisi04} studied the slope and the zero-point of 
the theoretical Wesenheit function incorporating evolutionary
and pulsational properties for various metallicity and HB type.
They provided the theoretical Wesenheit function $W(BV)_{th}$,
\begin{equation}
W(BV)_{th} = W_{BV}^{-0.3} + b_{BV}^{W}(\log P_\mathrm{f} + 0.3),
\end{equation}
where $P_f$ is the fundamental period and 
$W_{BV}^{-0.3}$ is the $W(BV)$ value at $\log P_f = -0.3$.
By interpolating the data given in Table 3 of \citet{cassisi04},
we obtained $W_{BV}^{-0.3}$ =  $-$0.329 and $b_{BV}^{W}$  = $-$2.341
for \feh\ = $-$1.23 dex and the HB type of $-$0.08 for NGC~6723.
Then the true distance modulus, $(m-M)_0$, of NGC~6723 can be written as
\begin{eqnarray}
(m-M)_0 &=& W(BV) - W(BV)_{th} \nonumber \\
     &=& W(BV) + 2.341\log P_\mathrm{f} + 1.031,
\end{eqnarray}
at a given $\log P_\mathrm{f}$.
In Figure~\ref{fig:W}, we show a plot of $W(BV)$ of individual RRab variable
stars as a function of their $\log P_\mathrm{f}$.
We calculated the least square fit of the theoretical Wesenheit
function to the observed data with a fixed slope 
on the $\log P_\mathrm{f}$ versus $W(BV)$ plane.
We obtained the true distance modulus of 
\dmz\ = 14.531 $\pm$ 0.061 for all 35 RRab variable stars
and 14.518 $\pm$ 0.043 for 14 regular RRab variable stars\footnote{We do not
use NV14 in our calculation because it appears to suffer from larger
differential reddening effect than other RRLs in NGC~6723.
However, as can be seen in Figure~\ref{fig:W}, the location of NV14 agrees well
with other RRLs in NGC~6723.
Since the Wesenheit function is reddening free, the location of NV14
in Figure~\ref{fig:W} strongly suggests that NV14 is a member variable star.}.
As shown in Figure~\ref{fig:W}, the regular RRab variables V14 and V15
significantly deviate from the theoretical Wesenheit function shown with
the dashed line in the plot. Note that V14 and V15 have somewhat brighter
\vint\ than other regular RRab type variable stars.
If we exclude V14, V15, V16 and NV2 (see Appendix~\ref{ap:s:notes}), 
our true distance modulus of NGC~6723
based on the theoretical Wesenheit function by \citet{cassisi04} becomes
\dmz\ = 14.530 $\pm$ 0.017 from 10 RRab variable stars
and the distance from the Sun is 8.05 $\pm$ 0.06 kpc.

\begin{figure}
\epsscale{1}
\figurenum{16}
\plotone{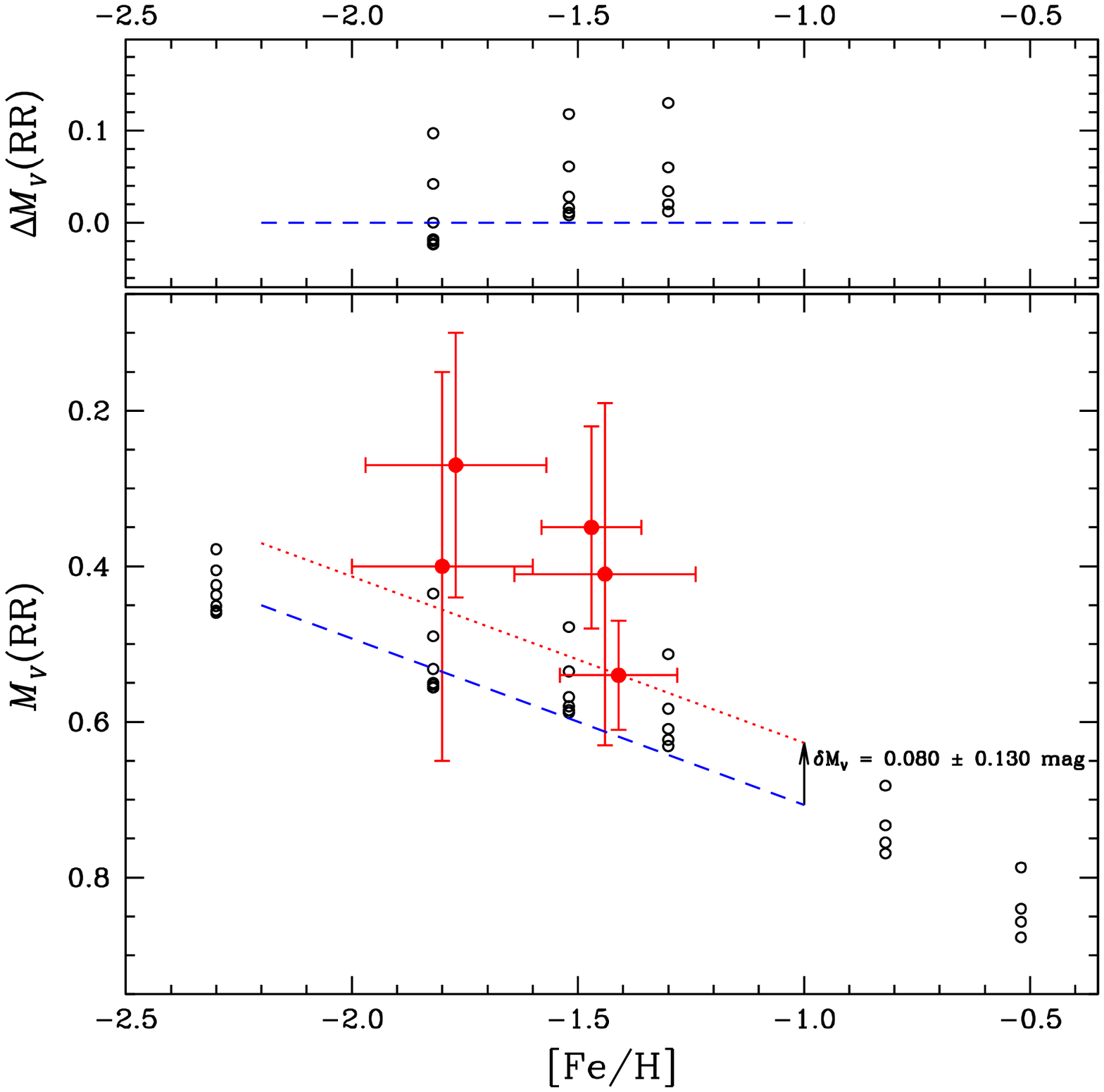}
\caption{(Bottom) A plot of \mvrr\ as a function of metallicity.
The filled red dots represent five RRLs with absolute
trigonometric parallaxes studied by Benedict et al.\ (2011) and
the open circles represent the theoretical mean absolute visual magnitudes of
RRLs with different HB types at fixed metallicity
(Cassisi et al.\ 2004). 
The red dotted line is our revised metallicity-luminosity relation
based on RR Lyr by Benedict et al.\ (2011),
\mvrr\ = 0.214(\feh\ + 1.50) + (0.52 $\pm$ 0.13).
As can be seen in the figure, the linear fit
does not match with the lower envelope of the absolute visual magnitudes of RRLs
from theoretical model predictions by Cassisi et al.
The blue dashed line is the least-square fit to the theoretical models
using a fixed slope adopted by Benedict et al. 
The difference in magnitude between the two fitted lines is 0.080 $\pm$ 0.130 mag.
(Upper) Residuals to the fit.}
\label{fig:cassisi}
\end{figure}

The theoretical Wesenheit function of RRL can provide a powerful means
to determine the distance of GCs because it is reddening-free and, furthermore,
it incorporates evolutionary and pulsational properties for various metallicities
and HB types of stellar populations.
However, our distance modulus of NGC~6723 
from the theoretical Wesenheit function by \citet{cassisi04} appears 
to be slightly smaller than those from other methods (see Table~\ref{tab:dm}).
In Figure~\ref{fig:cassisi}, we show a plot of \mvrr\
from the absolute trigonometric parallaxes by \citet{benedict11} 
as a function of metallicity along with our revised metallicity-luminosity relation 
as presented in Equation (\ref{eq:rev_hstmv}).
Also shown are the theoretical mean absolute visual magnitudes of
RRLs with different HB types at fixed metallicity
\citep{cassisi04}.
As can be seen in the figure, our revised linear fit to the measured $M_V(\mathrm{RR~Lyr})$ 
by \citet{benedict11} does not match with the absolute visual magnitudes of RRLs
from theoretical model predictions by \citet{cassisi04}.
We performed a least-square fit to the theoretical models
using a fixed slope adopted by \citet{benedict11}, 0.214,
and we obtained the offset value in absolute visual magnitude of
$\delta M_V$ = 0.080 $\pm$ 0.130 mag, in the sense that
the theoretical model predictions by \citet{cassisi04} are fainter.
\cite{cassisi04} also noted that their theoretical prediction for
\mvrr\ is in good agreement with $M_V(\mathrm{RR~Lyr})$ by \cite{benedict02},
whose $M_V(\mathrm{RR~Lyr})$ is 0.08 mag fainter than that of \cite{benedict11}. 
If we apply the zero-point correction term for the theoretical model
predictions by \citet{cassisi04}, the true distance modulus of NGC~6723
from the theoretical Wesenheit function becomes
\dmz\ = 14.610 $\pm$ 0.131 mag and the distance from the Sun is 8.36 $\pm$ 0.50 kpc.

It would be very desirable to perform detailed calculation of
the Wesenheit function from the theoretical models with 
an updated absolute visual magnitude scale of RRLs in the future.

\subsubsection{$M_V(RR)$ versus $V(RR)$}
Finally, we also derive the apparent and true distance moduli of NGC~6723
by using the average magnitude of the RRLs.
In Figure \ref{fig:cmdHB}, we show the CMD of the HB region
using the  magnitude-weighted integrate colors and 
the intensity-weighted integrated magnitudes for RRLs.
As shown in the Figure, the average magnitude of RRLs
(excluding V14, V15, V16, NV1, NV2 and NV14; see Appendix~\ref{ap:s:notes}) is
$\langle V\mathrm{(RR)}\rangle_\mathrm{int}$  = 15.459 $\pm$ 0.055 mag.
If we use the absolute visual magnitude of RRLs 
from our revised metallicity-luminosity relation
presented in Equation (\ref{eq:rev_hstmv}),
$M_V\mathrm{(RR)}$ = 0.578 $\pm$ 0.133 mag at \feh\ = $-$1.23 $\pm$ 0.11 dex,
and  the apparent distance modulus of NGC~6723 is
$(m-M)$ = 14.881 $\pm$ 0.143 mag.
Then the true distance modulus of NGC~6723 becomes
\dmz\ = 14.686 $\pm$ 0.143 mag if we use \ebv\ = 0.063 mag,
corresponding to the distance from the Sun of 8.65 $\pm$ 0.57 kpc.

\medskip
In Table~\ref{tab:dm}, we summarize our derived true distance
modulus and the distance from the Sun for NGC~6723 using various methods.
Our \dmz\ ranges from 14.57 to 14.69 mag with the unweighted mean value
of 14.64 $\pm$ 0.05 mag and 
the distance of NGC~6723 from the Sun ranges from 8.21 to 8.65 kpc
with the unweighted mean value of 8.47 $\pm$ 0.19 kpc
(the errors are those of the mean).

\section{$V(RR)$ versus RADIAL DISTANCE}\label{s:vhb}
Recently, \citet{majaess12} claimed that photometry of RRLs
in GCs depends on the radial distance from the center
and the photometric contamination in the crowded central region
of GCs may affect the inferred parameters, such as distance,
absolute magnitude, etc.
With the exception of M3, the difference in the distance modulus
derived from all RRLs and using only those in the outskirt
of the cluster can be as large as 0.25 mag,
in the sense that the RRLs in the outskirt of the GCs
are fainter than those in the crowded central regions.
They found an average difference for the five GCs in their Table~1 of
about 0.1 mag.

We explore the variation of magnitude against the radial distance from
the center for all the RRLs found in NGC~6723.
We calculated the angular distance $r$ of each RRL
from the center of the cluster using the Spherical law of cosine,
\begin{equation}
r = \arccos[\sin\delta_0\sin\delta + \cos\delta_0\cos\delta\cos(\alpha_0 - \alpha)],
\end{equation}
where $\alpha_0$ and $\delta_0$ are the right ascension and the declination
of the center of NGC~6723 and $\alpha$ and $\delta$ are those of individual RRLs.
For our calculations, we adopted the coordinates for the cluster center
measured by \citet{glodsbury};
$\alpha_0$ = 18:59:33.15 and $\delta_0$ = $-$36:37:56.1.

\begin{figure}
\epsscale{1}
\figurenum{17}
\plotone{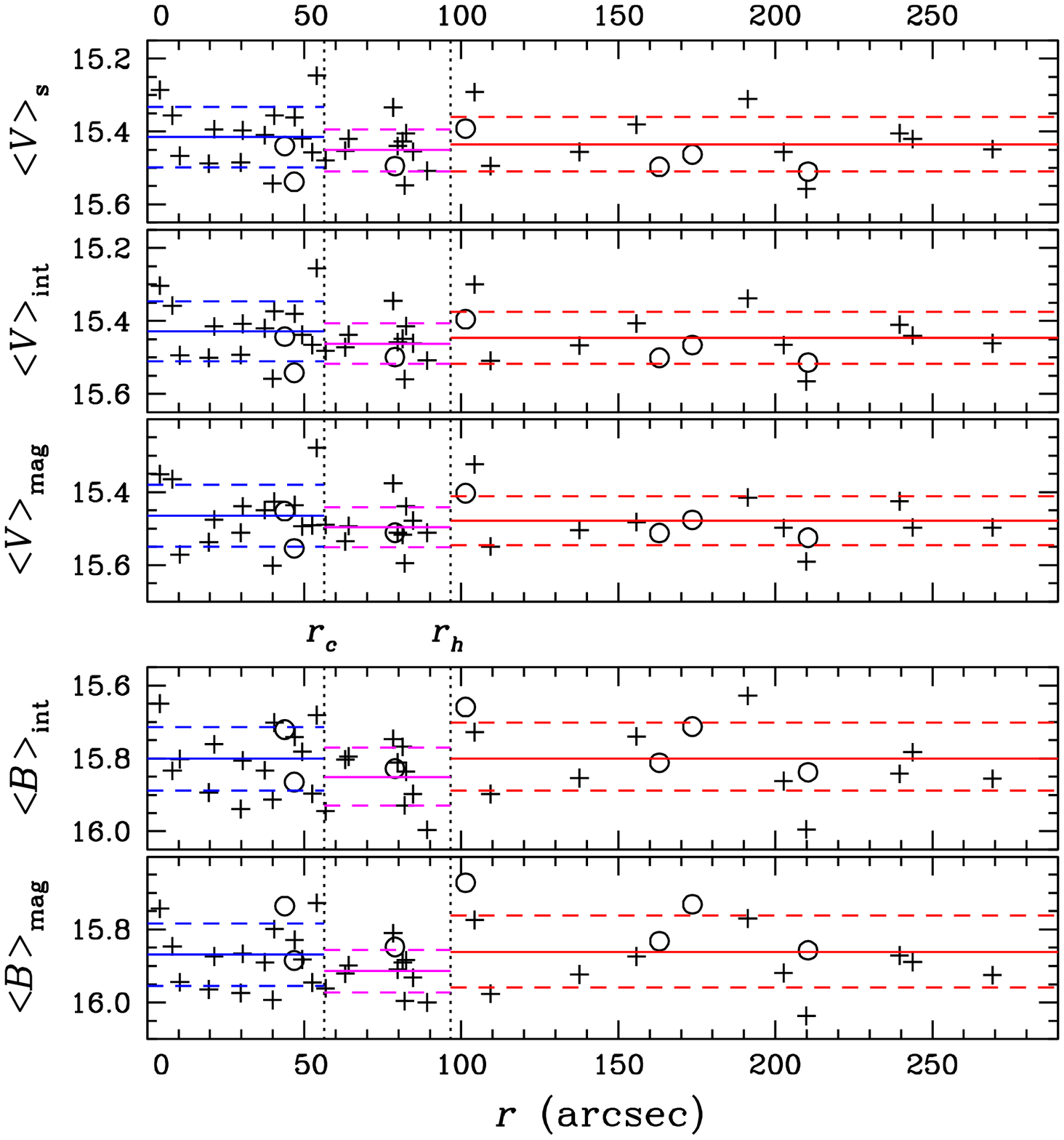}
\caption{RRL magnitudes as a function of the radial distance
from the center of the cluster.
The plus signs denote the RRab type variable stars and the open circles
denote the RRc type variable stars.
The vertical dotted lines denote the core and the half-mass radii of the cluster.
The horizontal solid lines are for the mean magnitudes and dashed lines show
the $\pm\sigma$ levels.
The $BV$ magnitudes from three different regions agree within measurement
errors and the radial gradient cannot be seen.}
\label{fig:magVSrad}
\end{figure}

In Figure~\ref{fig:magVSrad}, we show plots of $BV$ magnitudes of
all RRLs in NGC~6723 as functions of the radial distance.
Also shown are the average magnitudes and the dispersions in the magnitude
($\pm\sigma$ levels) in the central region ($r$ $\leq$ $r_c$, the core radius),
in the intermediate region ($r_c$ $\leq$ $r$ $\leq$ $r_h$, the half-light radius)
and in the outskirt region ($r$ $\geq$ $r_h$) of the cluster.
The $BV$ magnitudes of RRLs 
from three different regions agree well within measurement
errors and the radial gradient does not appear to exist, suggesting that
the photometric contamination in the central part of NGC~6723 is not severe.

\begin{figure}
\epsscale{1}
\figurenum{18}
\plotone{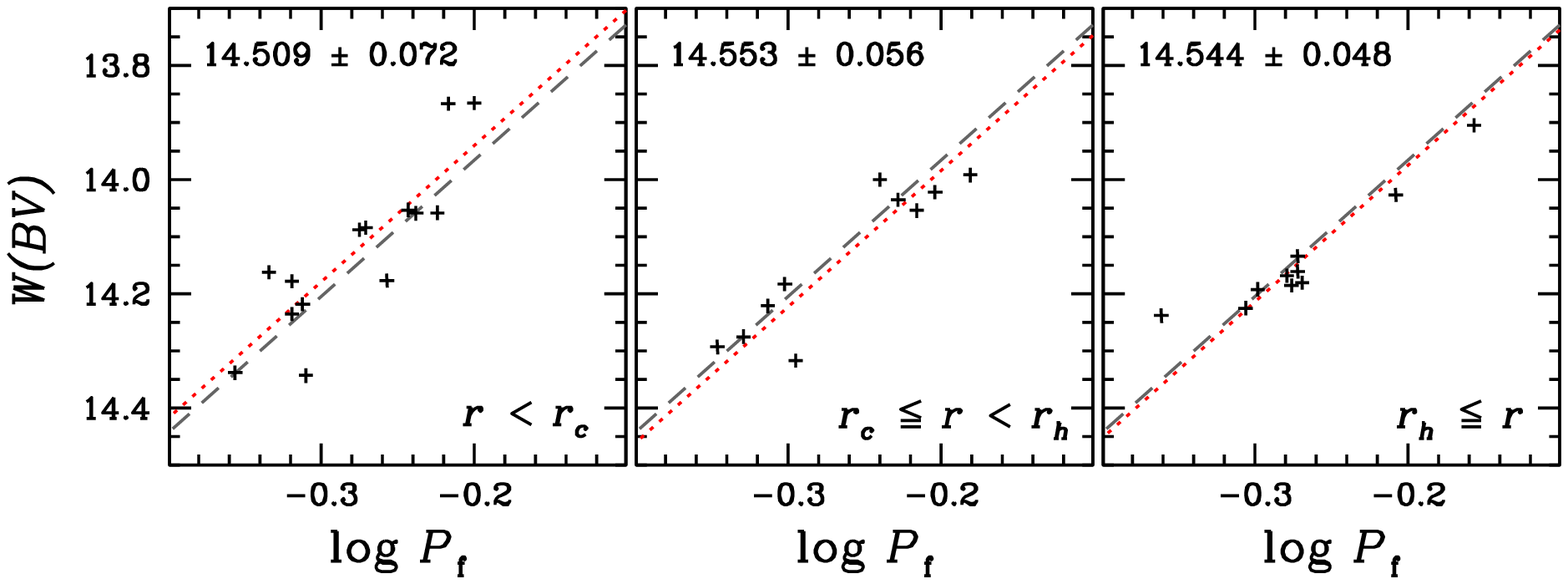}
\caption{Plots of  $W(BV)$ versus $\log P_f$ at three different radial zones.
The grey dashed lines in each panel are the theoretical Wesenheit function
with \dmz\ = 14.531 $\pm$ 0.061 mag using all 35 RRab type variable stars in NGC~6723.
The red dotted lines are the theoretical Wesenheit function with distance modulus
measured in each radial zone. 
The distance moduli measured from the theoretical Wesenheit function
from three different regions agree within measurement
errors and the radial gradient cannot be seen.}
\label{fig:Wrad}
\end{figure}

An alternative approach to examine a potential radial gradient
of RRL magnitudes can be found in Figure~\ref{fig:Wrad}, where we show 
plots of $W(BV)$ versus $\log P_f$ in three different radial regions.
We calculate the distance modulus in each radial zone
using the theoretical Wesenheit function with a fixed slope.
Also shown in the figure is the theoretical Wesenheit function
with \dmz\ = 14.531 $\pm$ 0.061 mag using all 35 RRab type variable stars
in NGC~6723 as a reference (see section \ref{ss:W_th}).
Again, the distance moduli measured from the theoretical Wesenheit function
from three different regions agree well within measurement errors
and the radial gradient cannot be seen.

\begin{figure}
\epsscale{1}
\figurenum{19}
\plotone{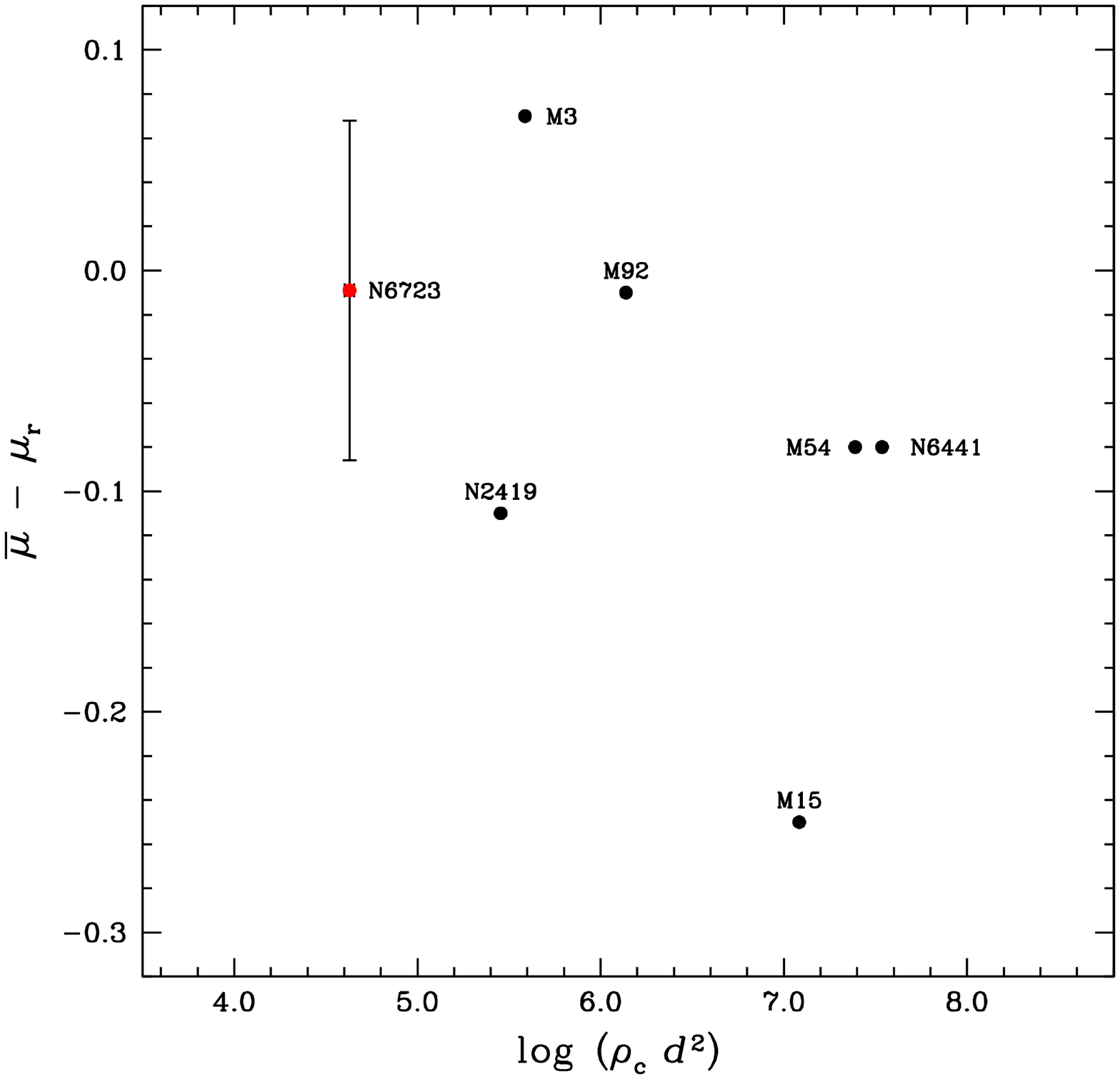}
\caption{A Plot of $\log \rho_c d^2$ versus $\bar{\mu} - \mu_r$
for 6 GCs from Majaess et al.\ (2012) and NGC~6723 from our current study,
where $\rho_c$ is the central luminosity density in units of solar luminosities
per cubic parsec, $d$ is the distance from the Sun in kpc, 
and $\bar{\mu} - \mu_r$ is defined to be the distance spread between
the average computed using all RRLs and
only those near the periphery (Majaess et al.\ 2012) and it is a measure
of the photometric contamination of GCs.
The figure strongly suggests that the degree of the photometric contamination
($\propto |\bar{\mu} - \mu_r|$) depends on the apparent crowdedness
($\propto \rho_c d^2$) of the globular cluster system.}
\label{fig:blend}
\end{figure}

To understand the radial gradient of RRL magnitude, 
in Figure~\ref{fig:blend} we show  
a plot of  $\log \rho_c d^2$ versus $\bar{\mu} - \mu_r$
for 6 GCs from Majaess et al.\ (2012) and NGC~6723 from our current study,
where $\rho_c$ and $d$ are the central luminosity density
in units of solar luminosities per cubic parsec and
the distance from the Sun in kpc, respectively, from \citet{harris96}.
We define $\rho_c d^2$ to be the apparent crowdedness of the central region
assuming that the mass-segregation is negligible in GCs under consideration.
Also $\bar{\mu} - \mu_r$ is defined to be the distance spread between
the average computed using all 35 RRLs and only those near 
the periphery (Majaess et al.\ 2012) and it is a measure
of the photometric contamination of the central region of a given GC system.
Since $\rho_c d^2$ is a measure of the apparent crowdedness
of the central region of GC systems,
it is natural to expect that as $\rho_c d^2$ increases,
the degree of photometric contamination, such as resulted from blending, increases.
Therefore one expects to have 
\begin{equation}\label{eq:blend}
|\bar{\mu} - \mu_r| \propto \log \rho_c d^2
\end{equation}
in GC systems as shown in Figure~\ref{fig:blend}.

Since the LSST, for example, can detect RRLs in the local group of galaxies
out to $\approx$ 1 Mpc scale in several years of time span \citep{lsst} 
and the future 30-m to 40-m class telescopes are expected to detect RRLs
in nearby groups of galaxies, the utility of RRLs
as a primary distance indicator will become more important.
A more thorough study of the relation between $|\bar{\mu} - \mu_r|$ and
$\log \rho_c d^2$ with an extended sample of GCs 
or high density regions would be desirable in the future, because if real, 
this relation can play a significant role to mitigate the discrepancy in 
establishing cosmic distance scale using RRLs.

\section{SUMMARY}
We have presented $BV$ CCD photometry for 54 variables in the metal-rich 
inner halo globular cluster NGC~6723, including 22 newly discovered variables.
We found that 13 RRLs and one SX Phe variable star are cluster member stars
and two $\delta$ Sct variable stars and five  W UMa type eclipsing binary stars
are most likely off-cluster field stars.

New light curves, periods and photometric parameters for all those variables
are presented. With the discovery of new RRLs, the mean periods of the RRab and
RRc variables in the cluster are $\langle$P$_{ab}$$\rangle$ = 0.541 $\pm$ 0.066 day 
from 35 RRab type variable stars and $\langle$P$_{c}$$\rangle$ = 0.292 $\pm$ 0.030 day
from seven RRc type variable stars, respectively.
The number ratio of the RRc type variable stars to the total number of
the RRLs is n(c)/n(ab+c) = 0.167.
Our new mean periods and the RRL distribution of NGC~6723 
are in good agreement with those of a typical Oosterhoff group I globular cluster.
We also obtained \vint\ = 15.459 $\pm$ 0.055 mag from 30 RRab type
and seven RRc type variable stars and it is in good agreement with
previous measurements by others.

We carried out the Fourier decomposition analysis for the RRLs in NGC~6723.
Using empirical transformation relations by \citet{jk96} and \citet{morgan07},
we obtained \feh$_\mathrm{ZW}$ = $-$1.27 $\pm$ 0.09 dex from RRab 
and $-$1.18 $\pm$ 0.15 dex from RRc variable stars.
The unweighted mean of the photometric metallicity of NGC~6723
in the Zinn \& West metallicity scale becomes
$\langle$[Fe/H]$_\mathrm{ZW}\rangle$ = $-$1.23 $\pm$ 0.11 dex,
consistent with the previous metallicity measurement of the cluster
from the high resolution spectroscopic study of red-giant branch stars
by \citet{fullton96}.
 
We investigated the interstellar reddening value by comparing 
the intrinsic color using empirical transformation relations from
the Fourier and the pulsational parameters and the observed color
of individual RRab type variable stars in NGC~6723.
We also provided a new calibration of the interstellar reddening values
on the HST ACS photometric system using the ACS GC survey by \citet{ACS}.
When compared to the interstellar reddening values from 
the HST ACS main-sequence photometry of NGC~6723,
the \ebv\ values from the Fourier and the pulsational parameters
of RRab type variable stars appear to slightly over-estimate
the foreground interstellar reddening value.
Furthermore, we found that the slope in the relation between
the intrinsic color derived from the Fourier and the pulsational parameters 
and the observed color of RRab type variable stars is not unity.
As a consequence, the interstellar reddening values
from the Fourier and the pulsational parameters 
have substantial gradient against the observed color
and compressed the derived \bvz\ distribution of RRab type variable stars.
The later effect may explain the compressed \bvz\ distribution of 
M3 RRLs noted by \citet{cacciari05} when they used
the empirical transformation relation by \citet{kovacs99}.
Our results also suggest that derived intrinsic colors of individual variable stars
from the Fourier and the pulsational parameters may not be reliable, confirming
the similar results by \citet{cacciari05}.
For the interstellar reddening value of NGC~6723, we adopted the unweighted mean 
of those  from the static color \bvs\ of RRLs of our current study
and from the isochrone fitting by \cite{ACS}, finding \ebv\ = 0.063 $\pm$ 0.015.

We discussed that RR Lyr may not be an evolved star, in contrast to
the recent Str\"omgren photometry study by \cite{catelan08} who
claimed that RR Lyr is about 0.06 mag brighter than the ZAHB.
The reason for having brighter luminosity for RR Lyr by \cite{catelan08}
is that their analysis was relied on the theoretical HB models by
\cite{catelan04}, whose HB absolute magnitude levels are 0.08 mag
fainter than the current standard value.

Using the recent HST absolute trigonometric parallax measurements of RR Lyr
by \cite{benedict11}, we obtained the following metallicity-luminosity
relation,
\begin{eqnarray}
M_V\mathrm{(RR)} &=& (0.214 \pm 0.047)(\mathrm{[Fe/H]} + 1.5) + (0.52 \pm 0.13). \nonumber
\end{eqnarray}
Our revised relation is in excellent agreement with the current
standard absolute visual magnitude of RRLs,
\mvrr\ = 0.52 mag at \feh\ = $-$1.50 dex.
Also our revised metallicity-luminosity relation for RRLs gives 
$(m-M)_0$ = 18.54 $\pm$ 0.13 mag for LMC, 
if we use the $V_0(\mathrm{RRL})$ value in LMC by \cite{clementini03}.

Our results for \mvrr\ of NGC~6723 confirmed that original transformation relations
of the absolute visual magnitude of RRLs by \citet{jurcsik98} and \citet{kovacs98}
are 0.270 $\pm$ 0.020 mag fainter than the current standard $M_V\mathrm{(RR)}$ value.
We calibrated the new zero-point constant of the empirical transformation 
to derive the absolute visual magnitude of RRab type variable star by \citet{kovacs01}.
Using our own zero-point constant, we obtained 
$\langle M_V\mathrm{(RR)}\rangle$ = 0.581 $\pm$ 0.038 mag for NGC~6723.
We then derived the distance modulus of NGC~6723 using 
(i) $M_V\mathrm{(RR)}$ from the Fourier parameters,
(ii) Wesenheit functions of the clusters,
and (iii) the $M_V\mathrm{(RR)}$-\feh\ relation.
Our true distance modulus of NGC~6723 ranges from 14.57 to 14.69 mag 
with the unweighted mean value of 14.65 $\pm$ 0.05 mag and 
the distance of NGC~6723 from the Sun ranges from 8.21 to 8.65 kpc
with the unweighted mean value of 8.47 $\pm$ 0.17 kpc.

Finally, we examined the RRL magnitude dependencies on the radial distance
from the center of the cluster, as claimed by \citet{majaess12}. 
We found no evidence of the radial gradient in the RRL magnitude in NGC~6723.
We showed that it is natural to expect to have more severe photometric
contamination with more severe degree of the apparent crowdedness.
We suggested that there appears to exist a relation between the degree
of photometric contamination and the apparent crowdedness of the central
region of globular cluster systems, 
\begin{eqnarray}
|\bar{\mu} - \mu_r| &\propto& \log \rho_c d^2. \nonumber
\end{eqnarray}
If real, this relation can play a significant role to mitigate the discrepancy 
in establishing cosmic distance scale using RRLs
in resolved stellar populations.

\acknowledgments
J.-W.L. acknowledges financial support from the Basic Science
Research Program (grant No. 2010-0024954) and the Center
for Galaxy Evolution Research through the National Research
Foundation of Korea.
J.-W.L. also thanks Dr.\ Kov\'acs for providing Fourier decomposition 
FORTRAN programs.
We thank the anonymous referee for a thorough and beneficial review.
Some of the data presented in this paper were based on observations
made with telescopes operated by the SMARTS consortium.

\appendix

\section{Notes on V14, V15, and V16}\label{ap:s:notes}
The period shift is the measure of the period difference of RRLs at 
fixed temperatures, the comparisons usually being made between
GCs' variables \citep{sandage81}.
The internal period shift analysis, which compares periods of RRLs at a fixed
temperature in a given GC, can also be useful to understand 
the physical properties of RRLs of the GC under consideration \citep{lc99b}. 
The trouble with the period shift analysis is that
the temperature of the pulsating RRLs is difficult to measure.
Fortunately, the blue amplitude $A_B$  known to be 
a good temperature indicator for RRab type variables
following the pioneering work by \cite{sandage81} \citep[see also][]{carney92}.

The period-density relation states that the period of a pulsating star
is inversely proportional to the square root of the mean density.
Assuming the same mass and the same effective temperature,
the less dense star would have a longer period.
Qualitatively, at a fixed temperature and mass, a lower density star
has a larger radius and hence a greater luminosity.
More precisely, one can make use of the relations given by \cite{vAB71}.
For the fundamental mode, the period of an RR Lyrae variable is
\begin{equation}
\log P = -1.772 - 0.68 \log \frac{M} {M_{\sun}} +
0.84 \log \frac{L}{L_{\sun}} + 3.48 \log \frac{6500} {T_{eff}}. \label{eq:vAB}
\end{equation}
For example, by employing an internal period shift analysis,
\cite{lc99b} showed that the RRab type variable V10 in GC NGC~7089(M2)
is brighter than other RRLs in the cluster from 
its longer period at a fixed temperature.

\begin{figure}
\epsscale{1}
\figurenum{12}
\plotone{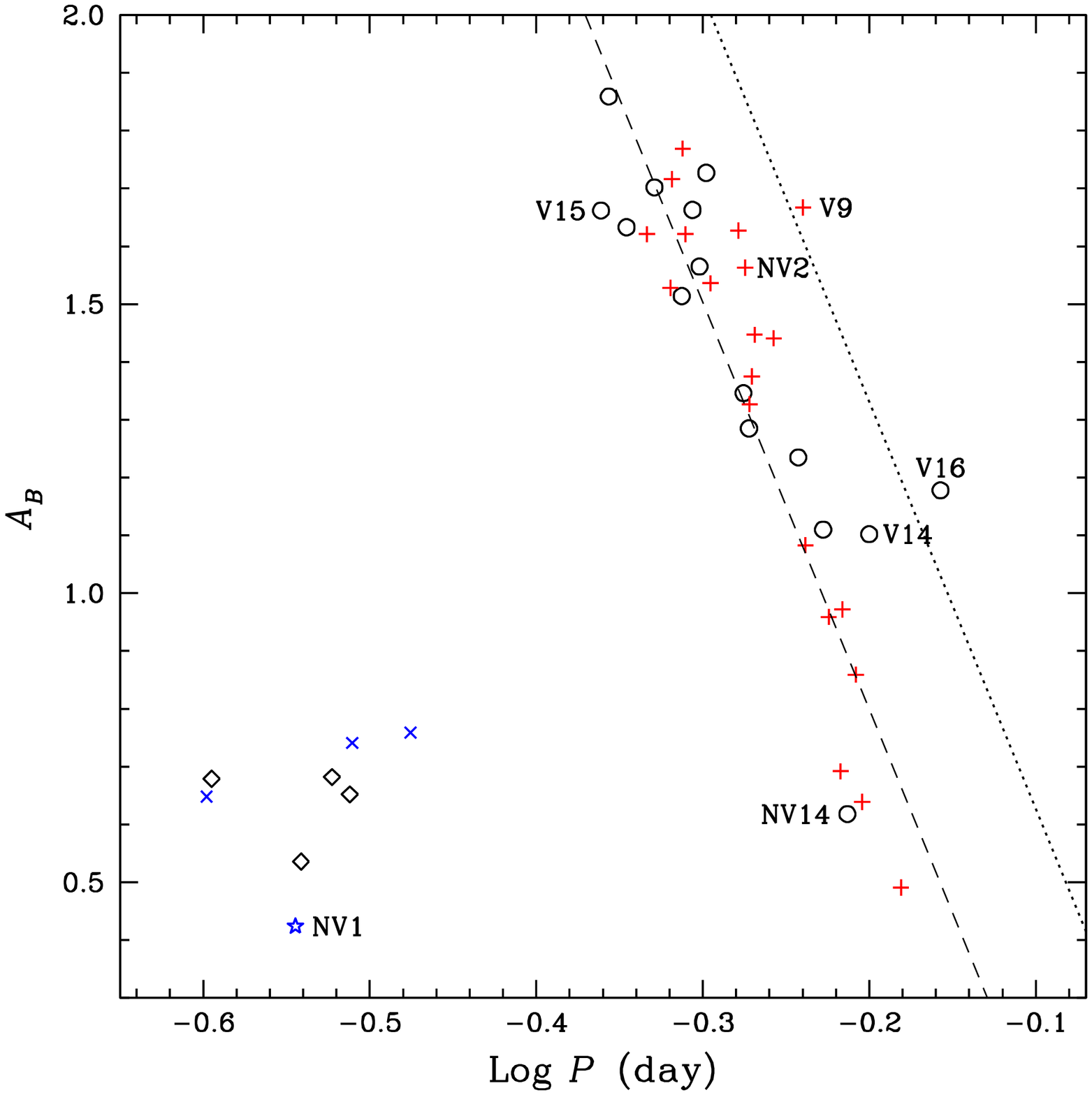}
\caption{The blue amplitude ($A_B$) as a function of $\log P$.
The regular RRab variable stars tagged from their light curves
are represented by open circles
and the Blazhko RRab variable stars by red plus signs.
The regular RRc variable stars are represented by open diamonds
and Blazhko RRc variable stars by blue crosses.
The dashed line is the period-amplitude relation for 
the Oosterhoff I globular cluster M3 and the dotted line for
the Oosterhoff II globular cluster M2 (Lee \& Carney 1999b).
The amplitude-period relation for NGC~6723 is in good agreement
with that of M3.
Note that the regular RRab variable V16 has a large $\log P$
at a given blue amplitude $A_B$, suggesting that V16 is more luminous
than other variables in the cluster (see Appendix~\ref{ap:s:notes}). }
\label{fig:PvsA_B}
\end{figure}

\begin{figure}
\epsscale{1}
\figurenum{A1}
\plotone{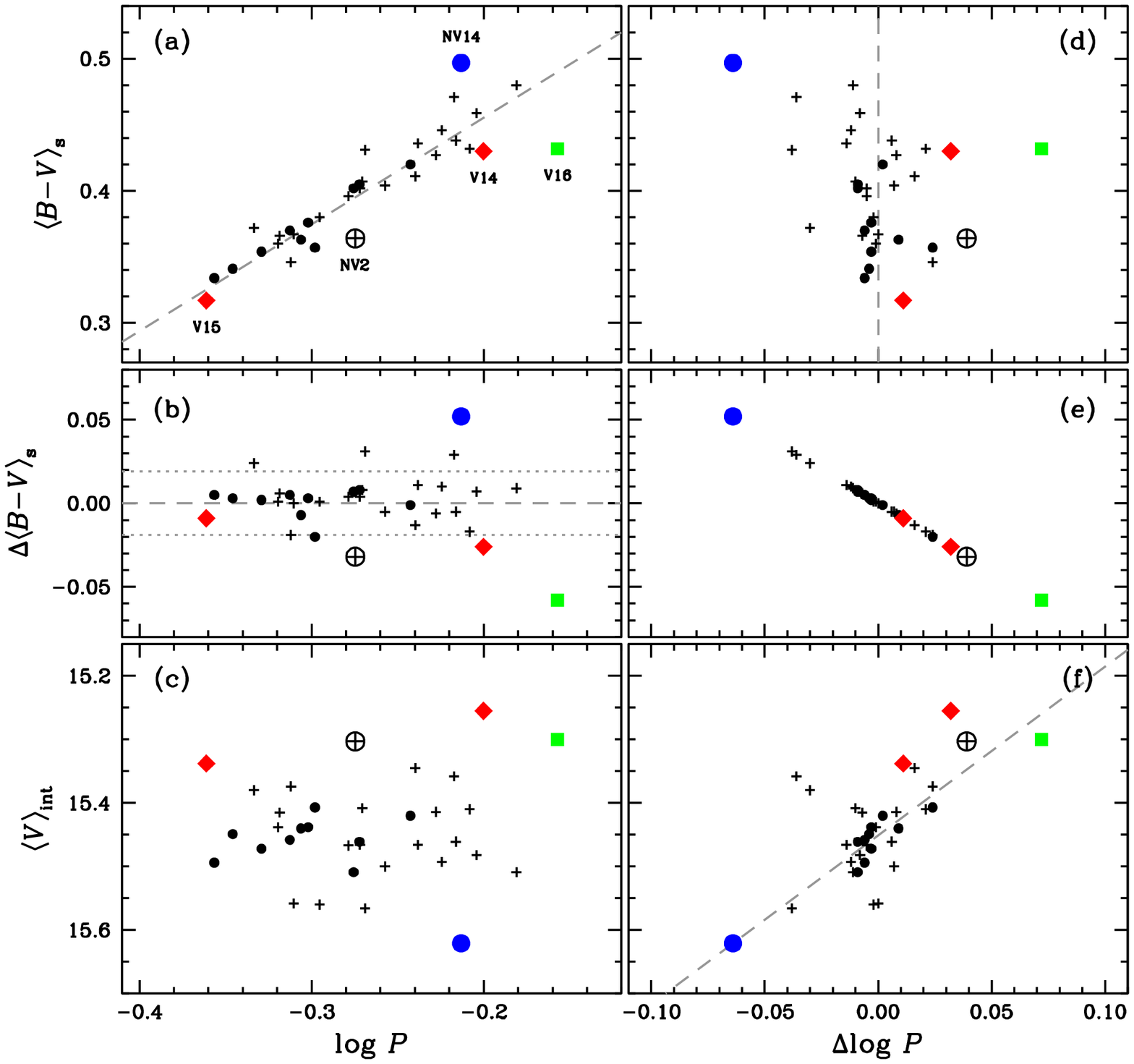}
\caption{(a) -- (c) Plots of \bvs, $\Delta$\bvs, and \vint\ of RRab type variables
as functions of $\log P$. (d) -- (f) Same as (a) -- (c) but for $\Delta \log P$.
The filled circles are for regular RRab variable stars and
the plus signs are for Blazhko RRab variable stars noted in Table~3.
The regular RRab variable stars with large deviation from the theoretical
Wesenheit function as in Figure~\ref{fig:W}, V14 and V15, are marked with
red diamonds. The filled green rectangle is for V16,
the filled blue circle for NV14, and the circle with the plus sign for NV2.
In panel (a), the grey dashed line represent the least-square fit to the data.
In panel (b), the grey dotted lines show the $\pm \sigma$ level.
Note that V16 is likely a more evolved RRL while V15 may suffer from
blending effect. V14 is likely an evolved RRL with photometric contamination.}
\label{fig:outlier}
\end{figure}

The regular RRab type variable V16 in NGC~6723 has a longer period than 
other variables in the cluster at the fixed $A_B$ 
(i.e.\ $\approx$ at the fixed temperature) 
as shown in Figure~\ref{fig:PvsA_B}.
Assuming the same mass and the same effective temperature 
in Equation~(\ref{eq:vAB}), the longer period of V16 indicates that
V16 is more luminous than other RRLs, consistent with our observations.
As shown in Table~\ref{tab:rrlyrae} or Figure~\ref{fig:PvsA_B},
V16 is 0.159 mag brighter than the mean visual magnitude of RRLs
in NGC~6723 [see also Figure~\ref{fig:outlier} (c)]. 
 
In Figure~\ref{fig:outlier} (a),
we show a plot of \bvs\ versus $\log P$ for RRab type variables in NGC~6723.
Although, the blue amplitude $A_B$ of RRab type variables is related to
the effective temperature as mentioned above, it is thought that
the mean color of the variables is more proper temperature indicator
than the blue amplitude $A_B$ is.
However, using \bvs\ has a disadvantage that it is not a reddening-free parameter,
such as $A_B$.
In the figure, we also show the linear fit to the data, finding
\begin{equation}
\langle B-V\rangle_\mathrm{s} = 0.617 (\pm 0.011) + 0.808 (\pm 0.038) \log P.
\end{equation}
We measure the period shift of $\Delta \log P$ = 0.07 for V16
at the fixed \bvs, resulting in $\Delta M_{\mathrm bol}$ = $-$0.21 mag
assuming the same mass and the same temperature in Equation~(\ref{eq:vAB}),
where $\Delta M_{\mathrm bol}$ is the difference in the bolometric
magnitude between the mean value and the RRL under consideration.
The magnitude difference from the period shift analysis 
is somewhat larger than that from \vint, $-$0.159 mag, but is compatible
with that from the calculated absolute visual magnitude.
In Table~\ref{tab:rrabpam}, we do not show physical parameters for V16
because it has a large compatibility condition parameter, $D_m$ = 10.9,
and, consequently, it is tagged as a potential Blazhko variable.
If we use Equation~(\ref{eq:mv:3}) and Fourier parameters for V16
in Table~\ref{tab:rrabfc}, the absolute visual magnitude of V16 becomes 
$M_V$ = 0.396 mag and it is 0.185 mag brighter than 
the mean absolute visual magnitude of the cluster.
From this exercise, one can show that V16 is intrinsically brighter
than other RRLs in the cluster and the photometric contamination, 
such as blending with nearby stars, may not be responsible for its high luminosity.
Figure~\ref{fig:outlier} (b) shows residual colors around the fitted line.
In the figure, V16 has a significantly smaller color, 
i.e.\ a higher temperature, at the fixed period. 
Again, assuming the same mass in Equation~(\ref{eq:vAB}),
one can show that a more luminous RRL has a higher temperature
at the fixed period. Therefore, all the evidences suggest that 
V16 may be a more evolved RRL in NGC~6723.
To a lesser extent, the same characteristics can also be applied to NV2,
as the referee of the paper pointed out.
But a caution should be advised that our period of this variable may be
slightly inaccurate or that is changing as we noted in Appendix~\ref{ap:s:var}.

For NV14 in Figure~\ref{fig:outlier} (a), the internal period shift is 
$\Delta \log P$ = $-$0.06, almost an opposite case of V16.
If we use Equation~(\ref{eq:vAB}) assuming the same mass and temperature,
it should be less luminous than other RRLs in NGC~6723 by 
$\Delta M_{\mathrm bol}$ = 0.18 mag.
Quantitively, the derived magnitude from the period shift analysis
appears to be in good agreement with the magnitude difference of NV14 from
the mean values, $\Delta$\vint\ = 0.162 mag.
However, it should be emphasized that \bvs\ in
Figure~\ref{fig:outlier} (a) is not a reddening free parameter. 
As shown in Figure~\ref{fig:PvsA_B}, NV14 has a normal period 
at the fixed blue amplitude.
Recall that both quantities are reddening free parameters.
Also shown in Table~\ref{tab:rrabpam}, the absolute visual magnitude
of NV14 from Fourier parameters using Equation~(\ref{eq:mv:3})
is in excellent agreement with the mean value of the cluster
to within 0.02 mag, suggesting that the fainter apparent visual
magnitude of  NV14 is most likely due to the excess differential 
foreground reddening effect by $\gtrsim$ 0.03 mag.

The RRab type variables V14 and V15 appears to be normal
in Figure~\ref{fig:PvsA_B} or Figure~\ref{fig:outlier} (a),
but significantly deviate from the mean value in Figure~\ref{fig:W}.
As shown in Figure~\ref{fig:outlier} (c) or (f),
V14 and V15 are brighter than other RRLs for their period.
Part of the bright nature of V14 can be explained by
employing the period shift analysis.
The period shift for V14 is $\Delta \log P$ = +0.03
at the fixed \bvs\ color, equivalent to 
$\Delta M_{\mathrm bol}$ = $-$0.09 mag.
As in Table~\ref{tab:rrabpam},
the absolute visual magnitude from the Fourier analysis of V14
is 0.08 mag brighter than the mean value of the cluster,
consistent with that from the period shift analysis.
However the period shift analysis can not fully explain
the bright apparent magnitude of V14.
As shown in Table~\ref{tab:rrlyrae}, V14 and V15 are
0.20 and 0.12 mag brighter than the mean \vint\ value of the cluster.

In Figure~\ref{fig:outlier}-(f), we show a plot of $\Delta\log P$ versus \vint. 
As can be seen in the figure, $\Delta\log P$ appears to be well correlated with
\vint. We perform a least square fit to the data, finding
\begin{equation}
\langle V\rangle_\mathrm{int} = 15.452 (\pm 0.005) -2.655 (\pm 0.351) \Delta\log P.
\end{equation}
The slope of the relation is in good agreement with that expected from 
Equation~(\ref{eq:vAB}) with the assumption of the same mass, temperature,
and bolometric correction, $-$2.976, to within a $\pm \sigma$ level.
The residual around the fitted line can not be explained with the theoretical
pulsational properties of RRLs given by \cite{vAB71}.
It is thought that blending with nearby faint objects would be 
responsible for the positive residual such as
can be seen in V14 and V15 in the figure. 
Using the above relation and $\Delta\log P$ for V14 and V15, 
0.032 and 0.011 respectively, the expected \vint\ values for each variable are
15.367 and 15.423 mag.
The calculate \vint\ values are 0.112 and 0.085 mag fainter than our measurements and
these amounts should be explained by the photometric contamination,
such as blending with nearby faint objects.
Note that these values are compatible with the residuals in $W(BV)$
as shown in Figure~\ref{fig:W}.
The residuals in $W(BV)$ for V14 and V15 are 0.101 and 0.104 mag, respectively.

We conclude that V14 and V16 are more evolved RRLs of NGC~6723.
In addition, V14 suffers from the blending effect with nearby faint
companions by $\approx$ 0.10 mag.
On the other hand, V15 is solely affected by the blending effect with
a similar amount.

\section{Red and $H_\alpha$ variables}
\begin{figure}
\epsscale{1}
\figurenum{A2}
\plotone{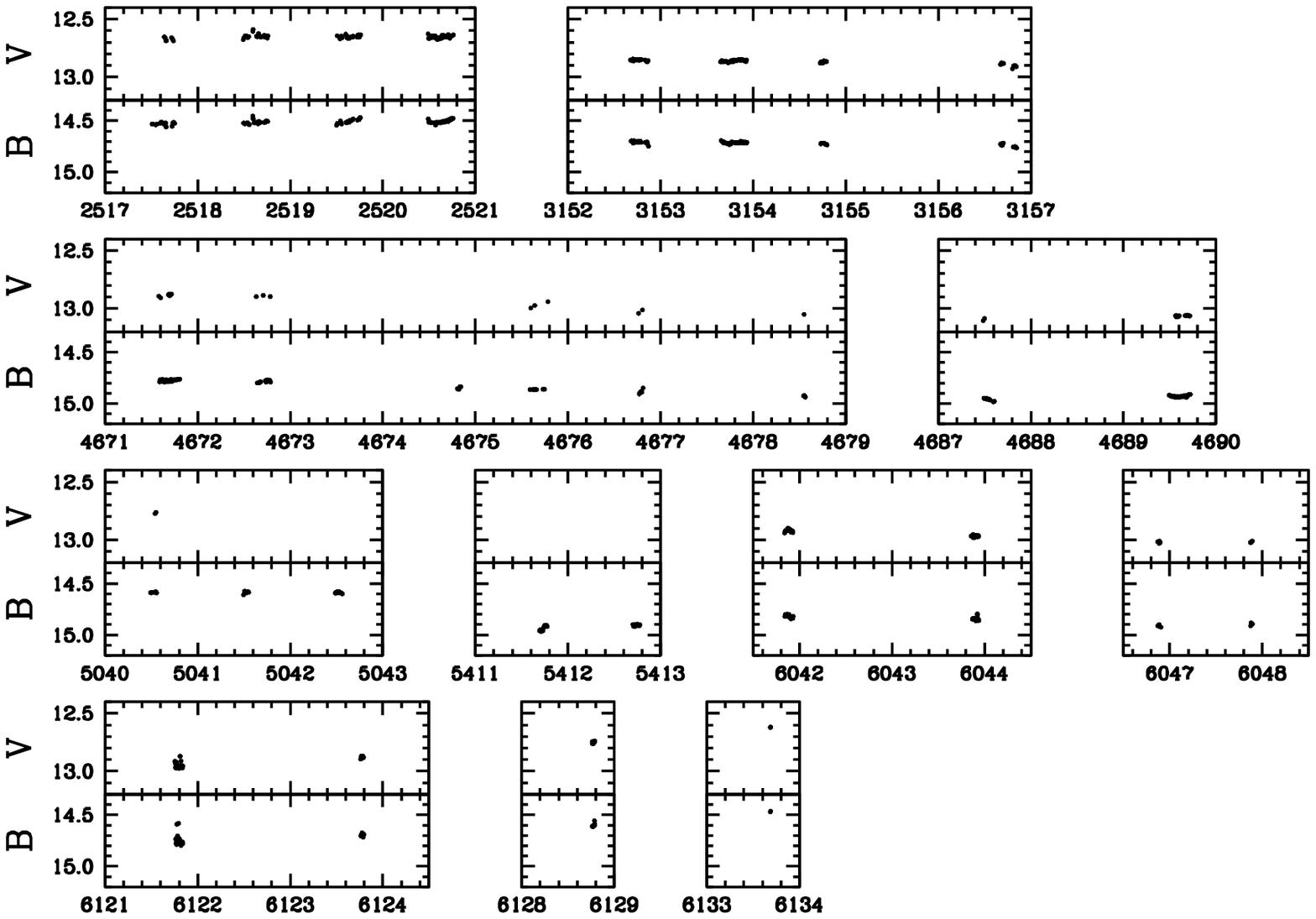}
\caption{Plots of $V$ and $B$ against the modified heliocentric Julian date 
(HJD $-$ 2,450,000) of the red variable V25.}
\label{fig:v25}
\end{figure}

\begin{figure}
\epsscale{1}
\figurenum{A3}
\plotone{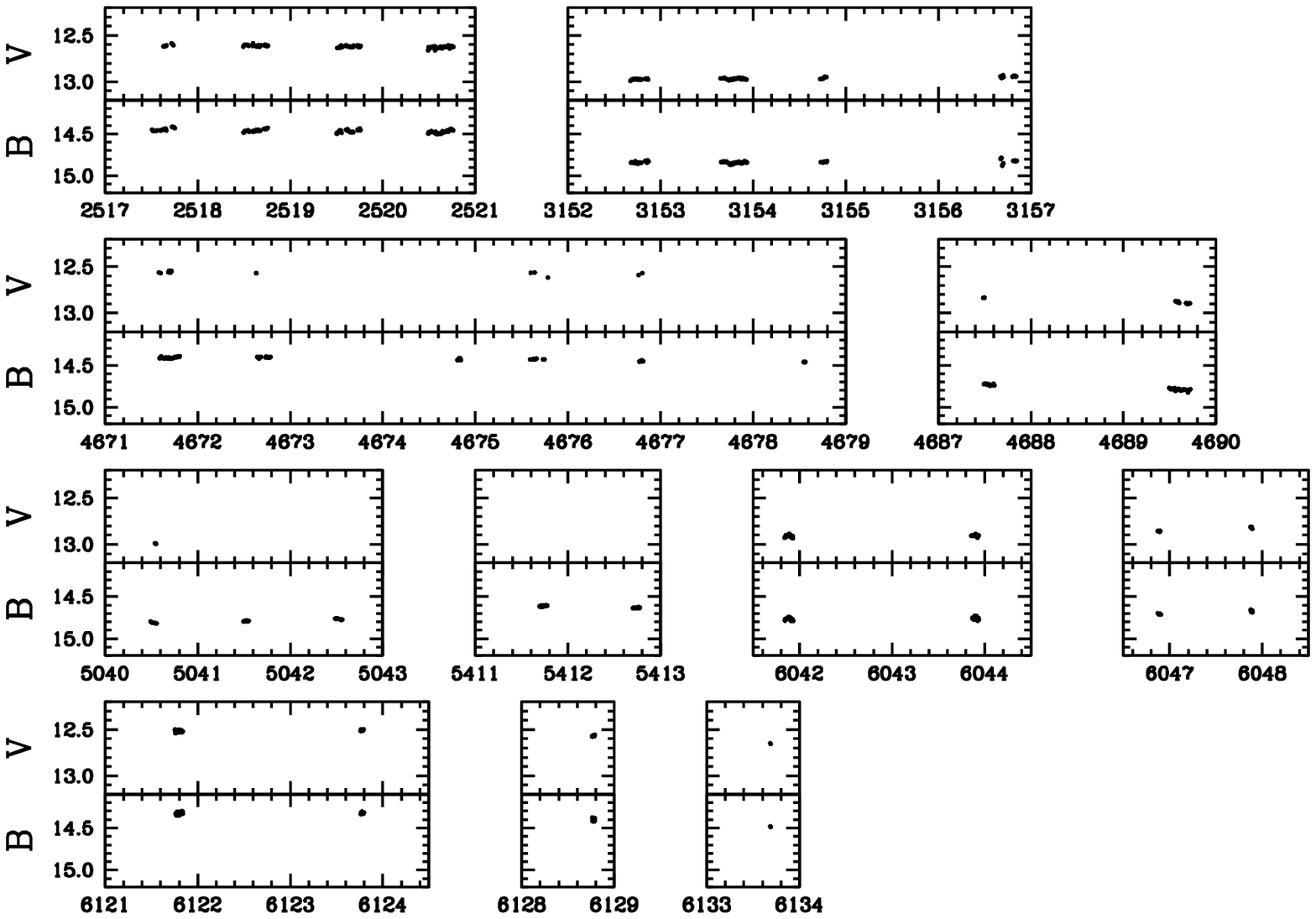}
\caption{Same as Figure~\ref{fig:v25}, but for the red variable V26.}
\label{fig:v26}
\end{figure}

\citet{menzies74} found two semi-regular red variables and 
one $H_\alpha$ variable in NGC~6723.
We show plots of $V$, $B$, and $(B-V)$ vs.\ HJD
for V25 and V26 in Figure~\ref{fig:v25} and \ref{fig:v26}, respectively.
Menzies pointed out that these two red variables are 0.2 mag redder 
than the reddest RGB stars in NGC~6723.
Our CMD shown in Figure~\ref{fig:cmd} also confirms this.
\citet{clement01} listed the period only for V25, $\approx$ 140 day,
but we were not able to derive the periods for both red variables.

\begin{figure}
\epsscale{1}
\figurenum{A4}
\plotone{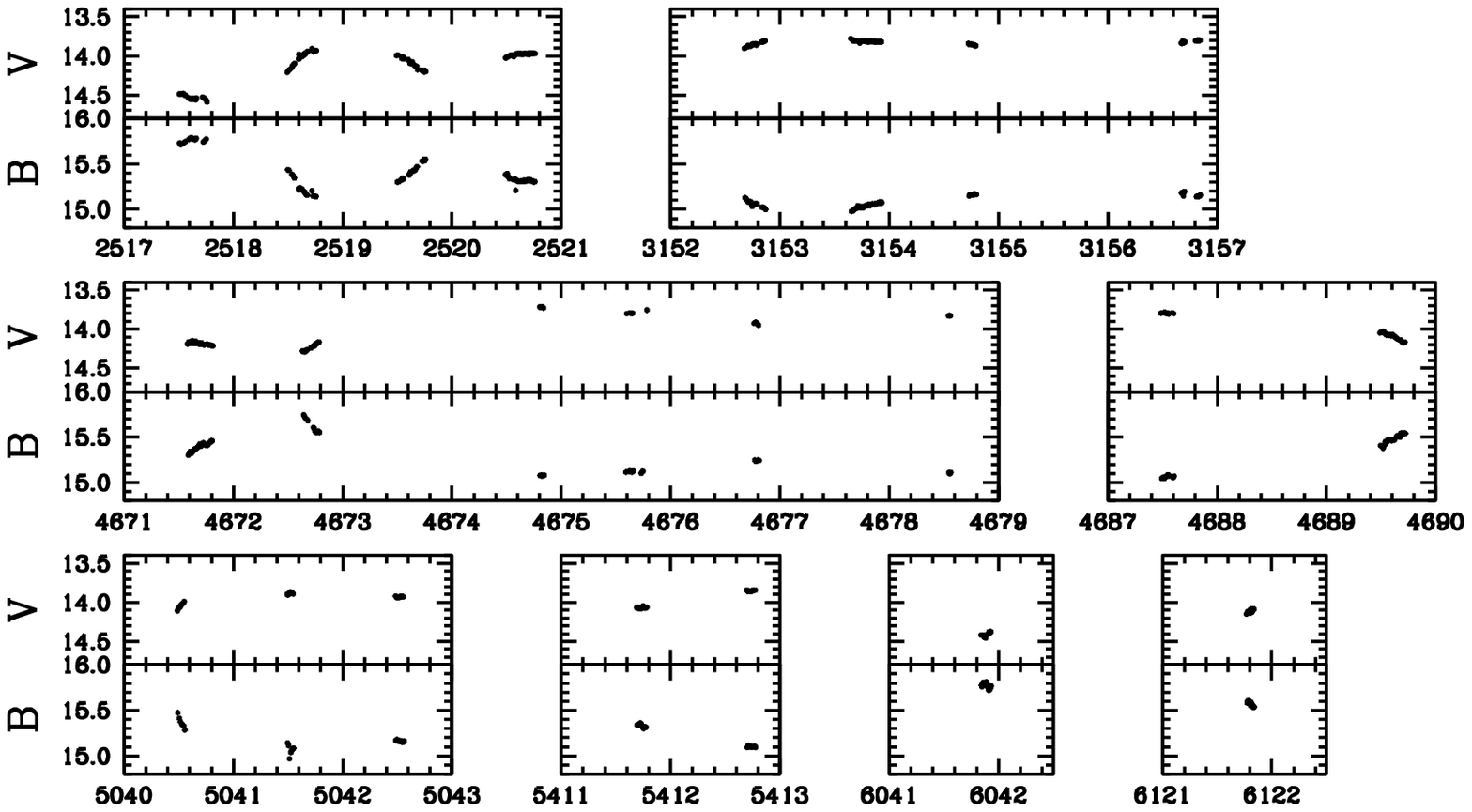}
\caption{Same as Figure~\ref{fig:v25}, but for the $H_\alpha$ variable V30.}
\label{fig:v30}
\end{figure}

NGC~6723 is within a degree from the Corona Australis dark cloud.
\citet{knacke73} reported two $H_\alpha$ emission objects
near NGC~6723 and \citet{menzies74} found that V30 is a $H_\alpha$ variable.
As Menzies noted this variable is probably associated with
the Corona Australis dark cloud, not with NGC~6723.
In Figure~\ref{fig:v30}, we show plots of $V$, $B$, and $(B-V)$ vs.\
HJD for V30.
Clement et al.\ (2001) did not list the period for V30
and we were not able to derive the period for this variable.

\section{SX Phe and $\delta$ Sct Stars}\label{ap:s:sxphe}
We investigate the membership status of newly found short-period
pulsating variables NV16, NV18 and NV19 based on the distance modulus 
of each pulsating variable.

\citet{mcnamara11} provided a period-luminosity-metallicity relation,
\begin{equation}
M_V = -2.90\log P - 0.19 \mathrm{[Fe/H]} - 1.27,
\end{equation}
while \citet{cohen12} provided a period-luminosity relation,
\begin{equation}
M_V = -3.389\log P - 1.640,
\end{equation}
for the SX Phe type variable.
The calculated absolute magnitudes from both relations are 
in agreement within the $\approx$ 0.03 mag level
and we adopt the average value from both transformation relations.
In Table~\ref{tab:sxphe}, we show the absolute visual magnitude of NV19,
$M_V$ = 2.667 mag.
\citet{mcnamara11} also provided the transformation relations
to calculate the intrinsic color of SX Phe in his Equations (6a) 
for \feh\ = 0.04 dex and (6b) for \feh\ = $-$1.91 dex.
Using the linear interpolation for \feh\ = $-$1.2 dex,
we obtain the intrinsic color for NV19, \bvz\ = 0.215,
and then we obtain the interstellar reddening value, \ebv\ = 0.109 mag,
by comparing the intrinsic color and the average magnitude-weighted
integrated color of NV19. The \ebv\ value for NV19 is slightly larger
than that from RRLs, \ebv\ = 0.063 $\pm$ 0.015 mag.
Finally, using the absolute visual magnitude and the interstellar reddening
value for NV19, we obtain the true distance modulus of 14.638 mag
and the true distance from the Sun of 8.46 kpc, consistent with those
from RRLs in the cluster, 8.47 $\pm$ 0.17 kpc.
Therefore, NV19 is truly a cluster member SX Phe type variable star.

We perform the same calculations for NV16 and NV18 and we obtain the true
distances of 4.67 kpc and 5.45 kpc for NV16 and NV18, respectively,
and they are not cluster members.
Therefore, NV16 and NV18 are likely $\delta$ Sct type variables.
Using a period-luminosity-metallicity relation for $\delta$ Sct type variable
stars by \citet{mcnamara11},
\begin{equation}
M_V = -2.89\log P - 1.31,
\end{equation}
we obtain the absolute visual magnitudes for NV16 and NV18.
Then, assuming the near solar metallicity, we use the Equation (6a)
from \citet{mcnamara11} and calculate the intrinsic colors for the variables.
Finally, we obtain the true distances from the Sun of 6.14 kpc and 6.31 kpc
for NV16 and NV18, respectively, 
and our results are shown in Table~\ref{tab:sxphe}.

\section{Eclipsing Binaries}\label{ap:s:eb}
We investigate the membership status of five W UMa type eclipsing binaries
based on (i) the radial distance from the center of NGC~6723 and
(ii) the distance modulus of each variable star.

In the second column of Table~\ref{tab:eb}, we show the radial distance
of each variable star from the center of NGC~6723. 
As can be seen, NV20, NV21 and NV22 lie beyond the tidal radius of the cluster,
$r_t$ = 631 arcsec \citep{harris96}, and they are not cluster member variables.

\citet{rucinski00} provided an empirical relation to calculate 
the absolute visual magnitude of W UMa type eclipsing binary,
\begin{equation}
M_V = -4.44 \log P + 3.02 (B-V)_0 + 0.12.
\end{equation}
Assuming \ebv\ = 0.063 mag, we calculate the absolute visual magnitude
for each W UMa type variable and we show our results in Table~\ref{tab:eb}.
During our calculations, we adopt the magnitude-weighted integrated color \bvmag.
Since the shape of the light curve of a typical W UMa type binary is symmetric and
sinusoidal, the average magnitude or the average color do not greatly depend on
the average scheme.
For our case, the mean difference in color from five variable stars is negligibly small,
\bvmag\ $-$ \bvint  = 0.000 $\pm$ 0.002 mag.
Then we derive the true distance modulus for each variable
by comparing the absolute visual magnitude with the de-reddened visual
intensity-weighted integrated magnitude.
As shown in Table~\ref{tab:eb}, the true distances of NV15 and NV17
are about 1.6 kpc and 3.1 kpc from the Sun, respectively, and they
are very small compared to the true distance of NGC~6723, 8.47 $\pm$ 0.17 kpc.
Therefore, unfortunately, neither NV15 nor NV17 are cluster members.
Also shown in the table are the true distances of NV20, NV21 and NV22,
suggesting again that they are not cluster members.

\section{Notes on Individual Variable Stars}\label{ap:s:var}
\noindent
V1 --- The light curve shows the Blazhko effect.

\noindent
V5 --- \citet{clement01} questioned a double mode RRL,
however, our light curve shows that it is a RRab showing the Blazhko effect..

\noindent
V6 --- The light curve shows the Blazhko effect.

\noindent
V8 --- The light curve shows the Blazhko effect.
The period of the variable by \citet{clement01} was $P$ = 0.53 day, 
but we have a shorter period for the variable, $P$ = 0.4803 day.
The mis-aligned maxima indicate that our period of the variable
may be slightly incorrect.

\noindent
V9 --- The light curve shows the Blazhko effect.

\noindent
V13 --- The light curve shows the Blazhko effect.

\noindent
V14 --- The light curve appears to show the Blazhko effect.
The period of the variable by \citet{clement01}
was $P$ = 0.619 day, but we have a longer period 
for the variable, $P$ = 0.6308 day.

\noindent
V18 --- The light curve shows the Blazhko effect.
The mis-aligned maxima indicate that our period of the variable
may be slightly incorrect.

\noindent
V18 --- The mis-aligned maxima indicate that our period of the variable
may be slightly incorrect.

\noindent
V20 --- The light curve shows the Blazhko effect.
The mis-aligned maxima indicate that our period of the variable
may be slightly incorrect.

\noindent
V23 --- The light curve appears to show the Blazhko effect.

\noindent
V29 --- The light curve appears to show the Blazhko effect.
The period of the variable by \citet{clement01}
was $P$ = 0.53 day, but we have a shorter period
for the variable, $P$ = 0.4989 day.

\noindent
V31 --- The light curve shows the Blazhko effect.

\noindent
NV1 --- RRc type variable. Our light curve suggests that the
period is slightly inaccurate or that it is changing.
Probable non-member variable.

\noindent
NV2 --- RRab type variable. Our light curve suggests that the
period is slightly inaccurate or that it is changing.

\noindent
NV3 --- RRab type variable.

\noindent
NV4 --- RRab type variable showing the Blazhko effect.

\noindent
NV5 --- RRab type variable showing the Blazhko effect.

\noindent
NV6 --- RRab type variable.

\noindent
NV7 --- RRc type variable.

\noindent
NV8 --- RRab type variable showing the Blazhko effect.

\noindent
NV9 --- RRab type variable showing the Blazhko effect.
It has the smallest $B$ amplitude, $A_B$ = 0.491 mag,
among all RRab type variables in NGC~6723.

\noindent
NV10 --- RRab type variable showing the Blazhko effect.

\noindent
NV11 --- RRc type variable showing the Blazhko effect.

\noindent
NV12 --- RRab type variable.

\noindent
NV13 --- RRab type variable showing the Blazhko effect.

\noindent
NV14 --- RRab type variable. It is suspected that NV14 suffers from
a large differential reddening effect than other RRLs
in NGC~6723 by \ebv\ $\gtrsim$ 0.03 mag.

\noindent
NV15 --- W Uma type eclipsing binary.
Probable non-member variable based on the true distance from the Sun.

\noindent
NV16 --- $\delta$ Sct type variable.
Probable non-member variable based on the true distance from the Sun.

\noindent
NV17 --- W Uma type eclipsing binary.
Probable non-member variable based on the true distance from the Sun.

\noindent
NV18 --- $\delta$ Sct type variable.
Probable non-member variable based on the true distance from the Sun.

\noindent
NV19 --- SX Phe type variable.

\noindent
NV20 --- W Uma type eclipsing binary.
Probable non-member variable based on the radial distance from the center
of NGC~6723 and the true distance from the Sun.

\noindent
NV21 --- W Uma type eclipsing binary.
Probable non-member variable based on the radial distance from the center
of NGC~6723 and the true distance from the Sun.

\noindent
NV22 --- W Uma type eclipsing binary.
Probable non-member variable based on the radial distance from the center
of NGC~6723 and the true distance from the Sun.


\clearpage

\begin{deluxetable}{l l c r}
\tablecaption{Journal of observations.\label{tab:obs}}
\tablenum{1}
\tablewidth{0pc} 
\tablehead{ \multicolumn{1}{c}{Date} &
\multicolumn{1}{c}{Telescope} & 
\multicolumn{1}{c}{n($B$)} &
\multicolumn{1}{c}{n($V$)} } 
\startdata
2002/08/30 & CTIO 0.9m & 16 & 22  \\
2002/08/31 & CTIO 0.9m & 21 & 24  \\
2002/09/01 & CTIO 0.9m & 22 & 24  \\
2002/09/02 & CTIO 0.9m & 28 & 29  \\

2004/05/26 & CTIO 0.9m & 18 & 19  \\
2004/05/27 & CTIO 0.9m & 31 & 31  \\
2004/05/28 & CTIO 0.9m & 7  &  7  \\
2004/05/30 & CTIO 0.9m & 9  & 10  \\

2008/07/23 & CTIO 1.0m & 27 & 28  \\
2008/07/24 & CTIO 1.0m & 12 & 14  \\
2008/07/26 & CTIO 1.0m &  5 &  5  \\
2008/07/27 & CTIO 1.0m &  7 &  6  \\
2008/07/28 & CTIO 1.0m &  5 &  5  \\
2008/07/30 & CTIO 1.0m &  3 &  3  \\
2008/08/08 & CTIO 1.0m &  9 &  9  \\
2008/08/10 & CTIO 1.0m & 21 & 21  \\

2009/07/27 & CTIO 1.0m &  7 &  7  \\
2009/07/28 & CTIO 1.0m &  7 &  7  \\
2009/07/29 & CTIO 1.0m &  9 &  9  \\

2010/08/02 & CTIO 1.0m & 12 & 13  \\
2010/08/03 & CTIO 1.0m & 10 & 10  \\

2012/04/23 & CTIO 0.9m & 24 & 24  \\
2012/04/25 & CTIO 0.9m & 17 & 17  \\
2012/04/28 & CTIO 0.9m &  5 &  5  \\
2012/04/29 & CTIO 0.9m &  4 &  4  \\

2012/07/12 & CTIO 0.9m & 25 & 27  \\
2012/07/14 & CTIO 0.9m & 10 & 10  \\
2012/07/19 & CTIO 0.9m & 10 & 10  \\
2012/07/24 & CTIO 0.9m &  2 &  2  \\
\enddata
\end{deluxetable}

\clearpage

\begin{landscape}
\begin{deluxetable}{c c r r r r r r r r r}
\tablecaption{Transformation coefficients and residuals.\label{tab:coeff}}
\tabletypesize{\footnotesize}
\tablenum{2}
\tablewidth{0pc} 
\tablehead{\multicolumn{1}{c}{Date} & \multicolumn{1}{c}{Tel.} &
\multicolumn{1}{c}{n} &
\multicolumn{1}{c}{$\alpha_v$}& 
\multicolumn{1}{c}{$\beta_v$} & \multicolumn{1}{c}{$\gamma_v$} &
\multicolumn{1}{c}{$\alpha_b$} & \multicolumn{1}{c}{$\beta_b$} & 
\multicolumn{1}{c}{$\gamma_b$} & \multicolumn{1}{c}{$\Delta V$} & 
\multicolumn{1}{c}{$\Delta (B-V)$}} 
\startdata
Aug 2002 & 0.9m & 27 & $-$0.031 & 0.135 & 3.029 & 0.080 & 0.231 & 3.342 & 0.002 $\pm$ 0.017 & $-$0.003 $\pm$ 0.018 \\
May 2004 & 0.9m & 35 & $-$0.019 & 0.112 & 3.258 & 0.106 & 0.238 & 3.596 & $-$0.002 $\pm$ 0.011 & 0.004 $\pm$ 0.015 \\
Jul 2012 & 0.9m & 36 & $-$0.022 & 0.111 & 2.825 & 0.101 & 0.204 & 3.172 & 0.001 $\pm$ 0.010 & $-$0.001 $\pm$ 0.008 \\
\enddata
\end{deluxetable}
\end{landscape}

\clearpage

\begin{landscape}
\begin{deluxetable}{l c c r r r r r r r r c l}
\tablecaption{Parameters for NGC~6723 Variables.\label{tab:rrlyrae}}
\tabletypesize{\scriptsize}
\tablenum{3}
\tablewidth{0pt} 
\tablehead{
\multicolumn{1}{c}{ID} & \multicolumn{1}{c}{$\alpha$}& \multicolumn{1}{c}{$\delta$} & 
\multicolumn{1}{c}{Period} & \multicolumn{1}{c}{$A_B$} & 
\multicolumn{1}{c}{$A_V$} &
\multicolumn{1}{c}{$\langle B \rangle_\mathrm{mag}$} &
\multicolumn{1}{c}{$\langle V \rangle_\mathrm{mag}$} & 
\multicolumn{1}{c}{$\langle B \rangle_\mathrm{int}$} &
\multicolumn{1}{c}{$\langle V \rangle_\mathrm{int}$} & 
\multicolumn{1}{c}{$\langle V \rangle_\mathrm{s}$} & 
\multicolumn{1}{c}{\bvs} & 
\multicolumn{1}{c}{Class}\\
\multicolumn{1}{c}{} & \multicolumn{1}{c}{(2000)} & \multicolumn{1}{c}{(2000)} & 
\multicolumn{1}{c}{(day)} & 
\multicolumn{1}{c}{(mag)} & \multicolumn{1}{c}{(mag)} & \multicolumn{1}{c}{(mag)} & 
\multicolumn{1}{c}{(mag)} & \multicolumn{1}{c}{(mag)} & \multicolumn{1}{c}{(mag)} &
\multicolumn{1}{c}{(mag)} & \multicolumn{1}{c}{(mag)} & \multicolumn{1}{c}{} \\
\multicolumn{1}{c}{(1)} & \multicolumn{1}{c}{(2)} & \multicolumn{1}{c}{(3)} & 
\multicolumn{1}{c}{(4)} & \multicolumn{1}{c}{(5)} & \multicolumn{1}{c}{(6)} &
\multicolumn{1}{c}{(7)} & \multicolumn{1}{c}{(8)} & \multicolumn{1}{c}{(9)} &
\multicolumn{1}{c}{(10)} & \multicolumn{1}{c}{(11)} & \multicolumn{1}{c}{(12)} &
\multicolumn{1}{c}{(13)} }
\startdata
V1    & 18:59:39.63 & $-3$6:41:09.1 & 0.5384105 &  1.447 &  1.127 & 16.037 & 15.590 & 15.995 & 15.566 & 15.558 & 0.431  & RRab Bl \\   
V2    & 18:59:44.56 & $-$36:39:08.8 & 0.5035381 &  1.727 &  1.342 & 15.874 & 15.482 & 15.740 & 15.407 & 15.381 & 0.357  & RRab \\           
V3    & 18:59:12.97 & $-$36:37:46.2 & 0.4940991 &  1.663 &  1.245 & 15.889 & 15.497 & 15.783 & 15.440 & 15.421 & 0.363  & RRab \\           
V4    & 18:59:34.71 & $-$36:36:34.7 & 0.4510559 &  1.633 &  1.288 & 15.890 & 15.517 & 15.768 & 15.449 & 15.427 & 0.341  & RRab \\           
V5    & 18:59:32.99 & $-$36:37:01.5 & 0.5776827 &  1.083 &  0.797 & 15.945 & 15.491 & 15.897 & 15.466 & 15.457 & 0.436  & RRab Bl? \\  
V6    & 18:59:33.91 & $-$36:37:05.3 & 0.4791830 &  1.528 &  1.170 & 15.882 & 15.494 & 15.782 & 15.438 & 15.419 & 0.360  & RRab Bl? \\  
V7    & 18:59:49.77 & $-$36:39:01.0 & 0.3076709 &  0.652 &  0.513 & 15.857 & 15.525 & 15.838 & 15.514 & 15.510 & 0.323  & RRc \\            
V8    & 18:59:34.68 & $-$36:37:42.4 & 0.4802776 &  1.716 &  1.326 & 15.874 & 15.475 & 15.761 & 15.415 & 15.394 & 0.366  & RRab Bl \\   
V9    & 18:59:39.50 & $-$36:37:34.5 & 0.5757466 &  1.667 &  1.230 & 15.810 & 15.376 & 15.747 & 15.345 & 15.334 & 0.411  & RRab Bl \\   
V10   & 18:59:45.70 & $-$36:36:27.4 & 0.2523218 &  0.648 &  0.471 & 15.731 & 15.476 & 15.713 & 15.466 & 15.463 & 0.247  & RRc~~~Bl \\            
V11   & 18:59:44.53 & $-$36:34:02.1 & 0.5342927 &  1.285 &  0.989 & 15.925 & 15.497 & 15.856 & 15.461 & 15.449 & 0.405  & RRab \\           
V12   & 18:59:37.08 & $-$36:38:36.5 & 0.4687582 &  1.702 &  1.261 & 15.921 & 15.535 & 15.803 & 15.472 & 15.453 & 0.354  & RRab \\           
V13   & 18:59:29.49 & $-$36:39:02.8 & 0.5065254 &  1.536 &  1.196 & 15.996 & 15.595 & 15.929 & 15.560 & 15.548 & 0.380  & RRab Bl \\   
V14\tablenotemark{b}   & 18:59:30.22 & $-$36:38:34.3 & 0.6308439 &  1.102 &  0.869 & 15.727 & 15.279 & 15.681 & 15.255 & 15.247 & 0.430  & RRab \\  
V15\tablenotemark{b}   & 18:59:25.57 & $-$36:35:06.3 & 0.4353677 &  1.662 &  1.268 & 15.770 & 15.415 & 15.628 & 15.338 & 15.311 & 0.317  & RRab \\           
V16\tablenotemark{b}   & 18:59:29.46 & $-$36:36:19.9 & 0.6962724 &  1.178 &  0.884 & 15.774 & 15.324 & 15.728 & 15.300 & 15.292 & 0.432  & RRab \\           
V17   & 18:59:36.96 & $-$36:39:33.5 & 0.5301725 &  1.346 &  1.026 & 15.976 & 15.549 & 15.898 & 15.509 & 15.495 & 0.402  & RRab \\           
V18   & 18:59:21.88 & $-$36:38:13.3 & 0.5264541 &  1.627 &  1.247 & 15.923 & 15.504 & 15.854 & 15.467 & 15.456 & 0.396  & RRab Bl \\   
V19   & 18:59:19.26 & $-$36:39:47.5 & 0.5347107 &  1.327 &  1.022 & 15.919 & 15.498 & 15.862 & 15.466 & 15.456 & 0.402  & RRab Bl \\   
V20   & 18:59:33.67 & $-$36:37:13.9 & 0.4874793 &  1.769 &  1.312 & 15.799 & 15.426 & 15.702 & 15.374 & 15.356 & 0.346  & RRab Bl \\   
V21   & 18:59:26.79 & $-$36:38:22.9 & 0.5919537 &  1.110 &  0.843 & 15.884 & 15.439 & 15.836 & 15.414 & 15.405 & 0.427  & RRab \\           
V22   & 18:59:27.39 & $-$36:37:17.4 & 0.3087057 &  0.741 &  0.561 & 15.848 & 15.511 & 15.828 & 15.499 & 15.495 & 0.328  & RRc~~~Bl \\            
V23   & 18:59:37.83 & $-$36:38:05.0 & 0.6248439 &  0.639 &  0.475 & 15.961 & 15.490 & 15.945 & 15.482 & 15.479 & 0.459  & RRab Bl? \\  
V24   & 18:59:43.13 & $-$36:39:45.0 & 0.3001437 &  0.682 &  0.511 & 15.832 & 15.512 & 15.812 & 15.501 & 15.497 & 0.311  & RRc \\            
V25   & 18:59:16.92 & $-$36:35:20.3 & \nodata   &\nodata &\nodata &\nodata &\nodata &\nodata &\nodata &\nodata &\nodata & LPV \\
V26   & 18:59:41.54 & $-$36:34:32.2 & \nodata   &\nodata &\nodata &\nodata &\nodata &\nodata &\nodata &\nodata &\nodata & LPV \\
V27   & 18:59:15.11 & $-$36:36:14.5 & 0.6192392 &  0.859 &  0.663 & 15.871 & 15.425 & 15.842 & 15.410 & 15.405 & 0.432  & RRab Bl? \\ 
V28   & 18:59:34.25 & $-$36:39:12.8 & 0.4868633 &  1.514 &  1.180 & 15.910 & 15.511 & 15.811 & 15.458 & 15.440 & 0.370  & RRab \\          
V29   & 18:59:34.39 & $-$36:36:51.5 & 0.4988830 &  1.565 &  1.188 & 15.898 & 15.493 & 15.795 & 15.438 & 15.420 & 0.376  & RRab \\  
V30   & 19:00:01.57 & $-$36:37:05.6 & \nodata   &\nodata &\nodata &\nodata &\nodata &\nodata &\nodata &\nodata &\nodata &  LPV \\
V31   & 18:59:28.07 & $-$36:38:51.9 & 0.6078313 &  0.972 &  0.704 & 15.932 & 15.478 & 15.898 & 15.461 & 15.455 & 0.438  & RRab Bl \\ 
V32   & 18:59:29.87 & $-$36:39:27.1 & 0.2875411 &  0.536 &  0.433 & 15.672 & 15.403 & 15.659 & 15.395 & 15.392 & 0.262  & RRc \\          
 & & & & & & & & & & & & \\                                                                                                         
NV1\tablenotemark{a}& 18:59:30.27 & $-$36:38:02.3 & 0.2853927 &  0.424 &  0.315 & 15.135 & 14.736 & 15.128 & 14.732 & 14.731 & 0.392  & RRc   \\
NV2   & 18:59:33.19 & $-$36:37:58.1 & 0.5314143 &  1.563 &  1.186 & 15.743 & 15.351 & 15.650 & 15.303 & 15.286 & 0.364  & RRab Bl \\
NV3   & 18:59:32.97 & $-$36:38:01.5 & 0.6064508 &  0.692 &  0.534 & 15.846 & 15.365 & 15.834 & 15.358 & 15.356 & 0.471  & RRab Bl \\
NV4   & 18:59:29.31 & $-$36:37:51.2 & 0.4639525 &  1.621 &  1.388 & 15.829 & 15.436 & 15.741 & 15.380 & 15.361 & 0.372  & RRab Bl \\
NV5   & 18:59:30.79 & $-$36:37:44.8 & 0.5363784 &  1.375 &  1.065 & 15.866 & 15.439 & 15.806 & 15.408 & 15.397 & 0.407  & RRab Bl \\
NV6   & 18:59:30.18 & $-$36:37:45.0 & 0.5719616 &  1.235 &  0.914 & 15.890 & 15.449 & 15.833 & 15.420 & 15.410 & 0.420  & RRab \\
NV7   & 18:59:35.79 & $-$36:37:23.2 & 0.2540772 &  0.679 &  0.481 & 15.736 & 15.452 & 15.721 & 15.443 & 15.440 & 0.277  & RRc  \\
NV8   & 18:59:35.67 & $-$36:37:52.2 & 0.5965565 &  0.959 &  0.810 & 15.973 & 15.511 & 15.939 & 15.493 & 15.485 & 0.446  & RRab Bl \\
NV9   & 18:59:40.50 & $-$36:38:08.9 & 0.6592001 &  0.491 &  0.399 & 16.000 & 15.512 & 15.996 & 15.509 & 15.508 & 0.480  & RRab Bl \\
NV10  & 18:59:32.03 & $-$36:37:40.4 & 0.5528632 &  1.441 &  1.109 & 15.963 & 15.536 & 15.894 & 15.500 & 15.488 & 0.404  & RRab Bl \\
NV11  & 18:59:30.98 & $-$36:38:32.4 & 0.3345803 &  0.759 &  0.653 & 15.885 & 15.554 & 15.865 & 15.542 & 15.538 & 0.323  & RRc~~~Bl \\
NV12  & 18:59:32.35 & $-$36:37:52.0 & 0.4400748 &  1.813 &  1.402 & 15.944 & 15.571 & 15.802 & 15.494 & 15.467 & 0.334  & RRab \\
NV13  & 18:59:35.38 & $-$36:38:24.1 & 0.4893353 &  1.621 &  1.245 & 15.993 & 15.601 & 15.913 & 15.558 & 15.543 & 0.367  & RRab Bl \\
NV14  & 18:59:55.06 & $-$36:37:31.0 & 0.6121545 &  0.618 &  0.468 & 16.138 & 15.630 & 16.122 & 15.621 & 15.618 & 0.497  & RRab \\
NV15\tablenotemark{a,c}  & 18:59:03.16 & $-$36:44:09.1 & 0.3096517 &  0.307 &  0.305 & 16.354 & 15.630 & 16.351 & 15.628 &\nodata &\nodata & EB  \\
NV16\tablenotemark{a,d}  & 18:59:24.93 & $-$36:45:58.4 & 0.0489877 &  0.413 &  0.290 & 16.723 & 16.418 & 16.718 & 16.415 &\nodata &\nodata & $\delta$ Sct  \\
NV17\tablenotemark{a,c}  & 18:59:51.09 & $-$36:42:06.5 & 0.3777837 &  0.225 &  0.229 & 17.297 & 16.591 & 17.295 & 16.590 &\nodata &\nodata & EB  \\
NV18\tablenotemark{a,d}  & 18:59:25.14 & $-$36:36:59.5 & 0.0520890 &  0.503 &  0.426 & 16.919 & 16.625 & 16.912 & 16.622 &\nodata &\nodata & $\delta$ Sct  \\
NV19  & 18:59:41.07 & $-$36:38:01.7 & 0.0531436 &  0.450 &  0.409 & 17.969 & 17.645 & 17.964 & 17.642 &\nodata &\nodata & SX Phe  \\
NV20\tablenotemark{a,c}  & 19:00:06.84 & $-$36:47:32.7 & 0.3605305 &  0.356 &  0.296 & 18.411 & 17.763 & 18.407 & 17.760 &\nodata &\nodata & EB  \\
NV21\tablenotemark{a,c}  & 19:00:07.87 & $-$36:47:52.3 & 0.2796234 &  0.723 &  0.589 & 19.620 & 18.448 & 19.605 & 18.437 &\nodata &\nodata & EB  \\
NV22\tablenotemark{a,c}  & 18:59:55.12 & $-$36:48:18.8 & 0.3274661 &  0.396 &  0.375 & 19.494 & 18.736 & 19.491 & 18.732 &\nodata &\nodata & EB  \\
\enddata
\tablenotetext{a}{Probable non-member.}
\tablenotetext{b}{See Appendix~\ref{ap:s:notes}}
\tablenotetext{c}{See Appendix~\ref{ap:s:eb}.}
\tablenotetext{d}{See Appendix~\ref{ap:s:sxphe}.}
\end{deluxetable}
\end{landscape}

\clearpage

\begin{deluxetable}{lcccccccr}
\tablecaption{Fourier Coefficients for NGC~6723 RR Lyrae Variables.\label{tab:rrabfc}}
\tablenum{4}
\tablewidth{0pc} 
\tablehead{
\multicolumn{1}{c}{ID} & 
\multicolumn{1}{c}{$A_1$} & \multicolumn{1}{c}{$R_{21}$} &
\multicolumn{1}{c}{$R_{31}$} & \multicolumn{1}{c}{$R_{41}$} &
\multicolumn{1}{c}{$\phi_{21}$} & \multicolumn{1}{c}{$\phi_{31}$} & 
\multicolumn{1}{c}{$\phi_{41}$} & \multicolumn{1}{c}{$D_{m}$} }
\startdata
\multicolumn{9}{c}{RRab}\\
V1     &  0.252  &  0.440  &  0.278  &  0.220  &  2.439  &  5.593 &   2.207  & 18.439  \\
V2     &  0.451  &  0.493  &  0.330  &  0.243  &  2.344  &  5.037 &   1.381  &  1.412  \\
V3     &  0.389  &  0.510  &  0.359  &  0.239  &  2.347  &  5.055 &   1.411  & 14.069  \\
V4     &  0.427  &  0.472  &  0.361  &  0.230  &  2.277  &  4.831 &   1.188  & 18.111  \\
V5     &  0.271  &  0.518  &  0.312  &  0.158  &  2.617  &  5.435 &   2.232  & 24.973  \\
V6     &  0.407  &  0.451  &  0.267  &  0.142  &  2.380  &  4.968 &   1.394  & 10.926  \\
V8     &  0.369  &  0.410  &  0.065  &  0.091  &  2.229  &  4.597 &   3.607  & 50.505  \\
V9     &  0.282  &  0.507  &  0.282  &  0.109  &  2.406  &  5.050 &   1.971  & 22.492  \\
V11    &  0.325  &  0.482  &  0.343  &  0.216  &  2.396  &  5.126 &   1.679  & 17.202  \\
V12    &  0.413  &  0.481  &  0.344  &  0.226  &  2.274  &  4.918 &   1.248  &  0.950  \\
V13    &  0.319  &  0.479  &  0.278  &  0.175  &  2.325  &  5.080 &   1.603  &  0.939  \\
V14    &  0.268  &  0.506  &  0.314  &  0.162  &  2.686  &  5.606 &   2.294  &  1.008  \\
V15    &  0.460  &  0.493  &  0.312  &  0.183  &  2.281  &  4.833 &   1.238  & 17.017  \\
V16    &  0.269  &  0.485  &  0.236  &  0.120  &  2.638  &  5.601 &   2.155  & 10.902  \\
V17    &  0.335  &  0.509  &  0.345  &  0.230  &  2.333  &  5.022 &   1.406  &  0.882  \\
V18    &  0.308  &  0.386  &  0.235  &  0.106  &  2.149  &  4.924 &   1.191  &  1.313  \\
V19    &  0.306  &  0.452  &  0.273  &  0.160  &  2.471  &  5.328 &   2.058  & 22.344  \\
V20    &  0.395  &  0.400  &  0.180  &  0.139  &  2.656  &  5.362 &   2.248  & 20.279  \\
V21    &  0.271  &  0.519  &  0.324  &  0.176  &  2.549  &  5.381 &   2.120  & 21.720  \\
V23    &  0.169  &  0.368  &  0.183  &  0.079  &  2.682  &  5.727 &   2.458  &  1.784  \\
V27    &  0.225  &  0.405  &  0.210  &  0.074  &  2.657  &  5.497 &   2.523  & 94.705  \\
V28    &  0.371  &  0.514  &  0.370  &  0.227  &  2.400  &  5.011 &   1.321  &  1.467  \\
V29    &  0.393  &  0.479  &  0.357  &  0.222  &  2.338  &  5.019 &   1.345  &  1.777  \\
V31    &  0.233  &  0.458  &  0.290  &  0.132  &  2.635  &  5.540 &   2.339  &  1.108  \\
NV2    &  0.377  &  0.448  &  0.239  &  0.165  &  2.487  &  5.296 &   1.769  &  2.522  \\
NV3    &  0.149  &  0.381  &  0.182  &  0.021  &  2.814  &  5.952 &   3.041  &  1.710  \\
NV4    &  0.380  &  0.466  &  0.204  &  0.106  &  2.470  &  4.859 &   1.775  & 16.479  \\
NV5    &  0.292  &  0.471  &  0.318  &  0.184  &  2.277  &  4.896 &   1.251  &  1.322  \\
NV6    &  0.291  &  0.505  &  0.306  &  0.187  &  2.551  &  5.361 &   2.046  &  1.142  \\
NV8    &  0.242  &  0.487  &  0.278  &  0.096  &  2.624  &  5.745 &   2.316  & 22.940  \\
NV10   &  0.318  &  0.530  &  0.347  &  0.200  &  2.492  &  5.332 &   1.804  &  0.920  \\
NV12   &  0.453  &  0.496  &  0.363  &  0.232  &  2.235  &  4.705 &   1.062  &  1.285  \\
NV13   &  0.349  &  0.421  &  0.194  &  0.140  &  2.298  &  5.078 &   1.883  &  2.858  \\
NV14   &  0.171  &  0.382  &  0.190  &  0.068  &  2.617  &  5.431 &   2.574  &  0.518  \\
\hline
\multicolumn{9}{c}{RRc}\\
V7     & 0.216   & 0.083   & 0.069   & 0.015   & 3.337   & 1.024  &  4.185  & \nodata \\
V10    & 0.203   & 0.134   & 0.058   & 0.011   & 3.164   & 5.612  &  2.531  & \nodata \\
V22    & 0.214   & 0.083   & 0.101   & 0.023   & 2.998   & 0.720  &  4.092  & \nodata \\
V24    & 0.220   & 0.117   & 0.066   & 0.037   & 3.048   & 0.401  &  4.066  & \nodata \\
V32    & 0.181   & 0.115   & 0.070   & 0.001   & 3.156   & 6.271  &  5.240  & \nodata \\
NV7    & 0.190   & 0.126   & 0.035   & 0.061   & 3.096   & 5.505  &  3.119  & \nodata \\
NV11   & 0.217   & 0.069   & 0.070   & 0.068   & 3.925   & 1.080  &  4.168  & \nodata \\
\enddata
\end{deluxetable}

\clearpage

\begin{landscape}
\begin{deluxetable}{lrccccccccc}
\tablecaption{Physical parameters for NGC~6723 RRab variables.\label{tab:rrabpam}}
\tabletypesize{\footnotesize}
\tablenum{5}
\tablewidth{0pc} 
\tablehead{
\multicolumn{1}{c}{ID} & \multicolumn{1}{c}{[Fe/H]\tablenotemark{a}} & 
\multicolumn{1}{c}{$\langle(B-V)_0\rangle$} & \multicolumn{1}{c}{$E(B-V)$\tablenotemark{b}} &
\multicolumn{1}{c}{$E(B-V)$\tablenotemark{c}} &
\multicolumn{1}{c}{$M_V$\tablenotemark{d}} & \multicolumn{1}{c}{$(m-M)_0$\tablenotemark{d}} &  
\multicolumn{1}{c}{$W_0$} &\multicolumn{1}{c}{$(m-M)_0$\tablenotemark{e}} &
\multicolumn{1}{c}{$M_V$\tablenotemark{f}} & \multicolumn{1}{c}{$(m-M)_0$\tablenotemark{f}} \\
\multicolumn{1}{c}{(1)} & \multicolumn{1}{c}{(2)} & 
\multicolumn{1}{c}{(3)} & \multicolumn{1}{c}{(4)} &
\multicolumn{1}{c}{(5)} & \multicolumn{1}{c}{(6)} &\multicolumn{1}{c}{(7)} &
\multicolumn{1}{c}{(8)} &  \multicolumn{1}{c}{(9)} &
\multicolumn{1}{c}{(10)} & \multicolumn{1}{c}{(11)} }
\startdata
 V2  & $-$1.299 & 0.306 & 0.087 & 0.051 & 0.822 & 14.390 & $-$0.066 & 14.255 & 0.519 & 14.690  \\ 
 V12 & $-$1.280 & 0.309 & 0.078 & 0.047 & 0.876 & 14.403 & $-$0.020 & 14.294 & 0.616 & 14.660  \\ 
 V13 & $-$1.270 & 0.331 & 0.071 & 0.049 & 0.885 & 14.480 & $-$0.099 & 14.412 & 0.618 & 14.744  \\ 
 V14\tablenotemark{g} & $-$1.244 & 0.362 & 0.086 & 0.070 & 0.790 & 14.270 & $-$0.309 & 14.176 & 0.494 & 14.563  \\ 
 V17 & $-$1.414 & 0.334 & 0.093 & 0.070 & 0.838 & 14.476 & $-$0.153 & 14.338 & 0.584 & 14.727  \\ 
 V18 & $-$1.492 & 0.334 & 0.084 & 0.060 & 0.846 & 14.427 & $-$0.169 & 14.341 & 0.585 & 14.684  \\ 
 V23 & $-$1.108 & 0.378 & 0.092 & 0.081 & 0.858 & 14.429 & $-$0.318 & 14.343 & 0.573 & 14.711  \\ 
 V28 & $-$1.261 & 0.321 & 0.078 & 0.052 & 0.880 & 14.383 & $-$0.063 & 14.284 & 0.630 & 14.630  \\ 
 V29 & $-$1.298 & 0.319 & 0.086 & 0.060 & 0.854 & 14.390 & $-$0.078 & 14.262 & 0.587 & 14.654  \\ 
 V31 & $-$1.220 & 0.365 & 0.088 & 0.074 & 0.832 & 14.434 & $-$0.273 & 14.329 & 0.551 & 14.711  \\ 
NV2  & $-$1.161 & 0.320 & 0.072 & 0.040 & 0.845 & 14.263 & $-$0.117 & 14.205 & 0.512 & 14.592  \\ 
NV3  & $-$0.827 & 0.380 & 0.101 & 0.092 & 0.916 & 14.247 & $-$0.248 & 14.115 & 0.617 & 14.542  \\ 
NV5  & $-$1.556 & 0.343 & 0.085 & 0.066 & 0.837 & 14.376 & $-$0.189 & 14.270 & 0.606 & 14.604  \\ 
NV6  & $-$1.253 & 0.348 & 0.093 & 0.074 & 0.836 & 14.389 & $-$0.204 & 14.257 & 0.551 & 14.670  \\ 
NV10 & $-$1.208 & 0.341 & 0.086 & 0.066 & 0.847 & 14.458 & $-$0.175 & 14.352 & 0.565 & 14.736  \\ 
NV12 & $-$1.372 & 0.297 & 0.074 & 0.039 & 0.875 & 14.424 & $-$0.036 & 14.308 & 0.639 & 14.656  \\ 
NV13 & $-$1.207 & 0.317 & 0.074 & 0.044 & 0.894 & 14.469 & $-$0.037 & 14.383 & 0.594 & 14.766  \\ 
NV14 & $-$1.339 & 0.376 & 0.133 & 0.122 & 0.844 & 14.582 & $-$0.284 & 14.328 & 0.589 & 14.834  \\ 
\hline
Mean     & $-$1.267 &       & 0.087 & 0.064 & 0.854 & 14.405 && 14.292  & 0.579 & 14.676   \\
$\sigma$ &    0.156 &       & 0.014 & 0.021 & 0.029 & ~0.083 && ~0.074  & 0.041 & ~0.074   \\
\hline
Mean\tablenotemark{h}     & $-$1.271 &       & 0.086 & 0.061 & 0.852 & 14.418 && 14.302  & 0.581 & 14.684  \\
$\sigma$\tablenotemark{h} &    0.086 &       & 0.007 & 0.014 & 0.020 & ~0.032 && ~0.037  & 0.038 & ~0.036  \\
\enddata
\tablenotetext{a}{Zinn \& West metallicity scale.}
\tablenotetext{b}{Using $\langle(B-V)_0\rangle_\mathrm{mag}$.}
\tablenotetext{c}{Using $\langle(B-V)_0\rangle_\mathrm{s}$.}
\tablenotetext{d}{Using Equation (\ref{eq:mv:1}).}
\tablenotetext{e}{Using Equation (\ref{eq:W0}).}
\tablenotetext{f}{Using Equation (\ref{eq:mv:3}) with $K$ = 0.36.}
\tablenotetext{g}{See Appendix~\ref{ap:s:notes}}
\tablenotetext{h}{Without V13, V14, V18, NV2, NV3, NV5, NV13 and NV14.}
\end{deluxetable}
\end{landscape}

\clearpage

\begin{deluxetable}{lrcccl}
\tablecaption{Physical parameters for NGC~6723 RRc variables.\label{tab:rrcpam}}
\tablenum{6}
\tablewidth{0pc} 
\tablehead{
\multicolumn{1}{c}{ID} & \multicolumn{1}{c}{[Fe/H]} & 
\multicolumn{1}{c}{$M_V$} &
\multicolumn{1}{c}{$(m-M)$} &
\multicolumn{1}{c}{$(m-M)_0$}}
\startdata
V7       &  $-$0.879  & 0.804 & 14.710 & 14.515 \\
V10      &  $-$1.216  & 0.870 & 14.596 & 14.401 \\
V22      &  $-$1.119  & 0.810 & 14.689 & 14.494 \\
V24      &  $-$1.217  & 0.802 & 14.699 & 14.504 \\
V32      &  $-$1.307  & 0.845 & 14.550 & 14.355 \\
NV7      &  $-$1.298  & 0.829 & 14.614 & 14.419 \\
NV11     &  $-$1.214  & 0.701 & 14.841 & 14.646 \\
\hline
Mean     &  $-$1.179  & 0.809 & 14.671 & 14.476 \\
$\sigma$ &     0.146  & 0.053 &  ~0.096 & ~0.096\\
\hline
Mean\tablenotemark{a}   & $-$1.175 & 0.820  & 14.643 & 14.448 \\
$\sigma$\tablenotemark{a} &  0.202 & 0.021  & ~0.076 & ~0.076 \\
\enddata
\tablenotetext{a}{Without V10, V22 and NV11.}
\end{deluxetable}

\clearpage

\begin{deluxetable}{lccc}
\tablecaption{The distance modulus of NGC~6723 from various methods.\label{tab:dm}}
\tablenum{7}
\tablewidth{0pc} 
\tablehead{
\multicolumn{1}{c}{Method} & 
\multicolumn{1}{c}{$(m-M)_0$} &
\multicolumn{1}{c}{$E(B-V)$} & 
\multicolumn{1}{c}{$d$ (kpc)}}
\startdata
Fourier (RRab)\tablenotemark{a} &  14.681 $\pm$ 0.038 & 0.063   & 8.63 $\pm$ 0.15 \\
Fourier (RRab)\tablenotemark{b} &  14.684 $\pm$ 0.036 & 0.063   & 8.65 $\pm$ 0.14 \\
Fourier (RRc)\tablenotemark{c}  &  14.648 $\pm$ 0.076 & 0.063   & 8.50 $\pm$ 0.31 \\
$W_0(BV)$\tablenotemark{a}      &  14.572 $\pm$ 0.042 & \nodata & 8.21 $\pm$ 0.16 \\
$W(BV)_{th}$\tablenotemark{d}   &  14.610 $\pm$ 0.131 & \nodata & 8.36 $\pm$ 0.50 \\
$\langle V\rangle_\mathrm{int} - M_V(\mathrm{RR})$ 
                   &  14.686 $\pm$ 0.143 & 0.063   & 8.65 $\pm$ 0.57 \\
\hline
Mean  & 14.647 $\pm$ 0.047 &  & 8.47 $\pm$ 0.17 \\
\enddata
\tablenotetext{a}{Applying the zero-point correction term, 0.270 $\pm$ 0.020 mag,
in Equation~(\ref{eq:mv:1}).}
\tablenotetext{b}{Applying the zero-point correction term using $K$ = 0.36 mag 
in Equation~(\ref{eq:mv:3}).}
\tablenotetext{c}{Applying the zero-point correction term by \cite{cacciari05}
0.20 $\pm$ 0.02 mag.}
\tablenotetext{d}{Applying the zero-point correction term, 0.080 $\pm$ 0.130 mag.}
\end{deluxetable}

\clearpage

\begin{deluxetable}{lccccccl}
\tablecaption{Parameters for the SX Phe and $\delta$ Sct stars.\label{tab:sxphe}}
\tablenum{A1}
\tablewidth{0pc} 
\tablehead{
\multicolumn{1}{c}{ID} & 
\multicolumn{1}{c}{$r$ (\arcsec)} & 
\multicolumn{1}{c}{$M_V$} & 
\multicolumn{1}{c}{$(B-V)_0$} & 
\multicolumn{1}{c}{$E(B-V)$} & 
\multicolumn{1}{c}{$(m-M)_0$} & 
\multicolumn{1}{c}{$d$ (kpc)} &
\multicolumn{1}{c}{Note} }
\startdata
NV19  &  ~95  &  2.667  &  0.215  &  0.109  &  14.638  &  8.46  & Member \\
\multicolumn{8}{c}{}\\
NV16  &  495  &  2.476  &  0.233  &  0.072  &  13.939  &  6.14  & Non-member \\
NV18  &  111  &  2.399  &  0.237  &  0.057  &  14.001  &  6.31  & Non-member \\
\enddata
\end{deluxetable}

\clearpage

\begin{deluxetable}{lccccl}
\tablecaption{Parameters for W UMa type eclipsing binaries.\label{tab:eb}}
\tablenum{A2}
\tablewidth{0pc} 
\tablehead{
\multicolumn{1}{c}{ID} & 
\multicolumn{1}{c}{$r$ (\arcsec)} & 
\multicolumn{1}{c}{$M_V$} & 
\multicolumn{1}{c}{$(m-M)_0$} & 
\multicolumn{1}{c}{$d$ (kpc)}  &
\multicolumn{1}{c}{Note} }
\startdata
NV15  &  521  &  4.377  &  11.056  &  1.63  & Non-member \\
NV17  &  332  &  3.939  &  12.456  &  3.10  & Non-member \\
NV20  &  706  &  3.854  &  13.711  &  5.52  & Non-member \\
NV21  &  729  &  5.926  &  12.315  &  2.90  & Non-member \\
NV22  &  678  &  4.372  &  14.165  &  6.80  & Non-member \\
\enddata
\end{deluxetable}

\clearpage

\clearpage

\clearpage

\clearpage

\clearpage

\clearpage

\begin{figure}
\epsscale{1}
\figurenum{6}
\plotone{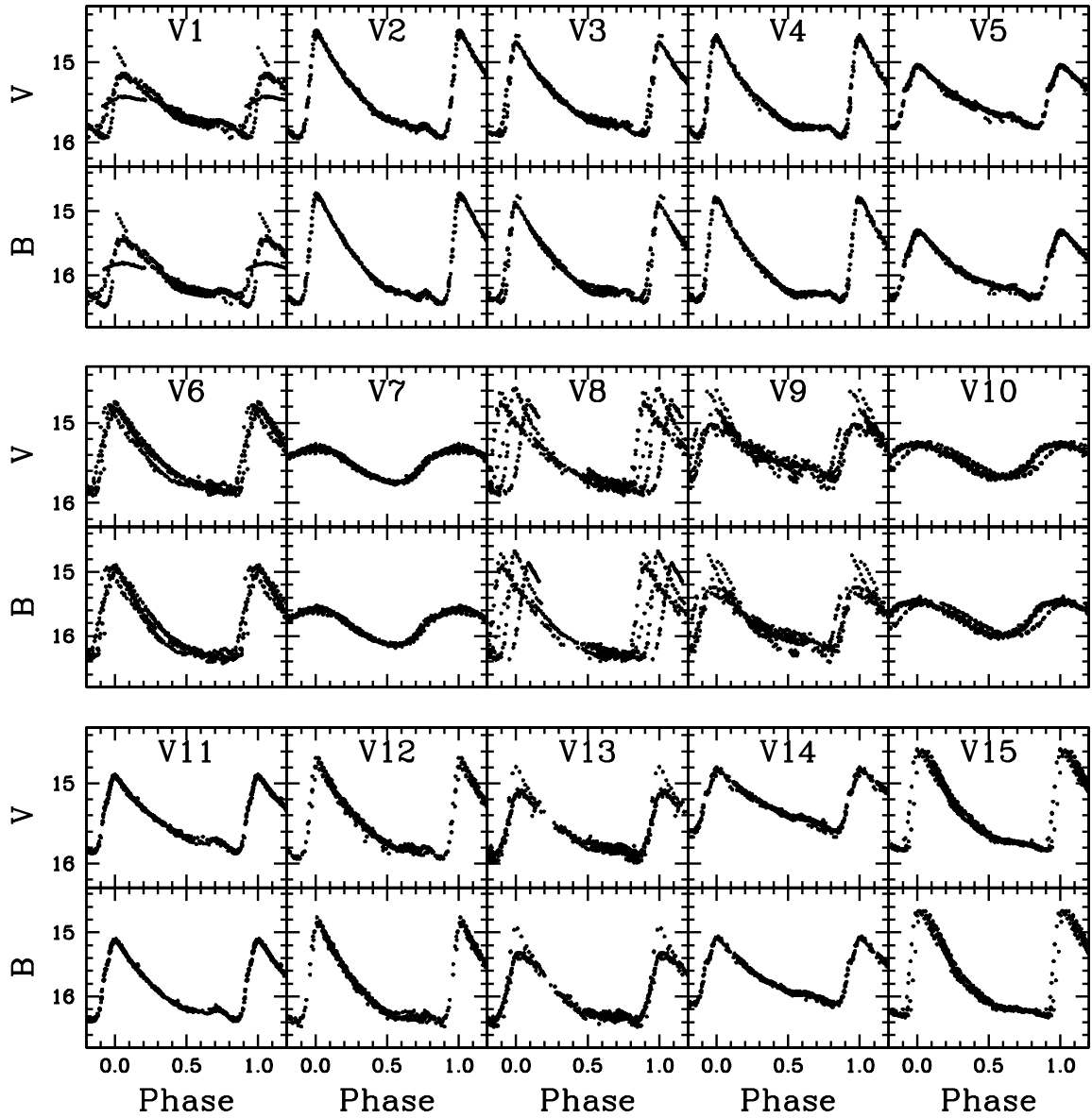}
\caption{Light curves for RR Lyrae variables.}
\label{fig:rrlc}
\end{figure}

\clearpage

\begin{figure}
\epsscale{1}
\figurenum{6}
\plotone{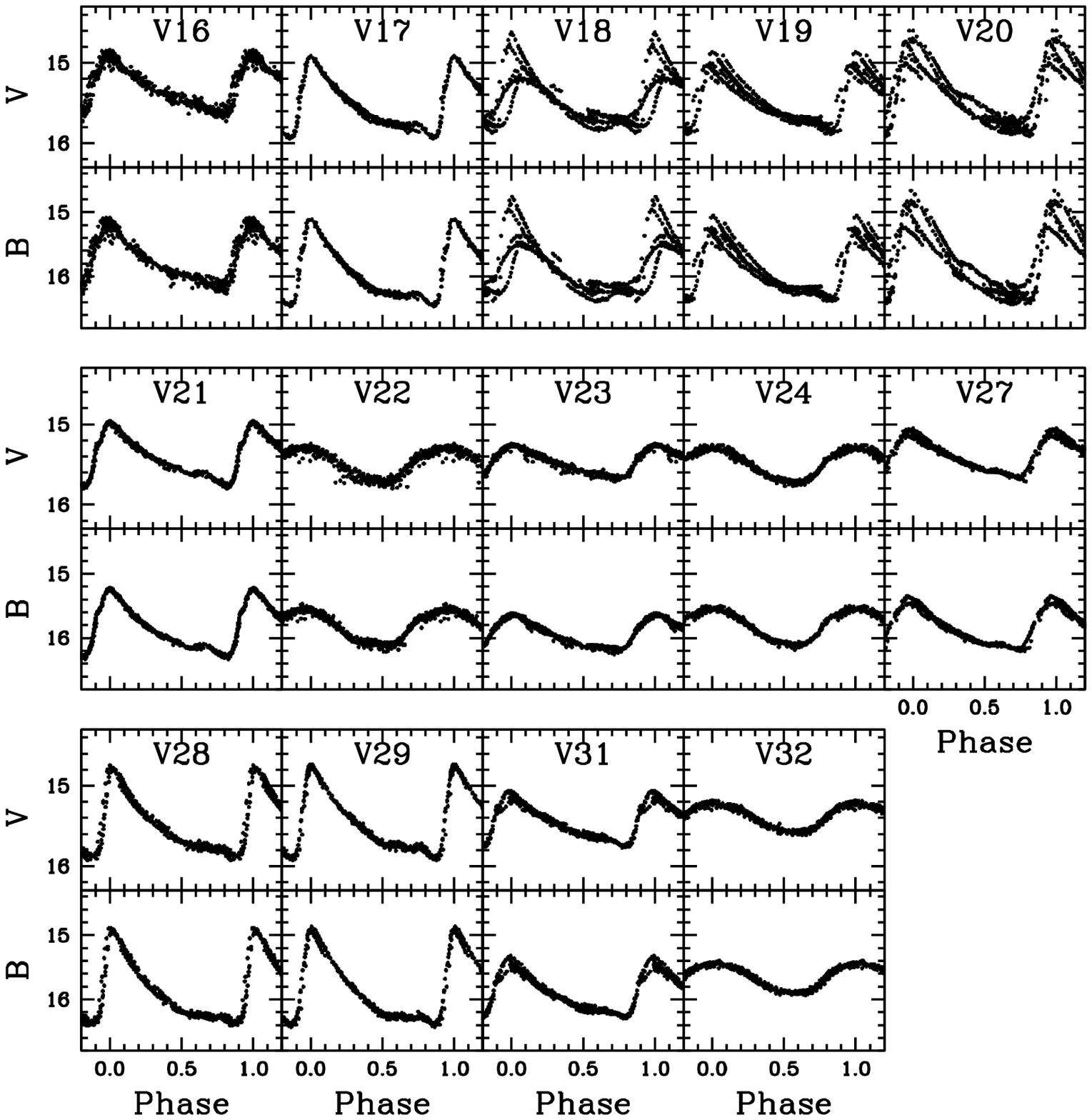}
\caption{Continued.}
\end{figure}

\clearpage

\clearpage

\clearpage

\begin{figure}
\epsscale{1}
\figurenum{9}
\plotone{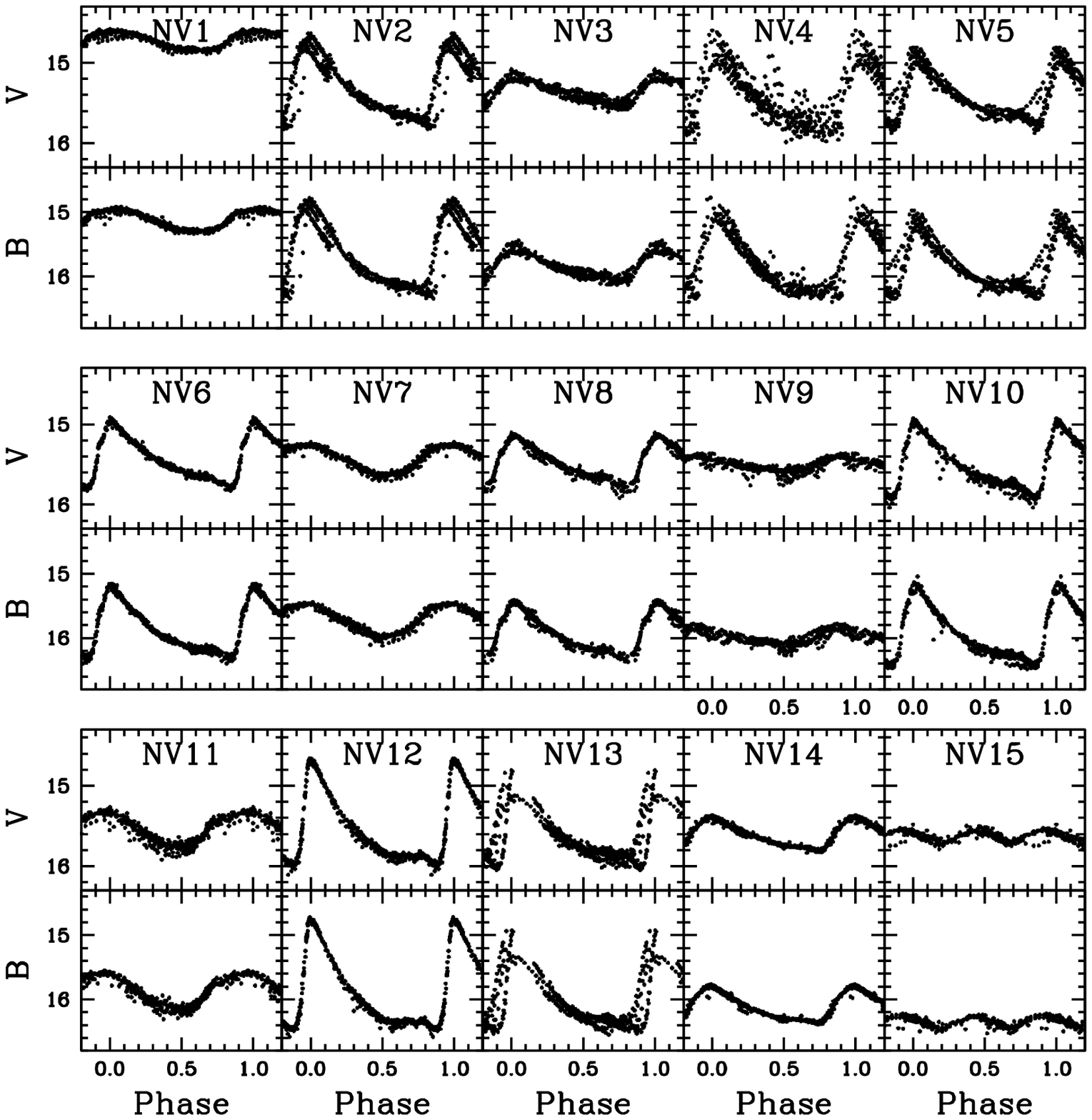}
\caption{Light curves for new variables discovered in this work.}
\label{fig:newrrlc}
\end{figure}

\clearpage

\begin{figure}
\epsscale{1}
\figurenum{9}
\plotone{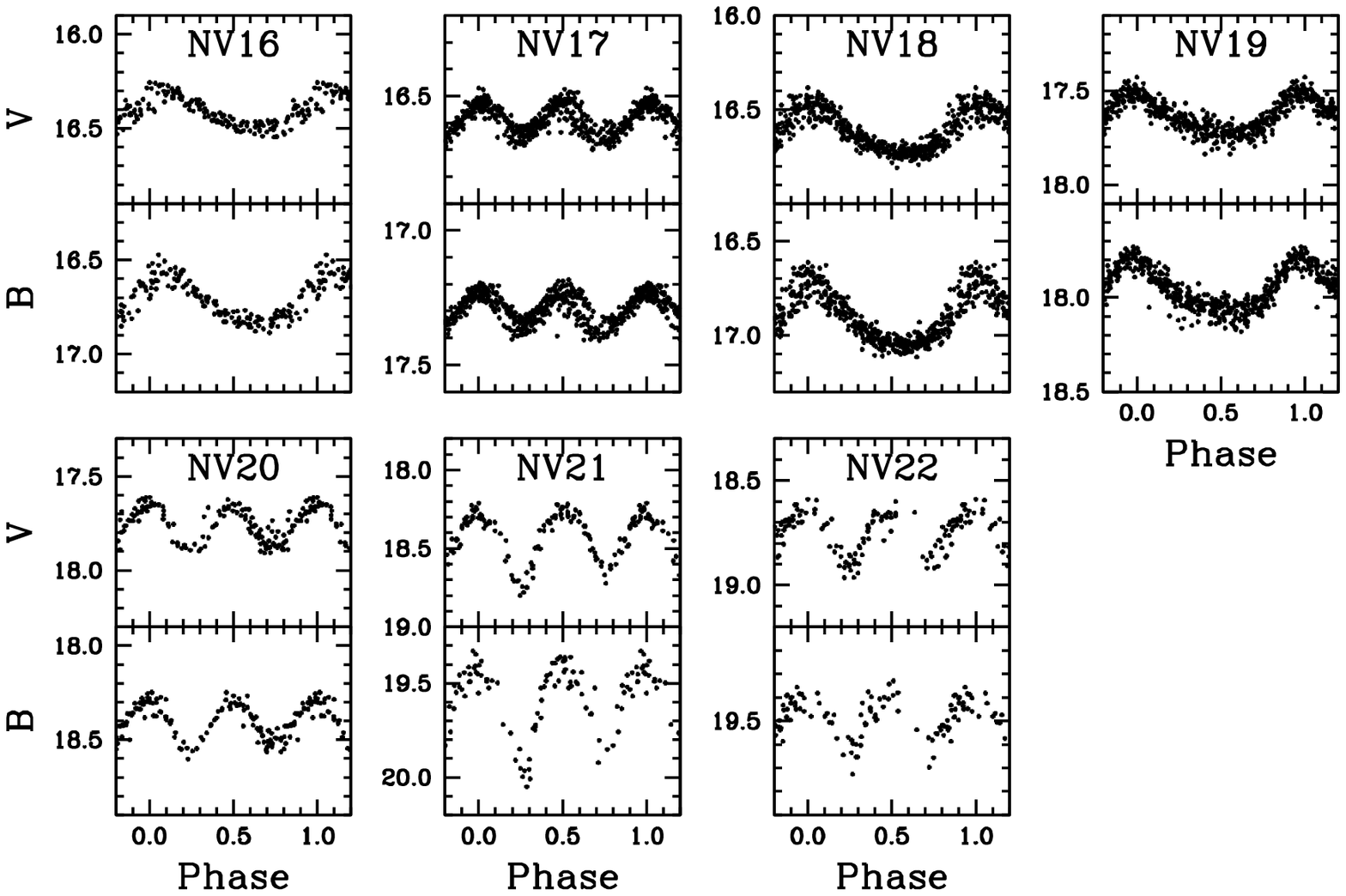}
\caption{Continued.}
\end{figure}

\clearpage

\clearpage

\clearpage

\clearpage

\clearpage

\clearpage

\clearpage

\clearpage

\clearpage

\clearpage

\clearpage

\clearpage

\clearpage

\clearpage

\end{document}